\documentclass[twocolumn]{aastex631}
\received{x}
\revised{x}
\accepted{x}
\submitjournal{ApJ}
\usepackage{fancyvrb} 
\usepackage{savesym}
\usepackage{amsmath}
\savesymbol{tablenum}
\usepackage{siunitx}
\restoresymbol{SIX}{tablenum}
\VerbatimFootnotes    

\begin{document}

\newcommand{\vdag}{(v)^\dagger}
\newcommand\aastex{AAS\TeX}
\newcommand\latex{La\TeX}

\newcommand\fj[1]{\textcolor{teal}{\textbf{[FJ: #1]}}}

\shorttitle{Precession-induced AGN variability}
\shortauthors{Britzen et al.}
\graphicspath{{./}{figures/}}

\title{Precession-induced Variability in AGN Jets and OJ\,287}

\author[0000-0001-9240-6734]{Silke Britzen}
\affiliation{Max-Planck-Institut für Radioastronomie,
Auf dem Hügel 69, 53121 Bonn, Germany}

\author[0000-0001-6450-1187]{Michal Zaja\v{c}ek}
\affiliation{Department of Theoretical Physics and Astrophysics, Faculty of Science, Masaryk University
 Kotl\'a\v{r}sk\'a 2, 611 37 Brno, Czech Republic}
 
\author[00000-0000-0000-0000]{Gopal-Krishna}
\affiliation{UM-DAE Centre for Excellence in Basic Sciences,
Vidyanagari, Mumbai-400098, India}

\author[0000-0002-3528-7625]{Christian Fendt}
\affiliation{Max Planck Institute for Astronomy,
K\"onigstuhl 17, 69117 Heidelberg, Germany}

\author[0000-0003-2769-3591]{Emma Kun}
\affiliation{Theoretical Physics IV: Plasma-Astroparticle Physics, Faculty for Physics \& Astronomy, Ruhr University Bochum, 44780 Bochum, Germany}
\affiliation{Ruhr Astroparticle And Plasma Physics Center (RAPP Center), Ruhr-Universität Bochum 44780 Bochum, Germany}
\affiliation{Astronomical Institute, Faculty for Physics \& Astronomy, Ruhr University Bochum, 44780 Bochum, Germany}
\affiliation{Konkoly Observatory, ELKH Research Centre for Astronomy and Earth Sciences, H-1121 Budapest, Konkoly Thege Miklós út 15-17., Hungary}
\affiliation{CSFK, MTA Centre of Excellence, Konkoly Thege Miklós út 15-17., Hungary}
\author[0000-0002-7449-0888]{Frédéric Jaron}
\affiliation{Institute of Geodesy and Geoinformation, University of Bonn,
Nu{\ss}allee 17, 53115 Bonn, Germany}
\affiliation{Department of Space, Earth and Environment, Chalmers University of Technology, Observatoriev\"agen 90, 439 92 Onsala, Sweden}
\affiliation{Department of Geodesy and Geoinformation, TU Wien,
 Wiedner Hauptstraße 8-10, 1040 Vienna, Austria}

\author[00000-0000-0000-0000]{Aimo Sillanp\"a\"a}
\affiliation{Tuorla Observatory, University of Turku,
Pikki\"o, Finland}

\author[0000-0001-6049-3132]{Andreas Eckart}
\affiliation{ I.Physikalisches Institut der Universität zu Köln,
Zülpicher Str. 77, 50937 Köln, Germany}

\begin{abstract}
 The combined study of the flaring of Active Galactic Nuclei (AGN) at radio wavelengths and pc-scale jet kinematics with Very Long Baseline Interferometry (VLBI) has
led to the view that i) the observed flares are associated with ejections of synchrotron blobs from the core, and ii)
most of the flaring would follow a one-to-one correlation with the component ejection.
Recent results have provided mounting evidence that the quasi-regular component injections into the relativistic
jet may not be the only cause of the flux variability. We propose that AGN flux variability and jet morphology changes can both be of deterministic nature, i.e. having a geometric/kinetic origin
linked to the time-variable Doppler beaming of the jet emission as its direction changes due to precession (and nutation).
The physics of the underlying jet leads to shocks, instabilities, or to ejections of plasmoids. The appearance (morphology, flux, etc.) of the jet can, however, be strongly affected and modulated by precession.
We demonstrate this modulating power of precession for OJ\,287. For the first time, we show that the spectral state of the Spectral Energy Distribution (SED) can be directly related to the jet's precession phase.
We model the SED evolution and reproduce the precession parameters. Further, we apply our precession model to eleven prominent AGN.
We show that for OJ~287 precession seems to dominate the long-term variability ($\gtrsim 1\,{\rm yr}$) of the AGN flux, SED spectral state, and jet morphology, while stochastic processes affect the variability on short timescales ($\lesssim 0.2\,{\rm yr}$).
\end{abstract}

\keywords{Active Galactic Nuclei (16) --- Blazars (164) --- Relativistic Jets (1390) --- Galaxy mergers (608) --- Radio interferometry (1346) --- Gravitational waves (678)}


\section{Introduction}
\label{sec:intro}

The term blazar encompasses BL Lac Objects and Flat-Spectrum Radio Quasars into a subgroup of active galactic nuclei (AGN). 
The identification of blazars as jets of relativistic plasma oriented close to our line-of-sight has helped to establish 
the current paradigm of AGN variability \citep{blandfordrees}. 
Many fundamental questions regarding the jet phenomena, however, remain unanswered. 
In particular, key questions remain concerning the physical connections between the innermost constituents of an AGN - the 
compact jet, the accretion disk, and the central SMBH (see \citealt{2021bhns.confE...1K} for a recent review and references 
therein).

The central engine is directly linked to the issue of the origin of the temporal flux variability and the emission observed in different
wavebands across the electromagnetic spectrum. 
In order to probe this, radio monitoring campaigns of blazars and other AGN with single-dish telescopes have been increasingly 
combined with detailed radio interferometric monitoring of the morphological changes witnessed in parsec-scale jets. While the former yields extensive information on radio flux variability on different time scales and frequencies, the latter 
reveals the kinematics of the radio-emitting components in the parsec-scale jet. It is indeed necessary to combine the temporal and the structural information to make significant claims on the deterministic nature and/or the quasi-periodicity of the radio flares since light curves alone often exhibit false periodicities that are purely of stochastic nature rather than deterministic \citep[see e.g., ][]{2016MNRAS.461.3145V,2020A&A...634A.120A,2020ApJS..250....1T}.  

OJ~287 is one of the most promising candidates for harbouring a supermassive binary black hole (SMBBH)
(see  e.g. \citealt{aimo, valtonen2009, Britzen2018, gomez2021}), although there are other plausible scenarios to account
for its characteristic periodicities in the optical domain ($\simeq 11$ years). 
\citet{Britzen2018} find that its radio jet is precessing on a time scale of $\simeq 22-23$yr. The data is also consistent with an additional jet-axis rotation on an approximately yearly time scale. Besides jet rotation, nutation of the jet could also explain the observed second-order motion of the jet axis. The bulk jet precession explains the variability of the total continuum radio flux density via viewing angle changes, which leads to variable Doppler beaming. An SMBBH or Lense-Thirring precession (misaligned disc around a single BH) seem to be required to explain the time scale of the precessing motion. Such a scenario intrinsically connects the radio and the optical periods, with the radio period being twice the optical period when viewed close to the rotation axis. \citet{zhao} recently published the first GMVA and ALMA observations of OJ 287 observed at 3.5 mm. The radio structure resembles a precessing jet in projection which seems to confirm the precessing jet proposed by \citet{Britzen2018}.

In this paper, we focus on investigating the radio flux variability in conjunction with the observed structural changes in the parsec-scale radio jets of OJ~287 and other prominent AGN.
This has led us to propose a possible alternative to the currently popular paradigm that the dominant cause of blazar flux 
variability is the occurrence of internal shocks within the relativistic nuclear jets (the so-called ``shock-in-jet" model, see e.g. \citealt{MarscherGear85, Hughes1985, Qian1991}).

Here we argue that a major, if not dominant contribution to flux variations of most blazars on longer timescales, may come from the precession
effects (which are, in principle, of deterministic type). 
A number of observational findings as well as recent results of general relativistic magneto-hydrodynamic simulations reinforce 
this proposed explanation for AGN radio variability. 

The present work argues for jet precession, which, however, does not explain by itself the physical nature of jet components. Radio knots could manifest themselves as shocks in jets (see e.g. \citealt{MarscherGear85, Hughes1985, Qian1991}, 
instabilities (see e.g. \citealt{ferrari, hardee}), 
or plasmoid injections due to magnetic reconnection (see e.g. \citealt{comisso, Vourellis2019, nathanail, ripperda}).
Both mechanisms - shock-in-jet and precession - can operate at the same time. 
However, shocks are more likely if injections are always in the same direction. 
In case of precession, components are moving in different directions and do not collide and thus are less likely to produce shocks.
While the jet might be made of shocks or plasmoids, its wiggly structure and flux-density variability of 
the various jet features are very likely modulated (boosted) by precession. 
Without precession, jets would appear more straight (ignoring for the moment jet interactions or other phenomena that could make a jet turn), AGN variability aperiodic, and less pronounced. 
Stochastic processes will add to the variability, which predominantly appears organized by a deterministic process.

The approach of a precessing jet model was already applied to several sources to explain the radio variability as well as 
smoothed jet morphological changes
\citep{Abraham,AbrahamCarrara,AbrahamRomero,CaproniAbrahama2004, CaproniAbrahamb2004,Caproni2007,Caproni2013,Caproni2017}. 
We have successfully applied this model to the ``Rosetta stone'' blazar OJ~287 \citep{Britzen2018}, the central galaxy of the 
Perseus cluster 3C84 \citep{britzen3c}, as well as to blazars exhibiting high-energy neutrinos \citep{britzentxs,britzen_1502}. 
In addition, jet precession has been invoked for jetted tidal disruption events, where the infalling star is generally 
misaligned with respect to the equatorial plane of a rotating supermassive black hole \citep{2012PhRvL.108f1302S,2016MNRAS.455.1946F}.
This shows that jet precession is likely an integral part of AGN engines, which is not surprising since the main trigger 
for disk-jet precession is a secondary massive black hole (BH) or a misaligned accretion disk \citep{Abraham2018}, both of which are integral and recurring features of galaxy evolution.

We will first briefly describe our precessing ``nozzle'' (jet base) model originally developed for the BL~Lac object OJ\,287. The model directly relates the evolving VLBI structure (radio components in the parsec-scale jet) to the flux variability seen in the integrated light-curve. Thus, in this framework, both phenomena - flux variability and jet evolution - are coupled and represent two manifestations of the same physical process.
For OJ\,287, we show that the crossing of the projected reversal point is the most important precessional phase of the VLBI jet and it is related to a new phase in the Spectral Energy Distribution (SED).

In the following, we shall compare and discuss the most prominent reported cases of AGN where a coupling between flux variability and jet evolution may be at work (our own work, supplemented with data taken from the literature). We shall also examine the precession scenario from the perspective of the multi-wavelength spectral energy distribution (SED). 

We will then summarise our results for the other blazars in our sample and recapitulate some mechanisms proposed for the jet precession - the most prominent of them being the presence of a secondary BH in the proximity of the central engine. Finally, we discuss some possible implications of our findings to the AGN physics and potential follow-up studies.
While our focus here remains on cm-wavelength observations (VLBI and single-dish), we shall also mention some recent multi-wavelength observations of OJ\,287, which offer support to our model. 

The geometric modelling of the jet evolution has parallels in other contexts related to blazars.
For instance, we recall the population of Extreme High-frequency Peaked BL Lac objects (EHBLs) which emit a substantial fraction of their power in the GeV-TeV range (e.g., \citet{costamante2001}). The synchrotron component in the SEDs of these relatively low-power blazars peaks in the EUV, X-ray, or even the MeV band \citep{ghisellini}. The origin of this high-energy emission is yet to be well understood before the basic difference of these extreme blazars from the more typical ones can be grasped. 
For the jets of these blazars, VLBI measurements commonly reveal low (i.e., at most mildly superluminal) apparent speeds (e.g., \citealt{piner2004, piner2018, giroletti2004, sauge2004}). Such speeds are in stark contrast to the large bulk Lorenz factors ($\geq$50) commonly deduced for their jets, e.g., from gamma-ray transparency arguments (\citealt{tavecchio, henri2006}). In order to explain this huge discrepancy, a geometrical argument, namely stratified (spine-sheath) jet structures (e.g., \citealt{swain1998, chiaberge2000, laing,  giroletti2004, britzen_1502}) has often been invoked. It has also been shown that the inferences about the jet kinematics and viewing angle, made from observations of apparent speed and brightness of the VLBI knots and flux variability, can depend quite sensitively on another geometrical factor, namely a conical shape of the jet \citep[in particular uniform or stratified, see e.g.,][]{gopal2004, gopal2006, gopal2007, boutelier2011}.

Thus, it seems quite plausible that geometric effects, such as those explored here, can play a vital role in defining the general phenomenology of blazar jets. The manuscript is structured as follows. In Section~\ref{section_modelling_precession_nutation}, we introduce the precession-nutation model and apply it to radio jet components of OJ~287 as well as to its radio light curves. In Section~\ref{sec_precession_sed}, we show that the detected change of the SED shape of OJ~287 can be addressed by the Doppler factor variation as the jet precesses. In Section~\ref{section_modelling_precession_nutation}, we analyze power density spectra and periodograms of OJ~287. We can detect potentially significant periods in periodograms that correspond approximately to precession-nutation periods as well as to the putative orbital period of the SMBH binary or its multiples. In addition, we list precession-model parameters for 12 sources in Subsection~\ref{sec_comparison} and we compare Doppler factors inferred from precession model with those based on the ``shock-in-jet'' model in Subsection~\ref{sec_doppler_fac_prec_shock}. We discuss our results in the broader context in Section~\ref{sec_discussion} and conclude with Section~\ref{sec_conclusions}. 

\begin{figure*}
\centering
\includegraphics[width=\textwidth]{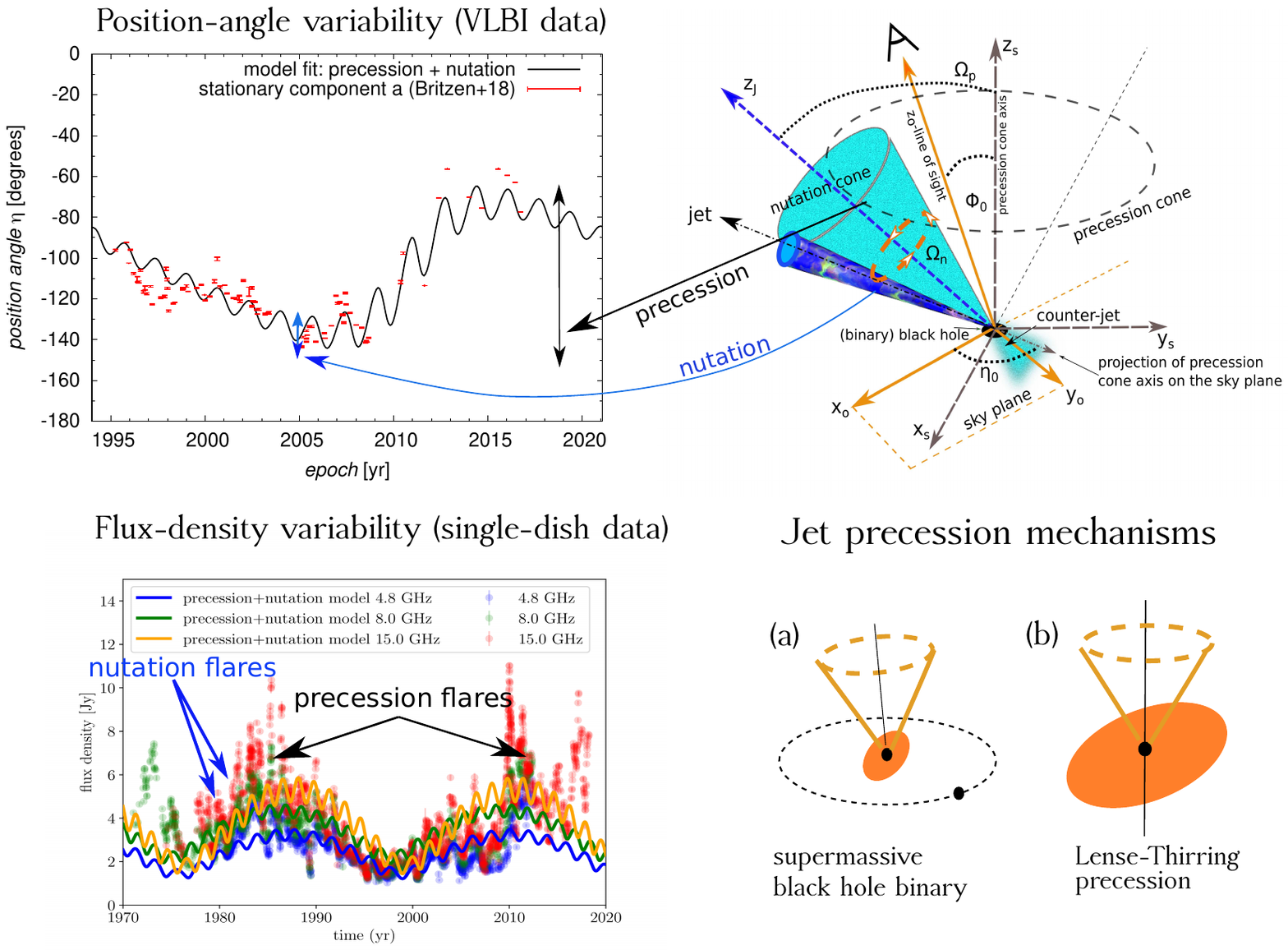}
\caption{This composite figure is based on results presented in \citet{Britzen2018} but modified to highlight the origin and
effects of precession and nutation. The top right figure indicates the kinds of jet motion that in combination produce the observed variability (precession + nutation). The top left panel shows the different timescales and effects on the position angle of one radio knot due to precession and nutation. The bottom left panel visualizes the effect of precession and nutation on the radio light-curves based on data obtained for OJ~287 (UMRAO single-dish data, combined with OVRO data at 15 GHz). The bottom right figure is a zoom-in of the central engine, showing potential drivers of the disk-jet precession (binary supermassive black hole or the Lense-Thirring precession due to the frame-dragging by the rotating Kerr black hole).}
\label{sketch}
\end{figure*}

\section{Modelling precession and nutation}
\label{section_modelling_precession_nutation}
Precession of AGN jets can generally be caused by a 
 misaligned binary supermassive black-hole system or the Lense-Thirring 
(hereafter LT) effect \citep{1918PhyZ...19...33T, 1918PhyZ...19..156L, Abraham2018}. 
The first effect is by nature classical -- the accretion disk orbiting the primary SMBH is perturbed by a secondary black
hole whose orbital plane is misaligned with respect to the accretion disk plane by angle $\Omega$. 
Due to induced torques, the accretion disk precesses which is followed by the jet precession along the precession cone 
with a half-opening angle of $\Omega$. 
The Lense-Thirring effect is by nature relativistic and is induced by the frame-dragging effect acting on the accretion disk 
misaligned with respect to the equatorial plane of the central Kerr black hole. 
Both effects can cause precession as well as nutation of the disk-jet system. 
By jet precession, we understand the periodic bulk motion of the jet with respect to the ambient medium, while nutation is 
understood as (i) the ``nodding motion" of the jet components with respect to the jet frame or (ii) as a secondary bulk motion of the jet with respect to the primary 
precession motion. The ``nodding motion" of the nutation and rotation may be hard to disentangle (in time and space).
Kinematically, these two cases of secondary, nutation-like motion are difficult to distinguish as they are manifested by 
the same modulation of the time-dependent Doppler-beaming factor with respect to the observer. 
The physical distinction depends on the type of intrinsic kinematics of jet components -- in case-(i) nutation components 
rotate in the jet frame, while in case-(ii) nutation components move rather linearly in the jet frame but they are rotated around the mean jet motion by an external perturbation, which leads to the same ``deferent--epicycle'' motion as in case (i).

Accreting binary systems composed of a compact object and a normal star can also be sources of jets \citep{Mirabel1999}. The microquasar SS433, with an orbital period of 13.1~days \citep{Crampton1980, Cherepashchuk1981}, is of particular interest in the context of our work, because the accretion disk and the jet in this source are known to exhibit both precession and nutation (see, e.g.,  \citealt{Fabrika2004} for a review).  Precession was first detected in the periodically Doppler shifted Balmer and He I emission lines \citep{Abell1979}. 

The precession-nutation model we will apply in this manuscript is of general kinematic nature and can be applied independently of the origin of the precession-nutation motion. Here we only discuss the kinematic model and do not investigate the reason of the precession nor do we consider different effects due to a BZ- \citep{znajek} or BP-\citep{payne}jet. We assume that the disk-jet system is intrinsically connected, and in particular that the disk precession leads to the jet precession. This assumption of the precessing disk-jet connection has recently been justified by the 3D general relativistic magnetohydrodynamic simulations \citep{liska}.

A concise mathematical description of our jet precession and nutation modelling was first applied in \citet{Britzen2018}. We here summarize the basic concept and refer the reader to \citet{Britzen2018} for details of the modeling. For a schematic view, we refer the reader to Fig.~\ref{sketch}, which illustrates the basic parameters of the model.
In order to model the jet precession and nutation in the observer's frame we apply a set of eight physical quantities (see Table~\ref{tab_parameters_precession_nutation}), specifically the precession period $P_{\rm p}$, the nutation period $P_{\rm n}$, the Lorentz factor $\gamma$, a half-opening angle $\Omega_{\rm p}$ of the precession cone, a half-opening angle $\Omega_{\rm n}$ of the nutation cone for case-(ii) nutation or it coincides with the jet half-opening angle for the jet rotation, the angle $\Phi_0$ between the line of sight and the precession-cone axis (viewing angle), and the position angle $\eta_0$ of the precession-cone axis projected on the sky plane. Finally, the jet position is expressed with respect to the reference epoch $t_0$ that is inferred from fitting the model to the data.

The parameters defining the geometry of the system -- in particular the angles 
$\Omega_{\rm p}$, $\Omega_{\rm n}$, $\Phi_0$, $\eta_0$ -- are depicted in Fig.~\ref{sketch} (top right panel). The bottom-right panel is a zoom-in into the central engine that depicts the two most common drivers of the jet precession -- (a) a secondary massive black hole whose orbital plane is misaligned with respect to the accretion-disk plane around the primary black hole and (b) the Lense-Thirring precession due to a frame dragging acting on a misaligned accretion disk with respect to the spin of a (single) SMBH \citep{Britzen2018}.

The nutation of the jet as a second-order precession effect can be observed as an extra rotation-like motion of jet components in the jet frame.
Therefore, the possible existence of nutation in precessing jets
was motivated by a rotation-like motion of the so-called stationary component \textbf{a} (see Fig.~5a in \citeauthor{Britzen2018}, \citeyear{Britzen2018}, for the plot on the motion perpendicular to the jet ridge line). The term \textit{stationary} is usually attributed to jet components which do not show any motion over longer periods of time, as in Mrk 501 (see, e.g., \citet{edwards}, \citet{britzen_mrk501}. For clarity, we call component \textbf{a} \textit{quasi-stationary} component in the following text. 

\subsection{Modelling of the jet components in OJ~287}
To verify the applicability of the jet-precession/nutation model, we refit the OJ~287 jet component viewing and position angles $\Phi$ and $\eta$ inferred for the component ejection times by \citet{1989ApJ...336L..59G}, \citet{1999ApJ...520..627T}, and \citet{Britzen2018}. Specifically, we fit $\beta_{\rm app}(t)$ and $\eta(t)$ simultaneously to the temporal evolution of the jet apparent velocities and the position angles at the time of the component ejection. The best-fit parameters are inferred using the global least-square fitting with the concatenated $\chi^2$-statistic, $\chi^2=(\chi^2_{\beta}, \chi^2_{\eta})$, where $\chi^2_{\beta}$ is the $\chi^2$-statistic corresponding to the apparent velocities and $\chi^2_{\eta}$ is the $\chi^2$-statistic corresponding to the position angles. The fitted parameters converge well, however, with a certain degeneracy close to the minimum $\chi^2$, i.e. more solutions are possible with different $\chi^2$ values. The parameter values are, however, comparable within the uncertainties, which justifies the application of the precession model for describing the temporal evolution of the OJ~287 VLBI data for a time period spanning 40 years.

\begin{figure*}[h!]
    \centering
    \includegraphics[width=0.9\textwidth]{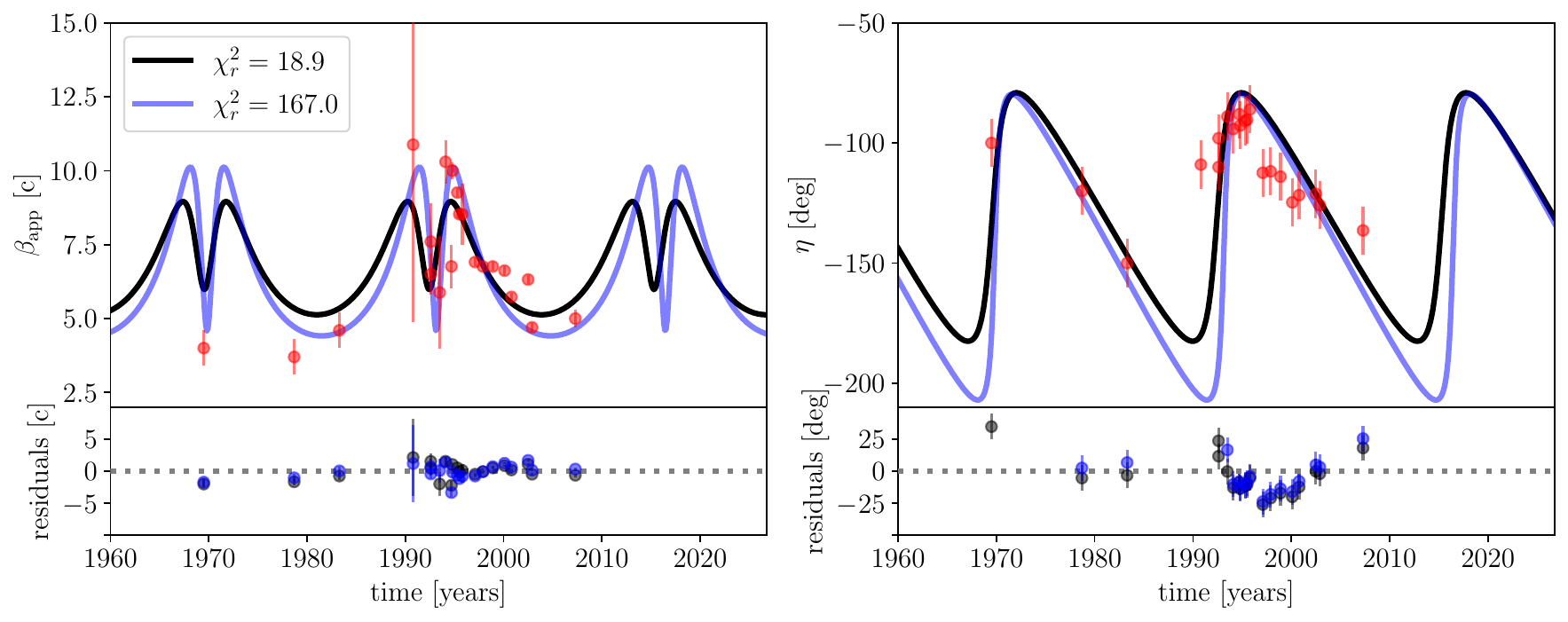}
    \caption{The best-fit solutions of the precession model applied to the apparent velocities (left panel; expressed in terms of the speed of light) and the position angles (right panel; expressed in degrees) of the jet components of OJ~287. One model solution with the reduced $\chi^2_{\rm r}=18.9$ ($\chi^2_{\rm r}=\chi^2_{\rm r,\beta}+\chi^2_{\rm r,\eta}=12.7+6.2$ with 15 and 16 degrees of freedom for $\beta$ and $\eta$ fitting, respectively) is depicted by a solid black line, while the other solution with $\chi^2_{\rm r}=167.0$ ($\chi^2_{\rm r}=\chi^2_{\rm r,\beta}+\chi^2_{\rm r,\eta}=25.4+141.6$ with 15 and 16 degrees of freedom for $\beta$ and $\eta$ fitting, respectively) is represented by a solid blue line. The bottom panels show the corresponding residuals (data-model). The best-fit parameters are listed in Table~\ref{tab_parameters_precession_nutation}. While the black solution provides a formally better fit, the residuals are comparable as well as the best-fit parameters within the uncertainties, see Table~\ref{tab_parameters_precession_nutation}}.
    \label{fig_precession_bappa_eta}
\end{figure*}
\begin{table*}[h!]
\caption{List of parameters for the precession-only and the combined precession-nutation jet model of OJ\,287 based on the refitted model in this work and the original analysis presented in \citet{Britzen2018}. For the precession-nutation model, we include the specific values of the parameters inferred for the quasi-stationary component \textbf{a} in parantheses.}
    \centering
    \begin{tabular}{c|c|c|c}
    \hline
    \hline
    Parameter & precession model $\chi_{\rm r}^2=18.9$ & precession model $\chi_{\rm r}^2=167.0$ & precession-nutation model \citep{Britzen2018} \\
    \hline
    $t_0$ [yr] & $1998.2 \pm 1.5$ & $1999.0 \pm 0.6$ & $1998.9\pm 0.9$ \\
    $P_{\rm p}$ [yr] & $22.9 \pm 3.4$  & $23.3 \pm 1.1$ & $23.6 \pm 2.5$ ($31.8 \pm 2.0$) \\
    $P_{\rm n}$ [yr]    & - & - & $1.6 \pm 0.1$ \\
    $\gamma$ & $9.0 \pm 1.4$  & $10.2 \pm 2.1$ & $9.6 \pm 1.6$\\
    $\Omega_{\rm p}$ [deg] & $8.8 \pm 2.9$  & $11.5 \pm 4.4$ & $10.3 \pm 3.6$ ($15.3 \pm 4.0$) \\
    $\Omega_{\rm n}$ [deg] & - & - & $2.7 \pm 1.0$ \\
    $\Phi_0$ [deg] & $11.3 \pm 2.9$ & $12.8 \pm 3.8$ & $12.2\pm 3.2$ ($27.7 \pm 7.6$)\\
    $\eta_0$ [deg] & $-130.8 \pm 46.3$  & $-143.3 \pm 63.2$ & $-148.9 \pm 52.2$ ($-99.1 \pm 7.6$)\\
    \hline
    \end{tabular}
    \label{tab_parameters_precession_nutation}
\end{table*}

To illustrate this better, we include two solutions, one with the reduced $\chi^2_{\rm r}=\chi^2_{\nu,\beta}+\chi^2_{\nu, \eta}$ of $18.9$ and the other with $\chi^2_{\rm r}=167.0$. The number of the degrees of freedom for the apparent velocities is 15, while for the position angles, it is 16. The application of the precession model to the data is graphically depicted in Fig.~\ref{fig_precession_bappa_eta}, where we also highlight the two exemplary fits according to the legend. The best-fit parameter values for both solutions are listed in Table~\ref{tab_parameters_precession_nutation}. These two fits, although differing considerably in terms of $\chi^2_{\rm r}$, have consistent precession parameters within 1$\sigma$ uncertainties.

\begin{figure*}[h!]
    \centering
    \includegraphics[width=0.8\textwidth]{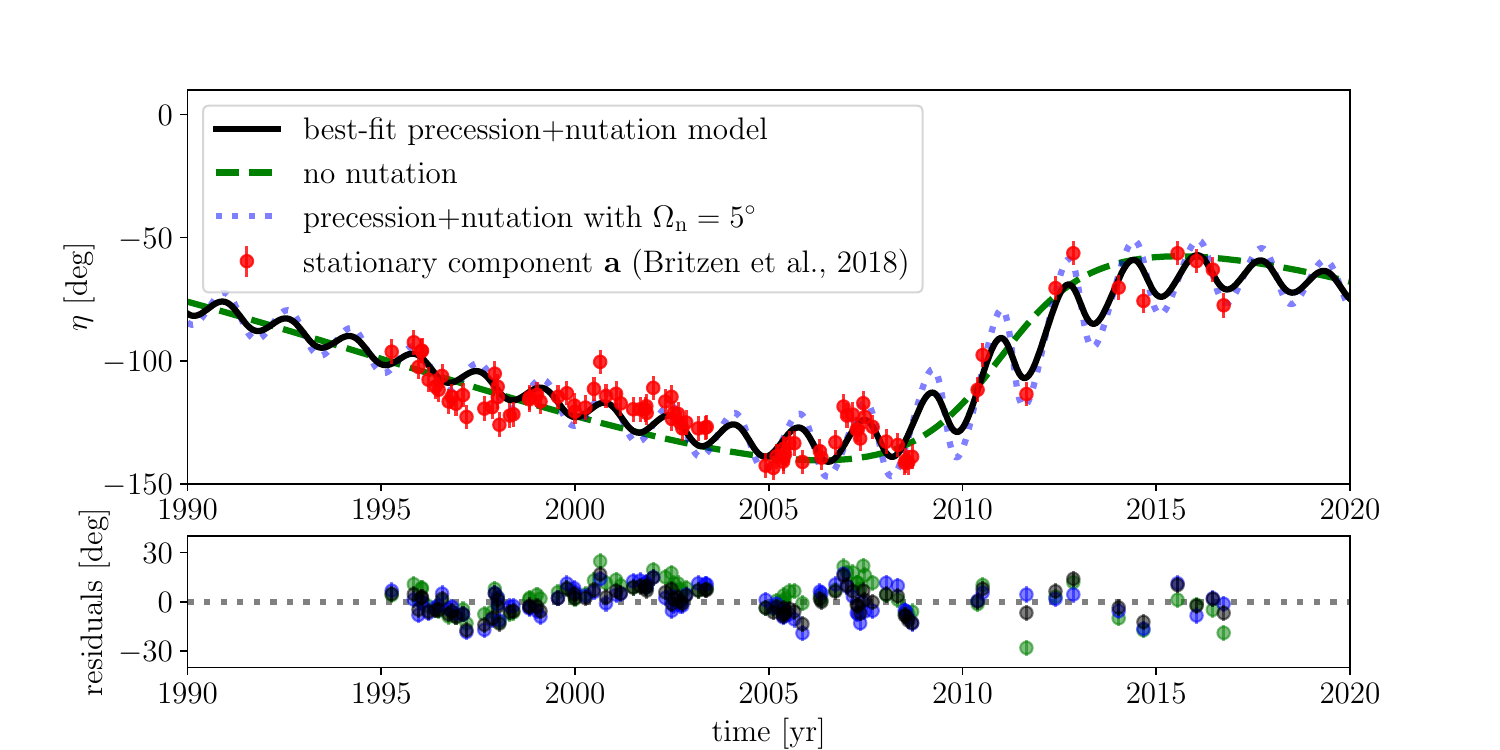}
    \caption{The precession-nutation model fitted to the position-angle evolution of the quasi-stationary component \textbf{a}. The reduced $\chi^2_{\rm r}$ of the presented model fit is $2.3$ for 86 degrees of freedom. The inferred best-fit precession period for this component is $P_{\rm p}=31.8 \pm 2.0$ years, while the nutation period is $P_{\rm n}=1.6 \pm 0.1$ years. For comparison, we also depict two other solutions -- a model with the same best-fit parameters but no nutation (green dashed line) and a model with the same best-fit parameters but a nearly double half-opening nutation-cone angle ($\Omega_{\rm n}=5^{\circ}$; blue dotted line). The bottom panel shows the fit residuals (data-model) for all three cases.}
    \label{fig_precession_nutation_a}
\end{figure*}

In Fig.~\ref{sketch} (top left panel), we use Eq.~(8) from \citet{Britzen2018} to fit the observed position-angle evolution of the quasi-stationary component $\mathbf{a}$ and we obtain $P_{\rm n}$ and $\Omega_{\rm n}$ (see Table~\ref{tab_parameters_precession_nutation}) in addition to the precession parameters that are consistent within uncertainties with the precession-only model \citep[see also][]{Britzen2018}, except for the precession period and the viewing angle of the precession axis that differ by more than 1$\sigma$. These differences can be interpreted by the lack of the velocity information that is not included in the fit since this component appears to be stationary with regard to the radial distance from the core. The positional-angle fit is shown along with the residuals in Fig.~\ref{fig_precession_nutation_a} in more detail. The best-fit $\chi^2_{\rm r}$ is $2.3$ for 86 degrees of freedom. When the evolution of the quasi-stationary component is fitted with the precession-only model, the residuals are clearly larger. The data appear to favour the half-opening angle of the nutation cone of at least $\sim 3^{\circ}$, potentially as much as $5^{\circ}$, see the model with $\Omega_{\rm n}=5^{\circ}$ in Fig.~\ref{fig_precession_nutation_a} (dotted blue line). To constrain the precession-nutation parameters more, the longer evolution of the component {\bf a} will be beneficial and/or detection of the nutation-like motion for more components.

Complementary to the combined least-square fitting, we apply the Bayesian analysis using Markov Chain Monte Carlo (MCMC) inference of precession and precession-nutation parameters. The advantage of the method is that based on the prior flat distributions of relevant parameters, we obtain marginalized posterior one-dimensional likelihood distributions and two-dimensional likelihood contours, which provide information about how tighly each parameter is constrained. To this goal, we maximize the likelihood function $\mathcal{L}$ using the form,
\begin{equation}
    \ln{\mathcal{L}}=-\frac{1}{2}\sum_{i=1}^{N}\left[ \frac{[q_{i}^{\rm obs}-q_{i}^{\rm theor}(\mathbf{p})]^2}{s_{\rm i}^2}+\ln{(2 \pi s_{i}^2)}\right]\,,
    \label{eq_likelihood_function}
\end{equation}
where $q_{i}^{\rm obs}$ are observed quantities (here the position angles and apparent velocities), $q_{i}^{\rm theor}(\mathbf{p})$ is the theoretical model that depends on precession and nutation parameters $\mathbf{p}$, $s_{i}^2=\sigma^2_{i,q}+\exp{(2\ln{f})}(q_{i}^{\rm theor})^2$ is the variance, which contains an underestimation factor $\ln{f}$ apart from the measurement errors $\sigma_{i,q}$. The quantity $q^{\rm theor}(\mathbf{p})$ depends on the precession-model parameters $\mathbf{p}$, which we already introduced. To find $\mathcal{L}_{\rm max}$ and infer $\mathbf{p}$, we apply the \texttt{python}-based MCMC solver \texttt{emcee}.

The non-zero flat priors and posterior likelihood distributions for each parameter are shown in Appendix~\ref{appendix_mcmc} separately for the case of the position angle and apparent velocity precession models. We also apply the MCMC method to infer precession-nutation parameters based on the temporal evolution of the quasi-stationary component \textbf{a}. The inferred posterior parameters for each case with 1$\sigma$ uncertainties are summarized in Table~\ref{tab_mcmc_summary}. We also list the obtained values of $\ln{\mathcal{L}_{\rm max}}$, the number of the model parameters $k$, the dataset size $N$, and Akaike and Bayesian information criteria defines as,
\begin{align}
   \text{AIC} &=-2\ln{\mathcal{L}_{\rm max}}+2 k\,,\notag\\
   \text{BIC} &=-2\ln{\mathcal{L}_{\rm max}}+k\ln{N}\,.
   \label{eqs_aic_bic}
\end{align}

Overall, the inferred precession-nutation parameters are consistent for all three datasets: the precession period is $\sim 20-30$ years, the half-opening angle of the precession cone is $\sim 10^{\circ}$, the viewing angle of the precession cone axis is close to the line of sight, $\Phi_0\sim 10^{\circ}-20^{\circ}$ or the similar value when subtracted from $180^{\circ}$, and the position angle of the precession cone axis less than $-100^{\circ}$. From the apparent velocities, we obtain a tight constraint on the Lorentz factor of the moving jet components, $\gamma\simeq 9$. From fitting the quasi-stationary component \textbf{a}, we obtain the nutation period of $P_{\rm n}\sim 1.6$ years and the half-opening angle of the nutation cone of $\Omega_{\rm n}\sim 2.2^{\circ}$. When we inspect the likelihood distributions in corner plots in Appendix~\ref{appendix_mcmc}, we see that the viewing angle $\Phi_0$ is not well determined for the position angle fitting. Both the viewing and the position angles are degenerate while fitting apparent velocities. On the other hand, when fitting the quasi-stationary component \textbf{a}, all the parameters are reasonably well determined. The prior values for the parameters were iteratively narrowed down during the comparison of median maximum-likelihood models with the actual data in Figs.~\ref{fig_mcmc_model_eta}, \ref{fig_mcmc_model_bapp}, and \ref{fig_mcmc_model_compa} when fitting position angles, apparent velocities, and the quasi-stationary component position angles, respectively.   

\begin{table*}[h!]
    \centering
    \caption{Summary of the MCMC-inferred precession and precession-nutation parameters based on the observed evolution of the position angles $\eta$, the apparent velocities $\beta_{\rm app}$, and the position angle evolution of the quasi-stationary component \textbf{a}. We also include the information about $\ln{\mathcal{L_{\rm max}}}$, AIC, and BIC statistical parameters.}
    \label{tab_mcmc_summary}
    \begin{tabular}{c|c|c|c}
    \hline
    \hline
    Parameter & $\eta\,[^{\circ}]$ & $\beta_{\rm app}\,[c]$ & $\eta_{a}\,[^{\circ}]$  \\
    \hline
    $t_{0}\,[{\rm yr}]$  & $1995.40^{+0.81}_{-0.99}$   & $2000.95^{+3.27}_{-29.68}$    &  $2010.91^{+1.18}_{-0.07}$     \\
    $P_{\rm p}\,[{\rm yr}]$  & $23.83^{+2.42}_{-1.81}$   & $34.52^{+4.13}_{-7.58}$     & $31.61^{+2.82}_{-1.25}$      \\ 
    $P_{\rm n}\,[{\rm yr}]$  &  -  &  -   &  $1.58^{+5.88}_{-0.02}$     \\  
    $\gamma$  & -   &  $8.96^{+2.89}_{-0.67}$   &  -     \\
    $\Omega_{\rm p}\,[^{\circ}]$  & $6.47^{+1.77}_{-1.55}$   & $12.36^{+8.37}_{-4.17}$    &  $11.73^{+11.06}_{-6.19}$     \\
    $\Omega_{\rm n}\,[^{\circ}]$  & -   &  -   &  $2.23^{+1.97}_{-1.04}$     \\
    $\Phi_{0}\,[^{\circ}]$  & $15.77^{+2.98}_{-3.70}$   &  $11.93^{+2.30}_{-1.53}$    &  $156.25^{+10.63}_{-29.38}$     \\  
    $\eta_{0}\,[^{\circ}]$  & $-120.22^{+3.87}_{-3.97}$   & $-124.48^{+17.41}_{-17.69}$     &  $-99.45^{+3.59}_{-1.97}$     \\   
     $\ln{f}$  & $-6.74^{+2.36}_{-2.19}$    &  $-1.96^{+0.30}_{-0.27}$   & $-2.98^{+0.21}_{-0.18}$      \\  
    \hline 
    $\ln{\mathcal{L}_{\rm max}}$  & $12.41$    & $-35.85$    & $33.72$   \\
    Parameter number $k$ (model parameters $+$ $\ln{f}$) & 5+1 & 6+1   & 7+1 \\
    Data point number $N$ & 21  & 21 & 92\\
    AIC  & $-12.81$   & $85.70$    & $-51.44$   \\ 
    BIC  & $-6.55$   &  $93.01$   &  $-31.26$   \\
    \hline
    \end{tabular}
\end{table*}

\subsection{Modelling radio light curves of OJ~287}
The time-variable jet kinematics that is predicted by the precession and nutation translates into the observed flux 
density variations via the Doppler-boosting factor
\begin{equation}
\delta(t,\gamma,\Phi) = \left[ \gamma \left( 1-\beta\cos\left(\Phi(t)\right) \right) \right]^{-1}\,.
\label{eq_doppler_boosting}
\end{equation}
Assuming that the underlying intrinsic jet emission can be described by synchrotron emission, 
$S_{\nu,0} \propto \nu^{-\alpha}$ with the spectral index~$\alpha$, 
the time-variable Doppler-boosted flux density can be expressed as $S_{\nu}(t)=S_{\nu,0}\delta^{p+\alpha}$,
where $p$ is the geometry factor for boosting, which adopts the value of $p\simeq 2$ for a uniform cylindrical jet emission and $p\simeq 3$ for discrete, isotropically radiating jet components \citep{Abraham,2001ASPC..227..108A}; for the original derivation of the Doppler beaming, see \citet{1972MNRAS.157..359M} and \citet{1979ApJ...232...34B}. The modulation of $S_{\nu}(t)$ by the Doppler-boosting factor $\delta(t,\gamma,\Phi)$ leads to the continuum outbursts with the characteristic periodicity of $P_{\rm p}$ and $P_{\rm n}$ that can potentially be traced in the radio light curves of OJ~287 (see Fig.~\ref{sketch}, bottom left panel).

\begin{figure*}[h!]
    \centering
    \includegraphics[width=0.32\textwidth]{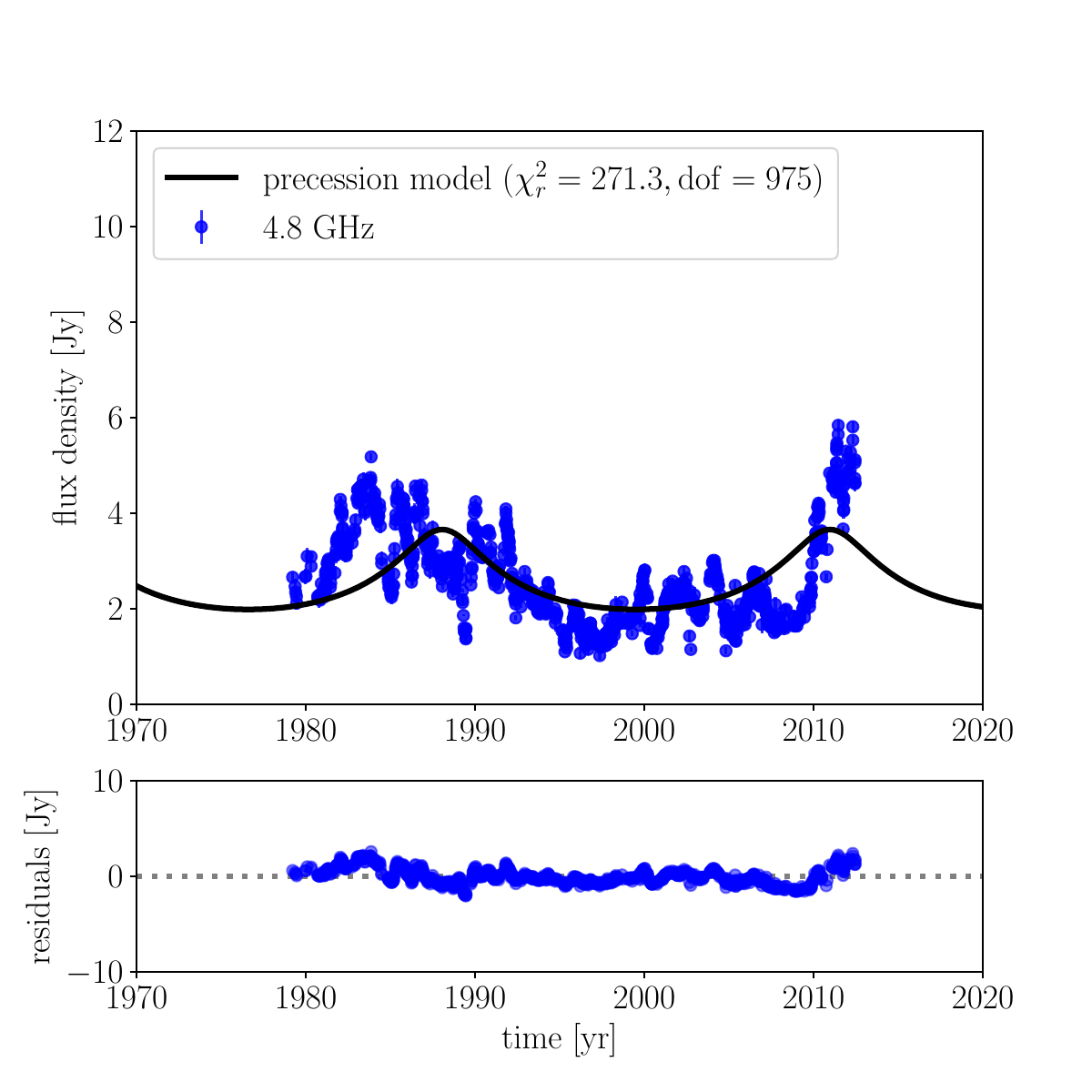}
    \includegraphics[width=0.32\textwidth]{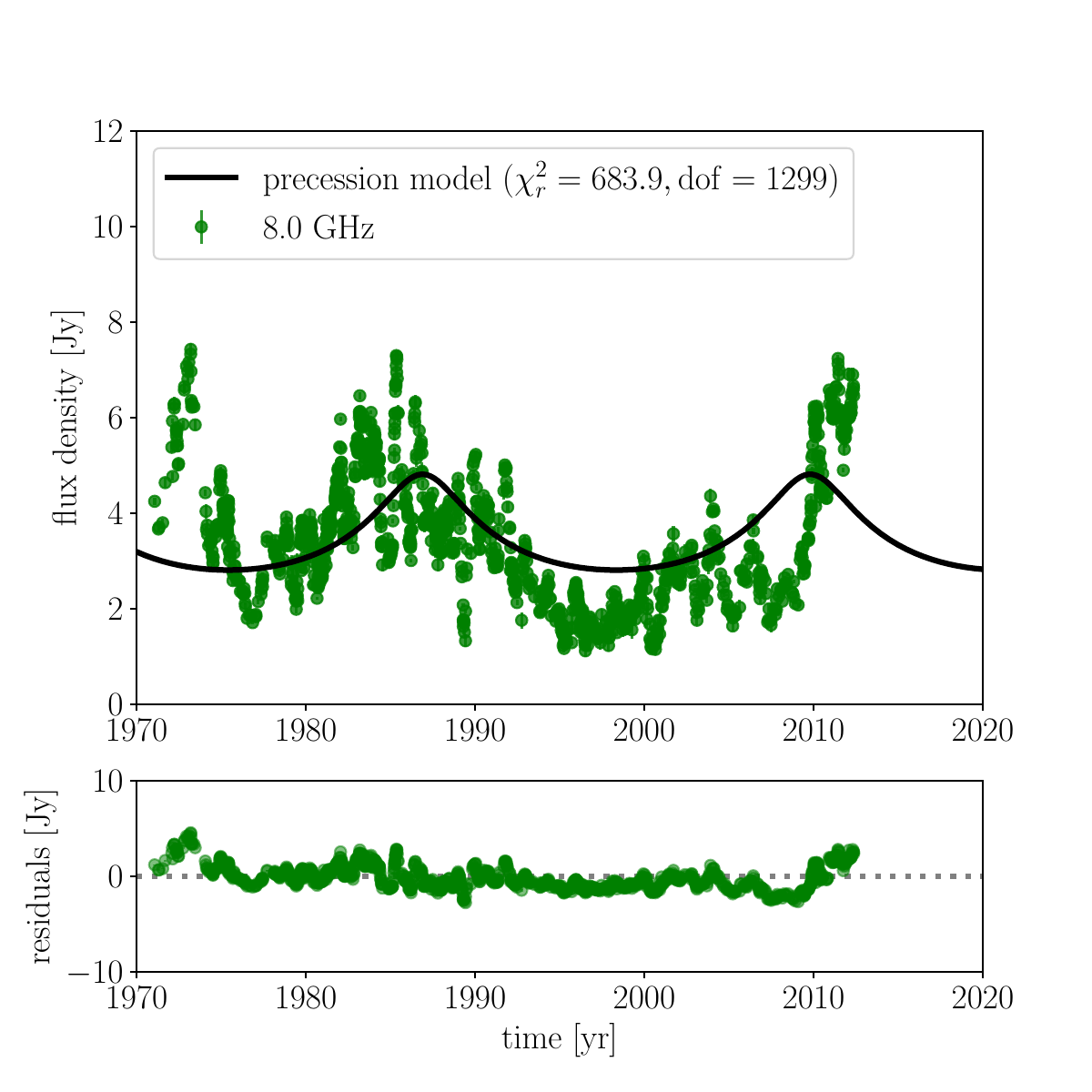}
    \includegraphics[width=0.32\textwidth]{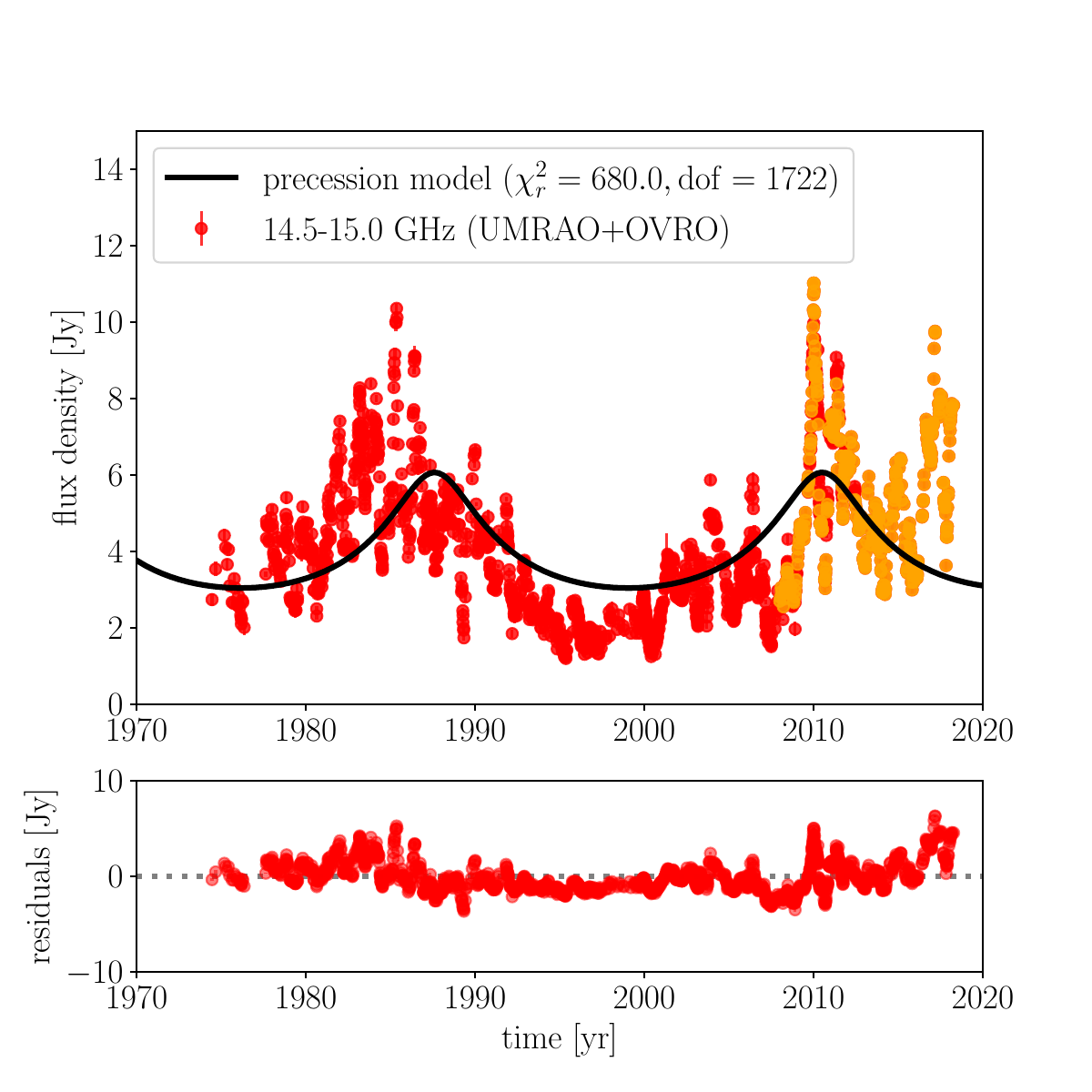}
    \caption{The precession model flux modulations $S_{\nu}=S_{\nu,0}\delta^{p+\alpha}$ fitted to the 4.8, 8.0, and 14.5-15.0 GHz continuum flux densities of OJ~287 (UMRAO measurements, with OVRO measurements -- orange points -- added to 15.0 GHz time series), which is depicted from the left to the right panels, respectively. The legends list the reduced $\chi^2$ values with the corresponding number of degrees of freedom. The best-fit parameters for each frequency band are listed in Table~\ref{tab_precession_flux}. The best-fit spectral indices are negative for all three frequencies, i.e. the spectrum is inverted, which is consistent with the increasing flux density for higher frequencies as is visible from the left to the right panels in the sequence of the increasing frequency.}
    \label{fig_flux_density_precession}
\end{figure*}

\begin{table}[h!]
    \centering
    \caption{Best-fit parameters based on fitting the precession-only model to the observed flux densities of OJ~287.}
    \begin{tabular}{c|c|c|c}
    \hline
    \hline
    Parameter  & $4.8\,{\rm GHz}$ & $8.0\,{\rm GHz}$ & $15.0\,{\rm GHz}$  \\
    \hline
    $S_{\nu,0}^{\rm prec}$ [Jy]  & $1.73 \pm 0.05$     &  $2.48 \pm 0.05$         &  $2.60 \pm 0.04$   \\
    $\alpha^{\rm prec}$  & $-1.73 \pm 0.05$    &  $-1.76 \pm 0.01$            & $-1.69 \pm 0.01$    \\
    $t_0^{\rm prec}$ [yr]    & $1993.8 \pm  0.2$        &  $1992.6 \pm 0.1$      & $1993.3 \pm 0.1$ \\
    $\chi^2_{\rm r}\,(\text{dof})$  & $271.3$ (975) & $683.9$ (1299) & $680.0$ (1722) \\
    \hline
    \hline
    \end{tabular}
    \label{tab_precession_flux}
\end{table}

First, we adopt the kinematical parameters corresponding to the best-fit precession-only model (with $\chi^2_r=18.9$; see Table~\ref{tab_parameters_precession_nutation}) as inferred from the joint least-square fitting of the apparent velocities and position angles. Then we use the time-variable flux-density function $S_{\nu}=S_{\nu,0}\delta^{p+\alpha}$ to fit the temporal evolution of the flux density of OJ~287 at 4.5, 8.0, and 14.5+15.0 GHz (combined UMRAO+OVRO radio data), where we fix $p=2$ since $p=3$ leads to unrealistically large negative spectral indices (steep inverted spectrum). The best-fit parameters $S_{\nu,0}^{\rm prec}$, $\alpha^{\rm prec}$, and $t_0^{\rm prec}$ are listed in Table~\ref{tab_precession_flux} for the three frequencies, while the precession-model flux densities vs. observed flux densities are shown in Fig.~\ref{fig_flux_density_precession}. The precession-only model reproduces well the main observed radio flares, i.e. the intrinsic jet flux density is clearly modulated by the bulk jet precession or at least it can address the increased radio flux density at $\sim 1980-1990$ and at $\sim 2005-2015$. However, there are residuals with the mean value of $0.64$, $0.99$, $1.16$ Jy for 4.5, 8.0, and 14.5+15.0 GHz data, respectively, that can be attributed partly to additional nutation-like jet motion and/or to stochastic accretion- or instability-driven red-noise variability.

\begin{figure*}[h!]
    \centering
    \includegraphics[width=0.28\textwidth]{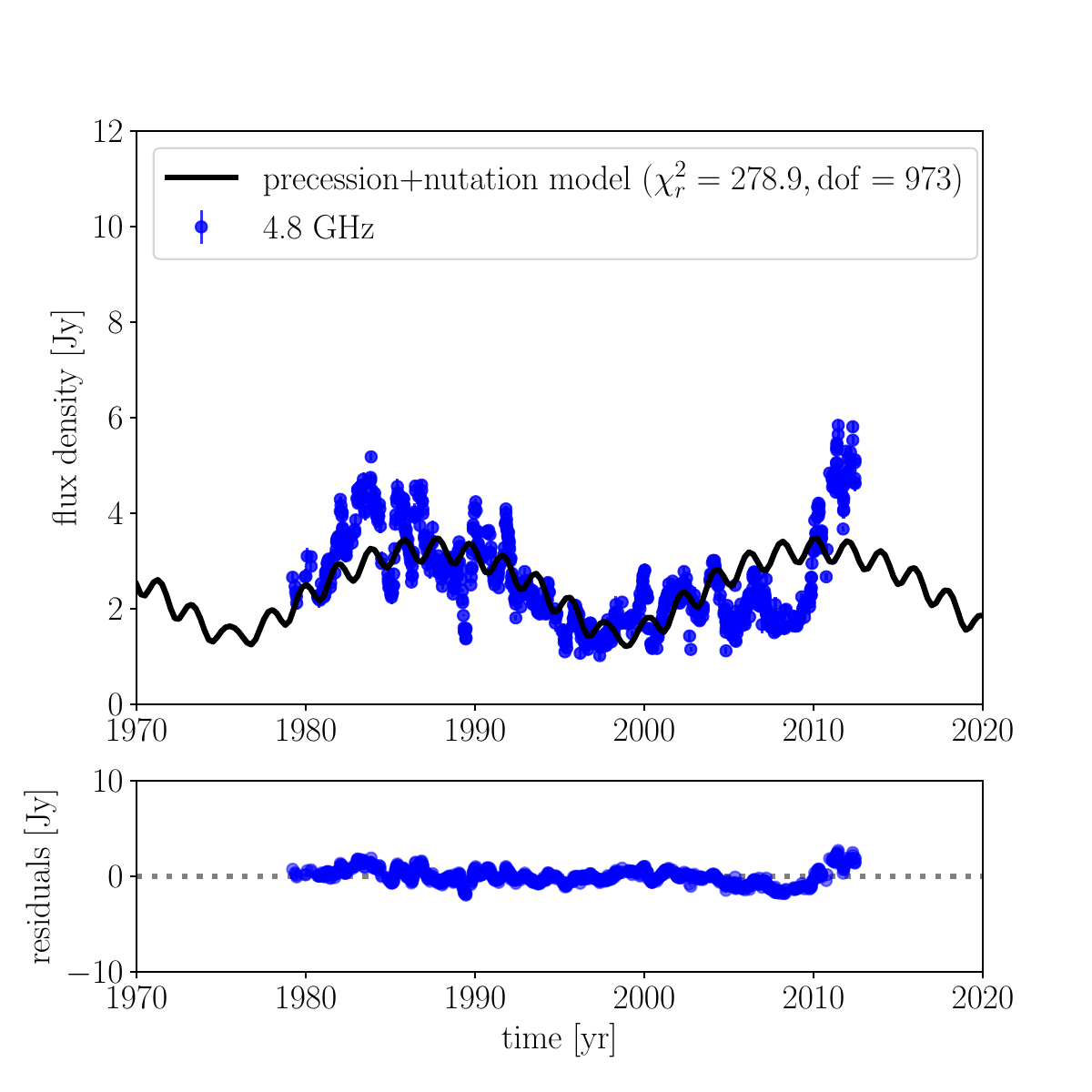}
    \includegraphics[width=0.28\textwidth]{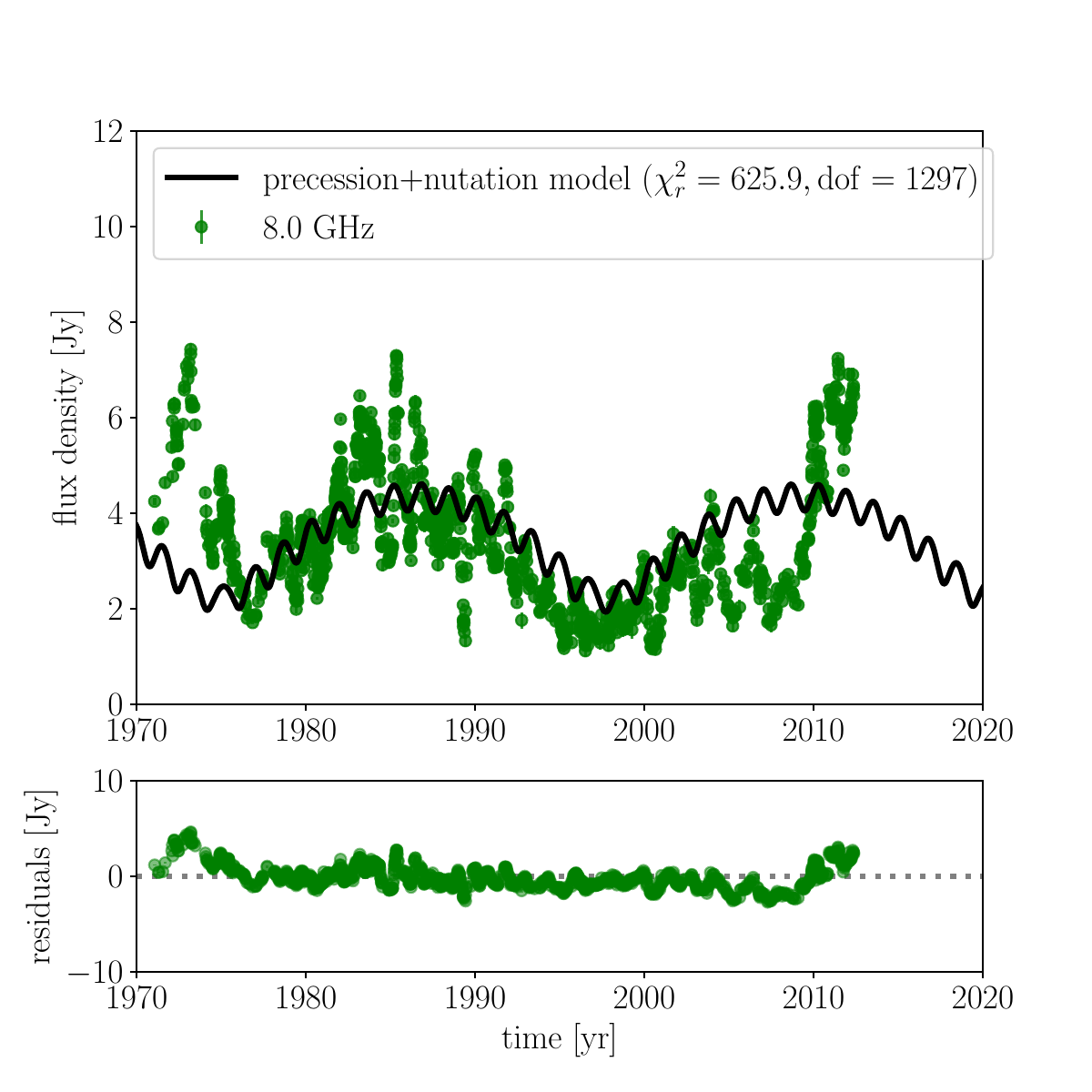}
    \includegraphics[width=0.28\textwidth]{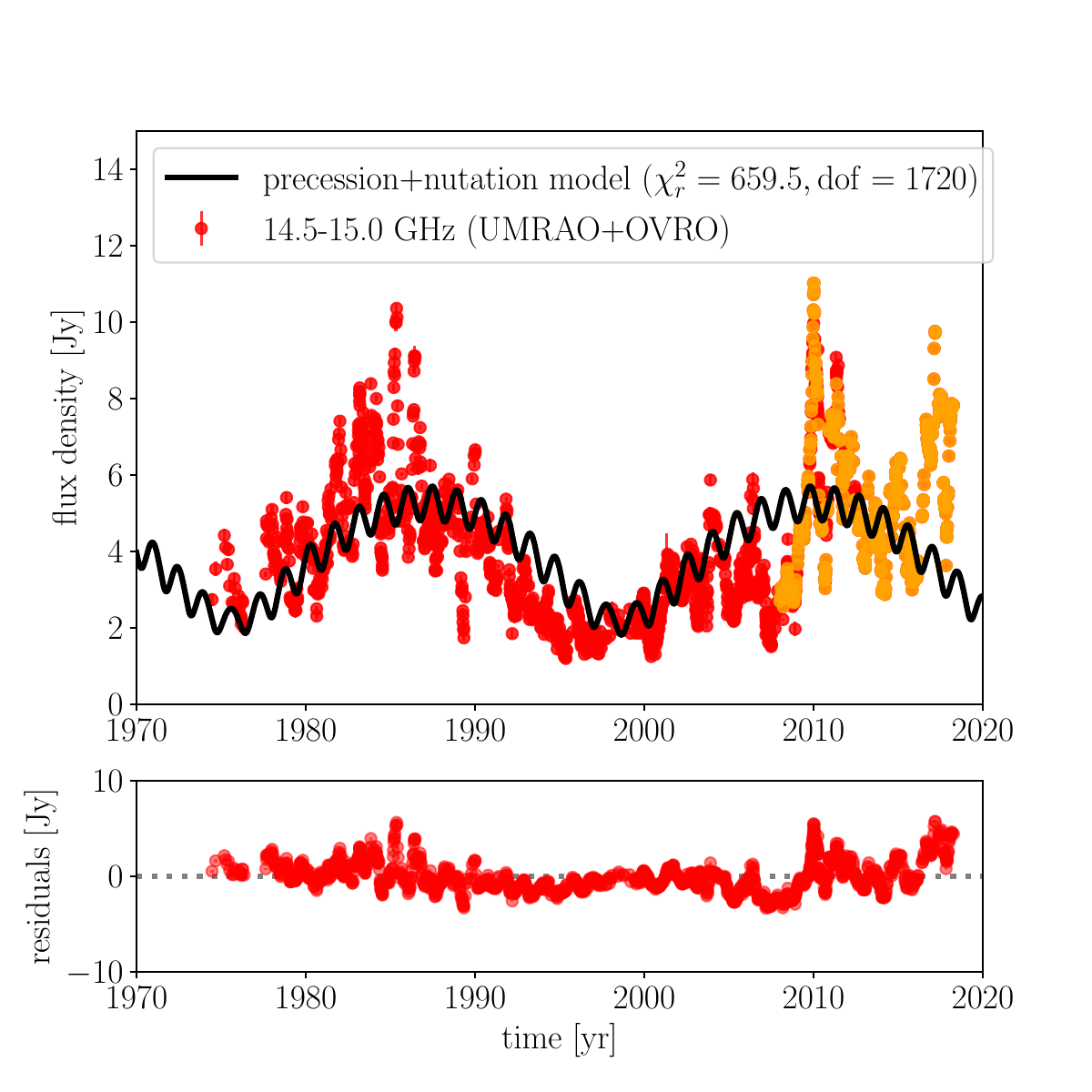}
    \caption{The precession$+$nutation model flux modulations $S_{\nu}=S_{\nu,0}\delta^{p+\alpha}$ fitted to the 4.8, 8.0, and 14.5-15.0 GHz continuum flux densities of OJ~287 (UMRAO measurements, with OVRO measurements added to 14.5 GHz data -- they are marked as orange points), which are shown in the panels from the left to the right, respectively. The legends list the reduced $\chi^2$ values with the corresponding number of degrees of freedom. The best-fit parameters for each frequency band are listed in Table~\ref{tab_precession+nutation_flux}. The best-fit spectral indices are negative for all three frequencies, i.e. the spectrum is inverted, which is consistent with the increasing flux density for higher frequencies as is see from the three panels above.}
    \label{fig_flux_density_precession+nutation}
\end{figure*}

\begin{table}[h!]
    \centering
        \caption{Best-fit parameters based on fitting the precession$+$nutation model to the observed flux densities of OJ~287.}
    \begin{tabular}{c|c|c|c}
    \hline
    \hline
    Parameter  & $4.8\,{\rm GHz}$ & $8.0\,{\rm GHz}$ & $15.0\,{\rm GHz}$  \\
    \hline
    $S_{\nu,0}^{\rm pn}$ [Jy]  &  $3.34 \pm 0.05$  &  $4.46 \pm 0.04$ &  $5.46 \pm 0.04$   \\
    $\alpha^{\rm pn}$  & $-2.36 \pm 0.02$   &  $-2.30 \pm 0.01$ &  $-2.39 \pm 0.01$ \\
    $P_{\rm n}$ [yr]  & $1.86 \pm 0.01$     &  $1.56 \pm 0.01$  &  $1.40 \pm 0.01$   \\
    $t_0^{\rm pn}$ [yr]    &  $1987.27\pm 0.07$       &  $1986.43 \pm 0.04$     & $1987.12 \pm 0.02$ \\
    $\chi^2_{\rm r}\,({\rm dof})$  &  $278.9$ (973)  &  $625.9$ (1297)   & $659.5$ (1720) \\
    \hline
    \hline
     \end{tabular}
    \label{tab_precession+nutation_flux}
\end{table}

Second, we add the second-order nutation-like motion to the bulk precession motion of the jet, see Fig.~\ref{sketch}, top right, for an illustration. This is motivated by significant residuals of the order of 1 Jy between the observed flux densities and the precession-only model, see Fig.~\ref{fig_flux_density_precession}. In addition, the quasi-stationary radio component {\bf a} of OJ~287 exhibits signs of a second-order, short-period motion, see Fig.~\ref{fig_precession_nutation_a}. For the precession kinematics, we fix the parameter values close to the best-fit precession model according to Table~\ref{tab_parameters_precession_nutation}, in particular $\gamma=9.0$, $P_{\rm p}=22.9\,{\rm yr}$, $\Omega_{\rm p}=10.3^{\circ}$, $\Phi_0=15^{\circ}$, and $\eta_0=-130.8^{\circ}$. The nutation-cone half-opening angle is set to $\Omega_{\rm n}=3^{\circ}$, which is within the uncertainties consistent with the best-fit value obtained from fitting the position-angle temporal variation of the quasi-stationary component {\bf a}, see Fig.~\ref{fig_precession_nutation_a}. The geometric boosting parameter is set to $p=2.0$ as in the precession-only model, which corresponds to the continuous (cylindrical) jet emission model. We fit the precession$+$nutation flux density modulation, $S_{\nu}=S_{\nu,0}\delta^{p+\alpha}$, to the observed continuum flux densities to infer the intrinsic jet base flux density $S_{\nu,0}^{\rm pn}$, spectral index $\alpha^{\rm pn}$, nutation period $P_{\rm n}$, and the reference epoch $t_0^{\rm pn}$. The best-fit parameters for 4.5, 8.0, and 14.5+15.0 GHz data are listed in Table~\ref{tab_precession+nutation_flux} and the best-fit models are graphically depicted in Fig.~\ref{fig_flux_density_precession+nutation}, including the residuals, for each frequency band. From the best-fit models, we obtain the inverted spectral index with the mean value of $\overline{\alpha}^{\rm pn}=-2.38 \pm 0.07$, which is consistent with the self-absorbed, inverted radio spectrum for the precession-only model. The mean period of the short-term nutation motion is $\overline{P}_{\rm n}=1.66\pm 0.14$ years, which is consistent with the nutation period of $1.6 \pm 0.1$ years inferred from fitting the position angle of component {\bf a}, see Table~\ref{tab_parameters_precession_nutation}. Adding the nutation to the bulk jet precession kinematics can temporally reproduce the short-term flux-density variability; however, the decrease of the mean values of residuals -- 0.61 (4.5 GHz), 0.93 (8.0 GHz), and 1.18 Jy (14.5 GHz) -- is only mild or essentially negligible, which is due to the fact that the nutation does not reproduce well the observed short-term flare amplitudes. Therefore, the short-term radio variability is not yet fully captured by the precession+nutation model. This may be due to the fact that the current model does not include stochastic radio variability, nor do we perform radiative transfer that can also modify the radio variability as such. Most of the residuals are due to a few large radio flares that exceed the Doppler-boosted flux density level at the peaks by about a factor of two. This is especially apparent towards higher frequencies. In Section~\ref{lightcurves}, we analyze in more detail the coexistence of purely stochastic radio variability with the deterministic, quasi-periodic kinematical effects, such as the precession and the nutation. 
\subsection{Precession and nutation in blazar jets: a case for binary systems} 
Since the orbital period of a putative supermassive binary black hole (SMBBH) (for example OJ\,287)
is not negligible as compared to the precession period of the accretion disk and the jet, the presence of a smaller nodding motion of the jet -- here referred to as nutation -- is generally expected, e.g., in analogy to the well known binary systems SS433 \citep{margon} and the accretion disk of the X-ray pulsar Hercules X-1, as proposed by \citet{Shakura1999}.

Essentially, the gravitational torque of the secondary BH acts on the accretion disk around the primary BH (with a jet).
When averaged over the binary orbital period, this yields a steady torque resulting in a mean counter-precession of the accretion disk
(opposite sense to the disk rotation). 
This steady torque is accompanied by a nutation torque that is periodic at the synodic period derived from the precession frequency and the binary's orbital frequency. 
This nutation torque has the same magnitude as the precession torque, but the amplitude of its effects is smaller by the ratio of the precession frequency and the binary's orbital frequency.

\citet{katz} developed a formalism to calculate the frequency as well as the amplitude of short-term nodding
motions in the precessing accretion flows in close binary systems. 
Here we apply their formalism for SMBBHs. 
The nutation frequency may be expressed as \citep[see also][]{Caproni2013},

\begin{equation}
    \omega_{\rm n}=2(\omega_{\rm orb}-\omega_{\rm p})\,,
    \label{eq_nutation_freq}
\end{equation}
where the angular frequency of the precession $\omega_{\rm p}$ should be negative with respect to the orbital angular frequency~$\omega_{\rm orb}$. 
Equation~(\ref{eq_nutation_freq}) can then be employed to derive the observed binary orbital period: 
\begin{equation}
    P_{\rm orb}^{\rm obs}=P_{\rm p}^{\rm obs}\left(\frac{P_{\rm p}^{\rm obs}}{2P_{\rm n}^{\rm obs}}-1 \right)^{-1}\,.
    \label{eq_orbital_period}
\end{equation}
 For the case of OJ\,287, the observed precession period is in the range of $P_{\rm p}^{\rm obs}\sim 20-30$ years and the nutation period is in the range of $P_{\rm n}^{\rm obs}\sim 1-1.6$ years, see Table~\ref{tab_parameters_precession_nutation}, which yields the observed SMBHB period in the range $P_{\rm orb}^{\rm obs}\simeq 2.1-3.8$ years.
This period is about three to six times shorter than the orbital period of 12 years that is adopted in most studies of OJ\,287, directly from the observations of (periodic) optical flares (see \citealt{valtonen}, and references therein).

We note that the inferred orbital period of $\sim 2-4$ years and the optical-flare periodicity of 12 years (e.g., \citealt{aimo}) are not mutually incompatible.
It is quite plausible that the optical flares of OJ\,287 occur at certain specific phases of the SMBBH orbital motion. 
Such a possibility has indeed been suggested to explain the 35-day periodicity of X-ray flares of the X-ray pulsar Her X-1, even though its binary has an orbital period of just 1.7 days. 
The flares would occur when the rapid nodding motion combines with the slower precession, effectively removing the absorbing matter from the line of sight \citep{katz}.
More detailed investigation of this effect is needed for OJ\,287, in combination with the polarimetric monitoring.
%
The amplitude of the nutation motion is then expected to be smaller than the precession amplitude by a factor of $P_{\rm p}^{\rm obs}/P_{\rm orb}^{\rm obs}\simeq 5.3-14.3$, applying the model of \citet{katz} to OJ\,287.

\begin{figure*}[h!]
\centering
\begin{minipage}{\columnwidth}
\includegraphics[width=\columnwidth]{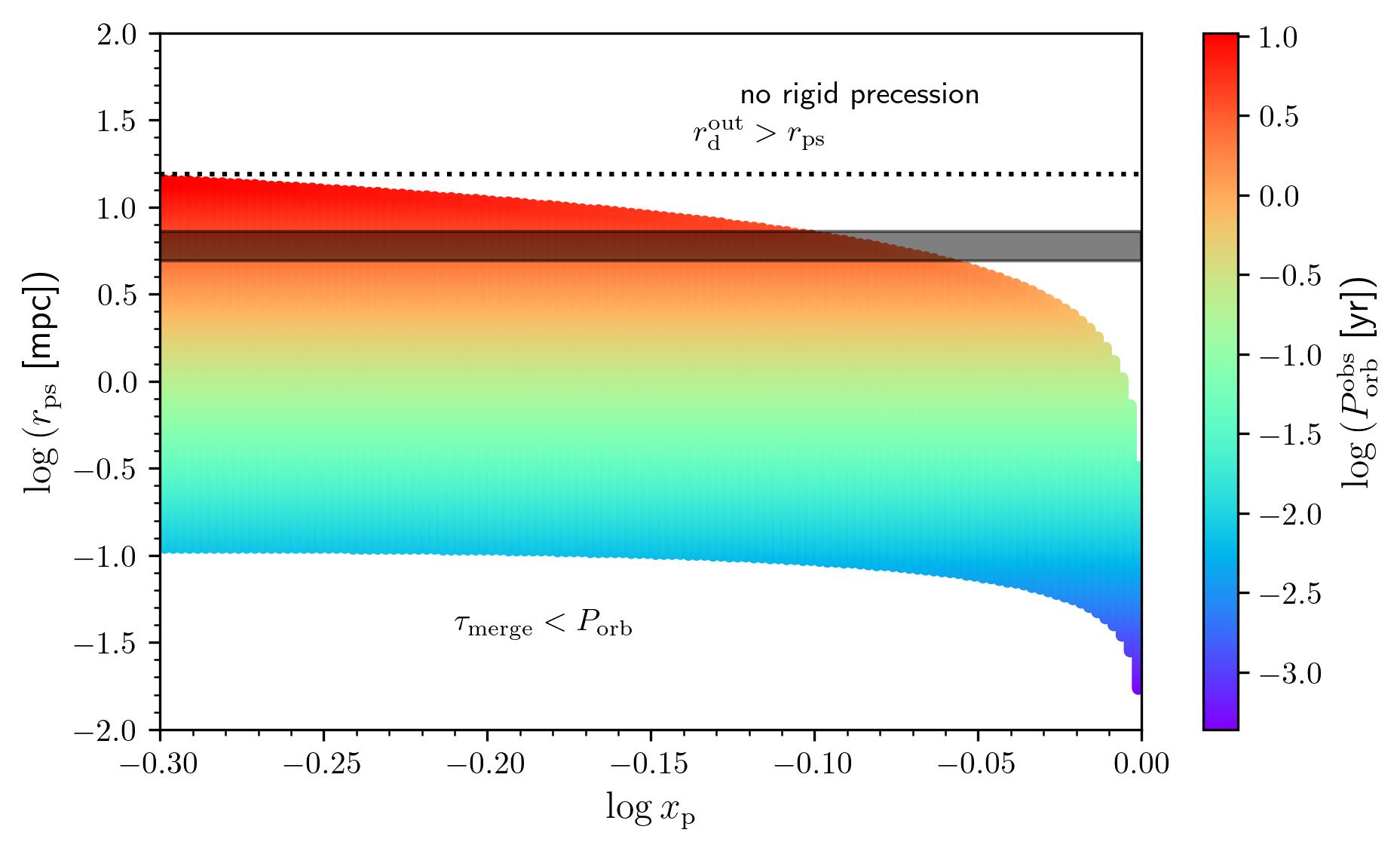}
[a]
\end{minipage}
\begin{minipage}{\columnwidth}
\includegraphics[width=\columnwidth]{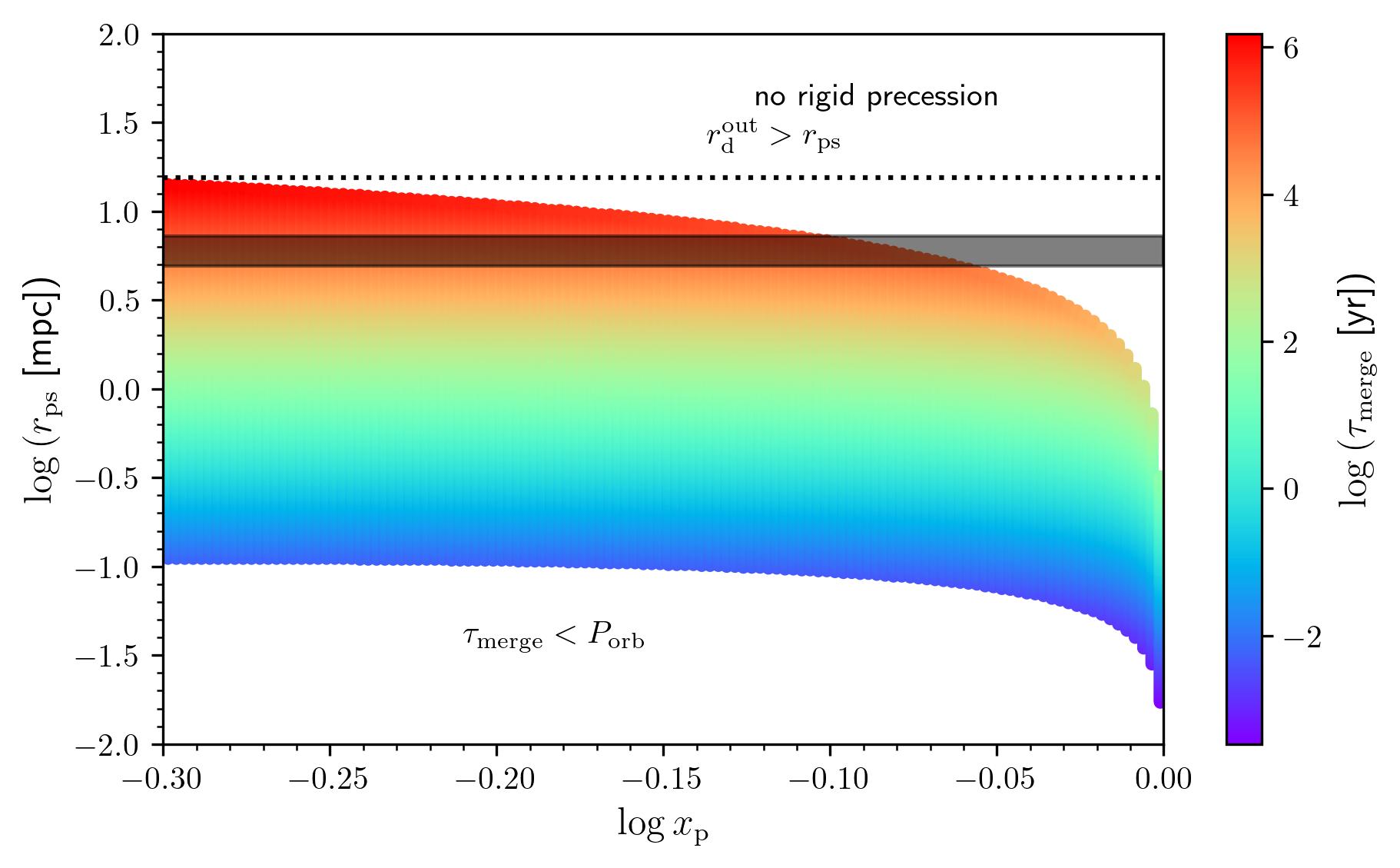}
[b]
\end{minipage}
\begin{minipage}{\columnwidth}
\includegraphics[width=\columnwidth]{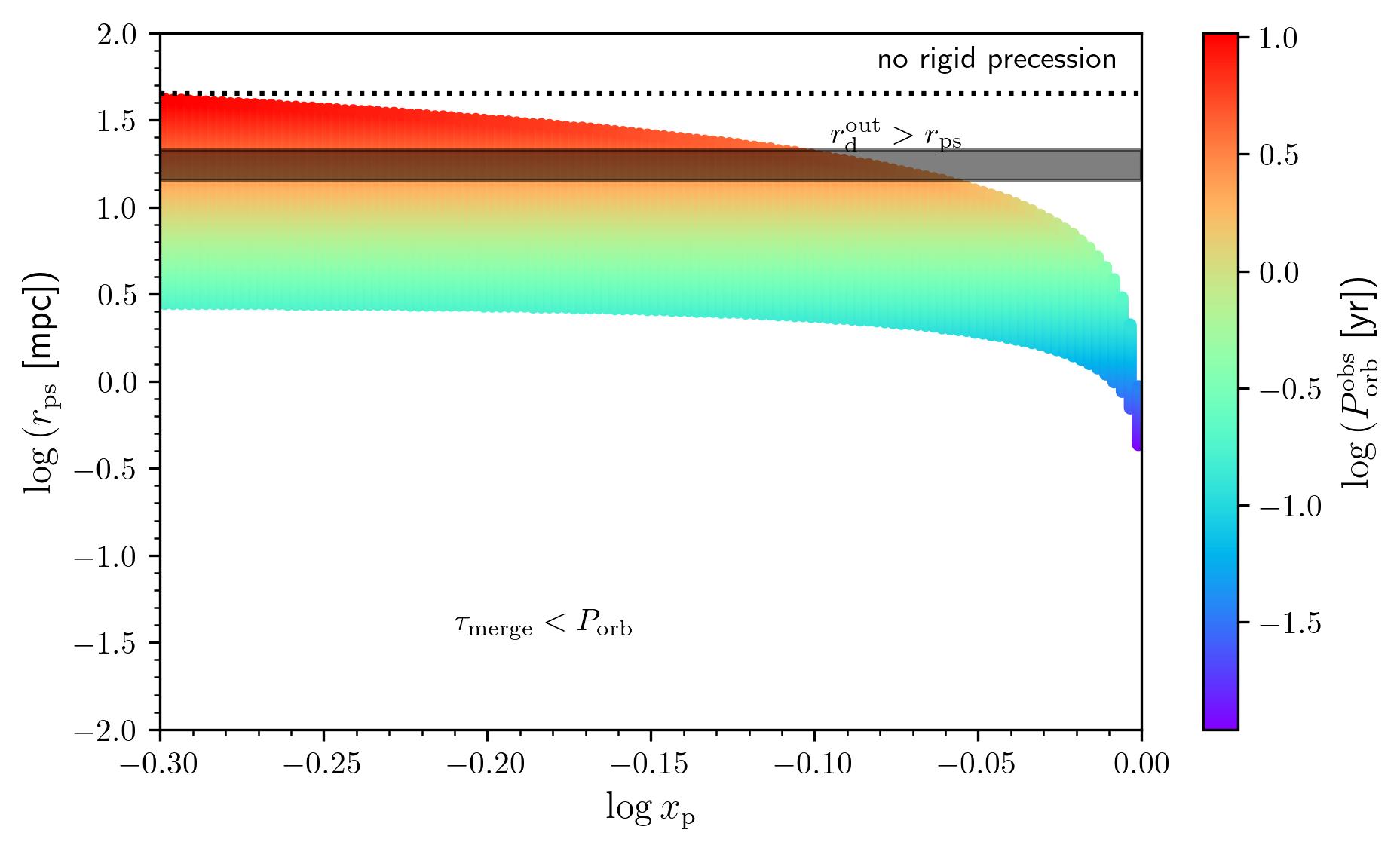}
[c]
\end{minipage}
\begin{minipage}{\columnwidth}
\includegraphics[width=\columnwidth]{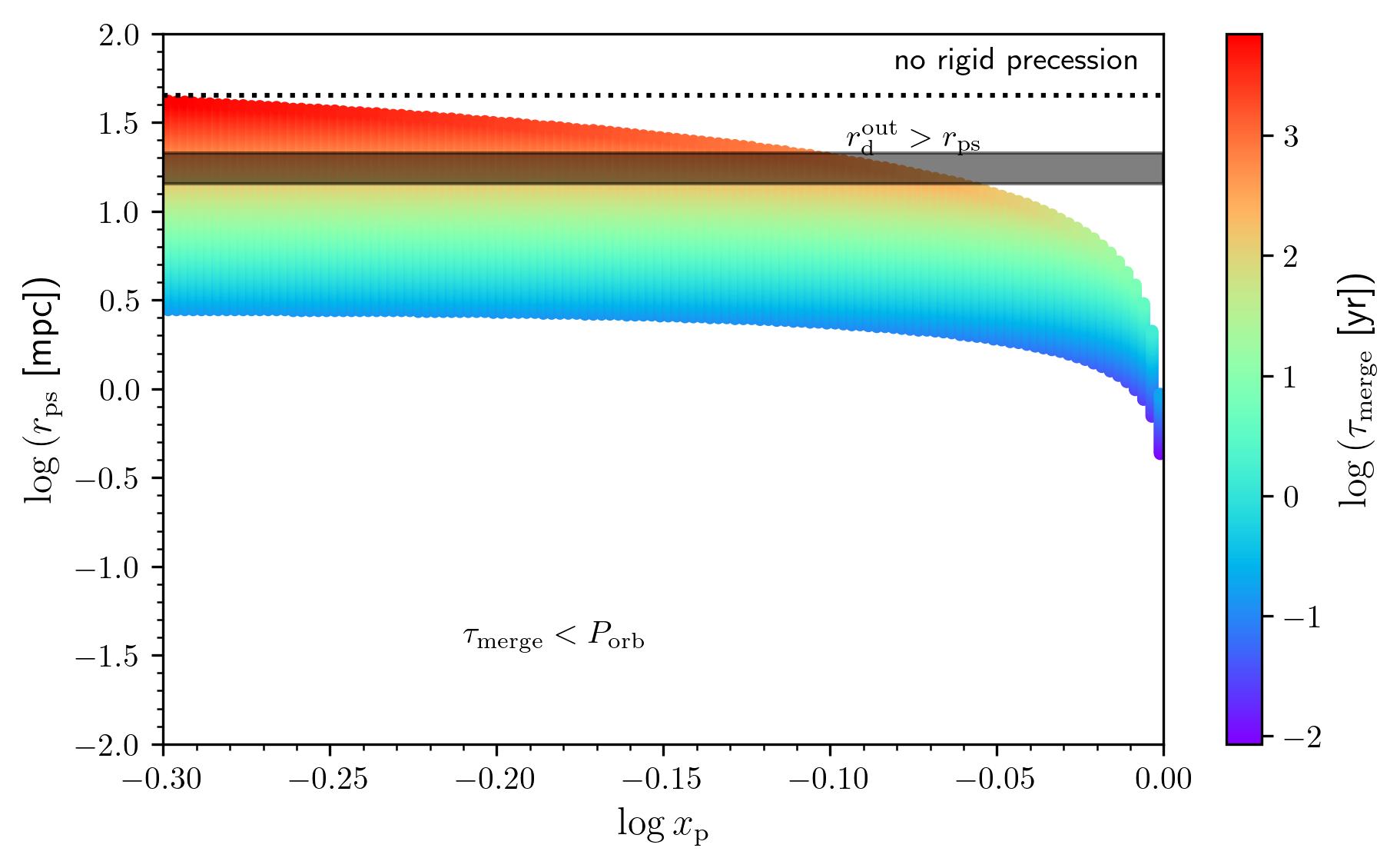}
[d]
\end{minipage}
\caption{Panel [a] shows the observed orbital period $P_{\rm orb}^{\rm obs}$ of the binary system of OJ\,287 as a function of the component distance in milliparsecs $r_{\rm ps}$ and the mass ratio $x_{\rm p}$ of the primary BH to the total SMBH binary mass (expressed as logarithms), i.e. $x_{\rm p}=m_{\rm p}/(m_{\rm p}+m_{\rm s})$. The black shaded rectangle marks the range of the putative orbital period of $2.1-3.8$ years as derived based on the nutation period and the dotted black line represents the orbital period of $11.7$ years based on the optical flares. The total BH mass was assumed to be $4\times 10^8\,M_{\odot}$. The upper white region represents the parameter space where the outer disc radius is larger than the primary-secondary BH distance, $r_{\rm d}^{out}>r_{\rm ps}$, and hence rigid precession cannot proceed. The lower white region represents the parameter space where the merger timescale is shorter than the orbital period of the SMBHB. Panel [b] displays the merger timescale as function of the component distance and the primary mass fraction (expressed as logarithms). The observed orbital periods are marked as in [a]. Panels [c] and [d] are analogous to Panels [a] and [b] but for the total SMBHB mass of $10^{10}\,M_{\odot}$.}
    \label{fig_precession_merger}
\end{figure*}

In Fig.~\ref{fig_precession_merger} [a] we display the observed period of the SMBBH system  OJ\,287 that yields the observed 
precession period of $23$ years, as a function of the component distance (in milliparsecs) 
and the mass ratio $x_{\rm p}$ of the primary to the total SMBH binary mass, i.e. $x_{\rm p}=m_{\rm p}/(m_{\rm p}+m_{\rm s})$. 
The SMBHB orbital period range of $2.1-3.8$ years in the observed frame is marked as a shaded black rectangle in  Fig.~\ref{fig_precession_merger}, while the previously discussed orbital period of $11.7$ years related to optical outbursts is marked as a dotted black line. We see that for the assumed total mass of $4\times 10^8\,M_{\odot}$ of the binary the primary mass fraction $x_{\rm p}$ is much broader for the shorter orbital period of 2-4 years, $x_{\rm p}\in (0.5,0.80-0.88)$, while for 11.7 years, the range shrinks to essentially two equal-mass components, which is unlikely. This situation is the same for the larger total mass of $10^{10}\,M_{\odot}$, see Panels [c] and [d] in Fig.~\ref{fig_precession_merger}. In addition, for the total mass of $m_{\rm tot}\simeq 10^{10}\,M_{\odot}$ of the SMBH binary, whose components have the observed period of $P_{\rm orb}^{\rm obs}=10$ years, the merger timescale decreases just to $\lesssim 10^4$ years according to \citet{1964PhRv..136.1224P}, 
\begin{align}
   \tau_{\rm merge}&=\frac{5c^5}{256 G^3}\frac{a_0^4}{x_{\rm p}x_{\rm s}m_{\rm tot}^3}\,,\notag\\
   &=\frac{5c^5}{(4\pi^2)^{4/3}256G^{5/3}}\frac{P_{\rm orb}^{8/3}}{x_{\rm p}x_{\rm s}m_{\rm tot}^{5/3}}\,,\notag\\
   &\sim 12886 \left[\frac{P_{\rm orb}^{\rm obs}}{10\,\text{yr}(1+z)} \right]^{8/3}\left(\frac{x_{\rm p}}{0.5}\right)^{-1} \left(\frac{x_{\rm s}}{0.5}\right)^{-1}\times \,\notag\\
   &\times \left(\frac{m_{\rm tot}}{10^{10}\,M_{\odot}} \right)^{-5/3}\,{\rm yr}\,,
 \end{align}
which appears to be statistically less likely, while for the total SMBH binary mass of $4\times 10^8\,M_{\odot}$, $\tau_{\rm merge}\simeq 10^6\,{\rm yr}$.

 For the above reasons, while doing estimates related to OJ\,287, we have assumed a total SMBBH mass of $4\times 10^8\,M_{\odot}$, which is preferred, e.g., in \citet{liu}, based on the SMBH-bulge mass correlation. 
A similar mass, $1.3\times 10^8\,M_{\odot}$, was derived by \citet{neronov}, based on the time-scales 
of fast $\gamma$-ray flares. 
For completeness, in Fig.~\ref{fig_precession_merger}[b] we also show the merger timescale as a function of a component separation and the 
SMBBH mass ratio. The same two orbital periods are depicted by thick grey lines as in Fig.~\ref{fig_precession_merger}[a]. 

\section{Precession-induced (temporal) variability of the OJ~287 SED}
\label{sec_precession_sed}

As discussed in this paper, radio variability of blazars is substantially, even predominantly, influenced by deterministic processes (i.e., precession). It seems plausible that the precession would then also influence the overall Spectral Energy Distribution (SED) of a blazar in a deterministic fashion. Turning this argument around, a predictable shift in the SED corresponding to a certain phase of the precession should strengthen the case for precession playing a prominent role in blazar variability. We examine this in this section. In essence, precession means a change of the Doppler factor due to a changing viewing angle, see Eq.~\eqref{eq_doppler_boosting}. A time-varying Doppler factor should also be reflected in temporal evolution of the SED. Both synchrotron emission and Inverse-Compton emission should get boosted (de-boosted) in tandem. 



\subsection{OJ~287 reveals an atypical SED-state}
\citet{kushwaha} discusses new spectral features which appeared during the 2015-2017 multi-wavelength high-activity state of OJ~287. In particular, a break in the NIR-optical spectrum and hardening of the MeV-GeV emission accompanied by a shift in the location of the peak, were observed. Fig.~1b in \citet{kushwaha} shows a flaring SED (MJD: 57359-57363), a typical SED during the VHE phase (MJD: 57786) and the quiescent state SED after the VHE activity (MJD: 57871).\\
According to \citet{kushwaha}, the new spectral features support the disk-impact binary SMBH model over the geometrical class of models.
The 2015-2017 multi-wavelength flare coincides with the time of the projected lower reversal point as deduced from VLBA studies of the jet kinematics and identified as the sign of precession by \citet{Britzen2018}. In the following, we explore whether the (geometrical) approach of changing just the viewing angle can reproduce the new SED-state discussed by \citet{kushwaha}.

\subsection{The SED variation of OJ\,287 during the passage of the lower projected precession reversal point}
\label{time_var_sed}
In the following, we assume a single emission zone generating the SED of OJ~287. We fit the SED observed at two epochs close in time to see if the change is consistent with a jet component following a typical trajectory reflecting the precession + nutation model.

\begin{figure*}[h!]
\centering
\includegraphics[width=0.45\textwidth]{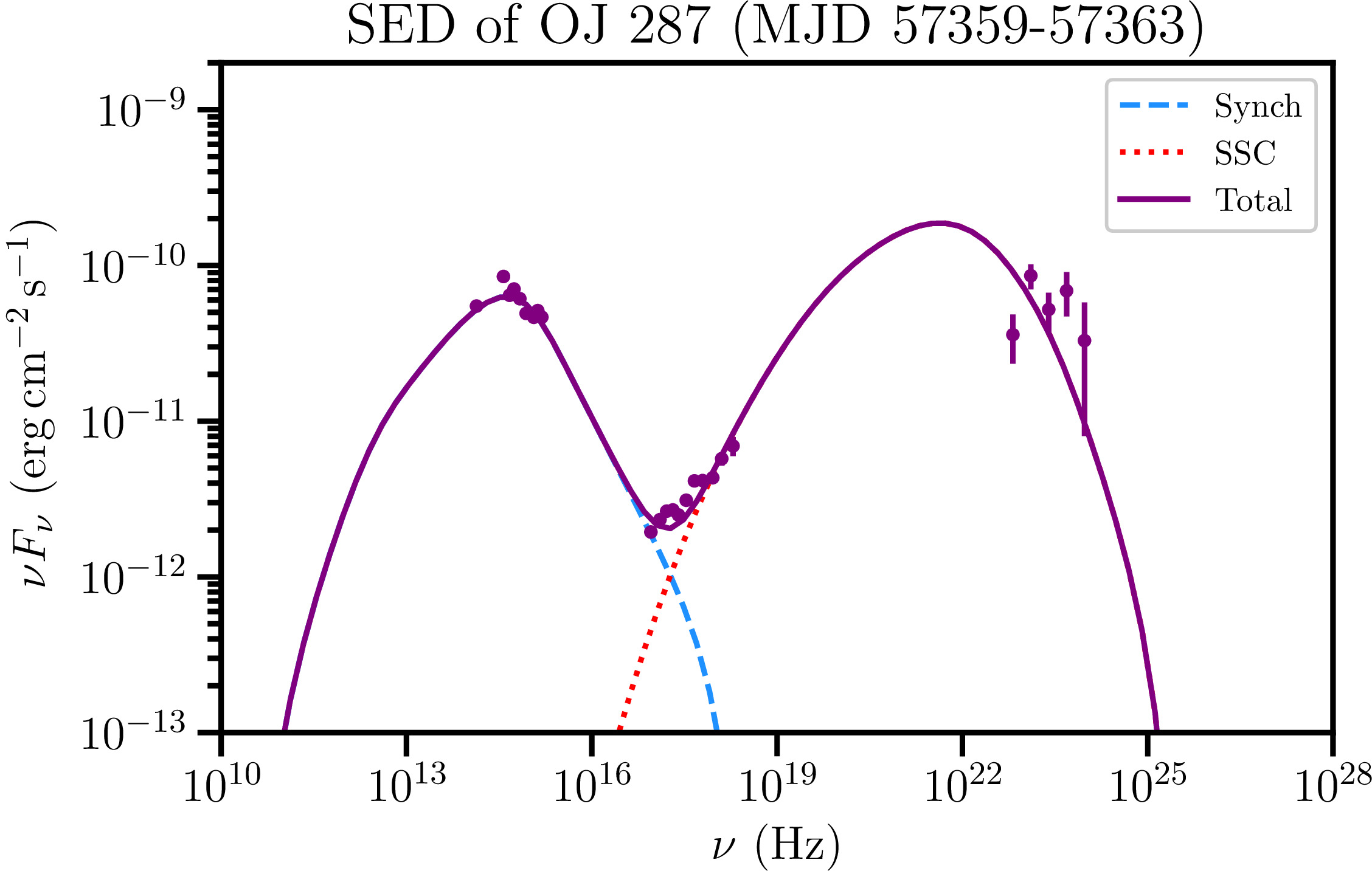}
\includegraphics[width=0.45\textwidth]{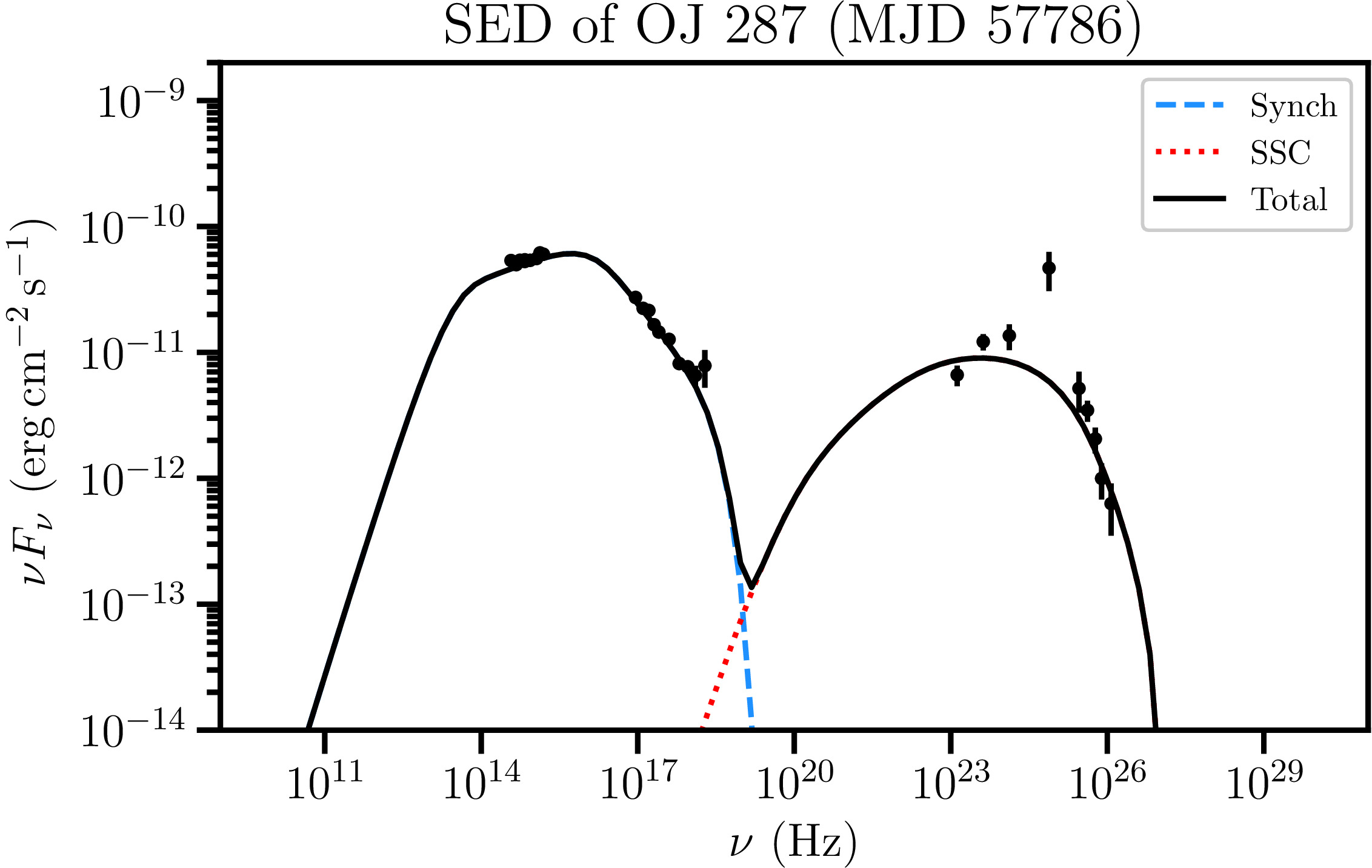}
\caption{Pure leptonic SEDs for OJ\,287 in different spectral states, where the SED points are adopted from \citet{kushwaha}. 
\textit{Left}: The dashed blue curve shows the (self-absorbed) synchrotron, the dotted red curve shows the SSC contribution to the 
total leptonic SED between MJD 57359-57363, that is shown by purple continuous line. 
\textit{Right.} The dashed blue curve shows the (self-absorbed) synchrotron, the dotted red curve shows the SSC contribution to the 
total leptonic SED at MJD 57786, that is shown by black continuous line. The main difference between the models applied in the figures is the viewing angle of the jet (left plot $\Phi \approx 12.1^\circ$, right plot $\Phi\leq2^\circ$) and the corresponding Doppler factor (left plot $\delta\approx3.7$, right plot $\delta\approx45.1$).}
\label{oj287_turning}
\end{figure*}

The fitting to the SED of OJ\,287 with self-absorbed synchrotron and synchrotron self-Compton (SSC) contributions is done by employing the AGNPY package \citep{Nigro2020}, version 0.0.8\footnote{\url{https://github.com/cosimoNigro/agnpy}}, written in python for numerically computing the photon spectra produced by leptonic radiative processes in radio-loud AGN. 
The electron energy spectrum (described by the spectral normalization factor), the low and high-energy spectral index, the minimum and maximum electron Lorentz factors, the Lorentz factor at the break energy, as well as the jet Doppler factor, inclination and the magnetic field strength were free in the fitting. The luminosity distance of OJ\,287 ($z=0.306$) is $d_L=4.9\times 10^{27}$ cm. The radius of the radiating blob is fixed at $R_b=3\times 10^{16}$ cm. We present the parameters of the fitted leptonic SEDs for MJD 57359-57363 and at MJD 57786 in Table \ref{table:fittedseds} and plot the computed SEDs in Fig.~\ref{oj287_turning}. We note that these results are possible realizations of a model with a large number of parameters.

In Table \ref{table:fittedseds} we compare the parameters obtained by SED modeling of the data at the end of 2015 (from 2015-12-13 to 2015-12-17, $\simeq 2015.92$, SED$_a$), in the so-called flaring state, with the values obtained for the VHE phase (SED$_b$).
The main difference lies in the viewing angle of the jet ($\sim12.1^\circ$ at the flaring phase, $\leq2^\circ$ at the VHE phase) and the corresponding Doppler factors (3.7 and 45.1 respectively).

It seems that the fit by AGNPY produces different values of the Doppler factor for SED$_a$ and SED$_b$, but in addition, other important quantities describing the synchrotron and the Compton emission might have changed according to the SED fits. We explore whether variations in: (i) both the component parameters and the global value of the jet's Doppler factor, (ii) just the global Doppler factor, or (iii) just the component parameters, can best model the observed SED. 
We made several SED fitting-runs for SED$_b$:
\begin{itemize}
    \item \textbf{case0:} the synchrotron and SSC parameters, as well as the Doppler factor are allowed to change during the fit (see the results in Table \ref{table:fittedseds}, column 'b');
    \item \textbf{case1:} the synchrotron and SSC parameters are allowed to change during the fit (initial parameters from SED$_a$), the Doppler factor is fixed to the value of SED$_a$ (see Table \ref{table:fittedseds}, column 'a');
    \item \textbf{case2:} the synchrotron and SSC parameters of SED$_b$ are fixed to the values of SED$_a$, the Doppler factor is allowed to change (initial value from SED$_b$);
    \item \textbf{case3:} the synchrotron and SSC parameters of SED$_b$ as well as the Doppler factor are fixed to the values of SED$_a$. 
     In this case there is no extra fitting, we will only calculate the $\chi^2$ between SEDa model and SEDb data.
\end{itemize}

The reduced chi-squares are the following: $\chi^2_r=2.06$ (case0, Table \ref{table:fittedseds}.), $\chi^2_r=27.7$ (case1), $\chi^2_r=377.5$ (case2), $\chi^2_r=450.39$ (case3). It seems that AGNPY does not find a good fit when either the electron energy parameters or the Doppler factor is fixed, such that case2 is worse compared to case0, much worse than case1 to case0. Since the observed synchrotron and SSC flux depend on $\delta^4$, we find that the changes in the Doppler factor have played a more decisive role between SED$_a$ and SED$_b$, compared to the parameters describing the electron energy spectrum. We further note that the increase of the Doppler factor between SED$_a$ and SED$_b$ cannot be evaded by increasing the spectral normalization factor ($k_e$) of SED$_b$, since the shape of the SED changes both along the frequency and the energy axes. Moreover, changing $k_e$ affects differently the synchrotron and SSC contributions, and the Doppler factor can only raise the SED amplitude by a factor that is independent of the frequency. 

By fitting the SED of OJ~287 for two epochs close in time and under the single-zone (leptonic) approximation, we find that the change of the shape of the SED of OJ~287 is compatible with a variation of the Doppler factor of a blob following a typical trajectory in the jet reflecting the precession and nutation motions.



\begin{table*}[h!]
\centering
\caption{a: MJD 57359-57363, b: 57786. $\Gamma=10$ ($\beta=0.995$c) fixed based on the present work.}
\label{table:fittedseds}
\begin{tabular}{l|lcc}
\hline 
Spectral state & & $a$ ($\chi^2_r\approx30$) & $b$ ($\chi^2_r\approx2$)\\
 \hline
Spectral normalisation factor &$k_e ({\rm cm^{-3}})$ & $(4.8^{+4.0}_{-2.2})\times 10^{-3}$& $(3.3^{+1.6}_{-1.1})\times 10^{-7}$ \\
Low-energy electron spectral index &$p_1$  & $1.91\pm0.09$ &  $2.67\pm0.06$\\
High-energy electron spectral index &$p_2$  & $4.56\pm0.06$ &  $4.05\pm0.09$\\
Minimum e- Lorentz factor &$\gamma_{\rm min}$  & $10^{2.3}$& $10^{3.0}$\\
Lorentz factor at break energy &$\gamma_{\rm b}$  & $10^{3.5}$ &  $10^{4.3}$\\
Maximum e- Lorentz factor&$\gamma_{\rm max}$ & $10^{5.0}$& $10^{5.5}$\\
Magnetic fields strength &B (G)& $4.56^{+2.98}_{-1.80}$ & $0.16^{+0.04}_{-0.03}$\\
Jet power in particles & $P_{\rm p} ({\rm erg/s})$  & $1.8\times 10^{45}$ & $2.5\times 10^{43}$\\
Jet power in magnetic field &$P_{\rm m} ({\rm erg/s})$  & $1.4\times 10^{46}$ & $1.8\times 10^{43}$ \\
Doppler factor &$\delta$ & $3.7\pm1.1$ &  $45.1\pm6.0$ \\
Viewing angle of the jet &$\Phi$ (deg) & $\sim12.1$& $<2$\\
\hline 
\end{tabular} 
\end{table*}

\subsection{New OJ~287 SED state correlates with special precessional phase}
\begin{figure*}[h!]
\centering
\includegraphics[width=\textwidth]{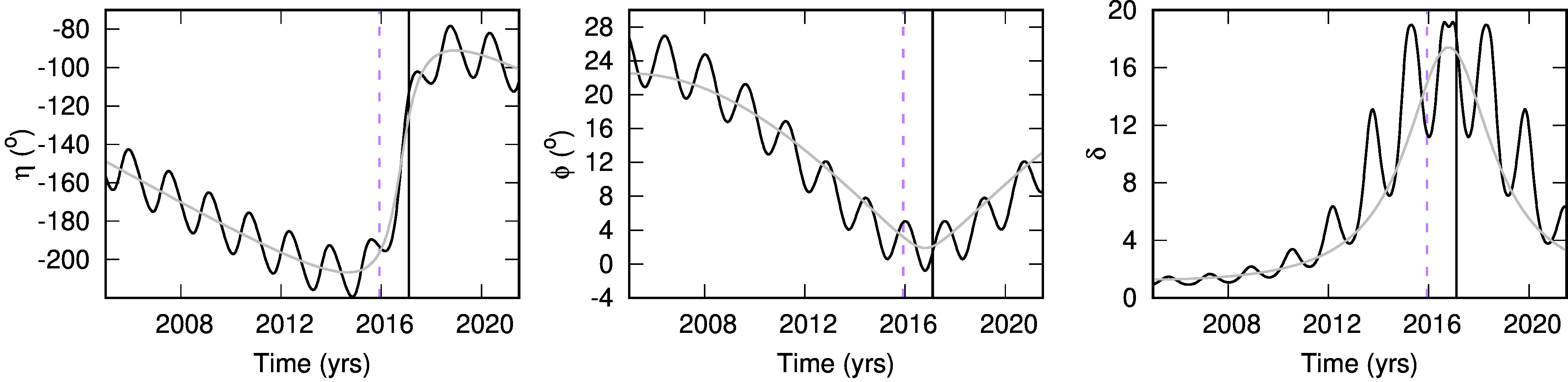}
\caption{The position angle (left panel), viewing angle (middle panel), and Doppler factor (left panel) of a plasmoid in the jet stream of OJ 287, based on the jet precession$+$nutation model of the present paper (black lines, see the parameters in Table \ref{tab_parameters_precession_nutation}. For comparison, the pure precession model is marked by gray lines. The epoch of the SEDa (SEDb) is marked  by a purple dashed (black continuous) vertical line.}
\label{fig_SB2018_z_d}
\end{figure*}

By fitting the jet kinematics of OJ~287, we derived a precession-only model with the precession period of $P_p \approx 23$~yr. In addition, we derived a precession + nutation model with a similar precession period ($P_p\approx 32$ yr) together with a much shorter nutation
period ($P_n\approx1.6$ yr). 
Changes in the observed apparent velocities of the identified jet components are attributed to changes in the viewing angle, hence changes in the Doppler factor. 
In Fig.~\ref{fig_SB2018_z_d} we show the position angle, viewing angle, and Doppler factor of a plasmoid in the jet stream of OJ 287, with fitted parameters presented in Table~\ref{tab_parameters_precession_nutation}.

SED$_a$ ($2015.92$) was observed close to a local minimum of the Doppler factor, while SED$_b$ ($2017.02$) was observed shortly after a global maximum of the Doppler factor. Meanwhile the position angle changed about 80 degrees. This suggests that during the time elapsed between the observation of SED$_a$ and SED$_b$, the VLBI structure of the inner jet changed drastically within a short time. 

During the observation time of SED$_{a}$, the viewing angle derived from the SED fitting as well as the viewing angle derived from the jet kinematics were larger than the corresponding values for SED$_b$. Thus, the two independent methods give consistent timing.

The peak value of the Doppler factor from modeling SED$_{b}$ ($\approx 45$), i.e. the SED which corresponds to the epoch closest to the lower projected precession reversal point, is more than twice of the peak Doppler factor derived from the jet precession+nutation model ($\approx18$, see Fig. \ref{fig_SB2018_z_d}). It is important to note that we derived an upper limit for the viewing angle of the jet for SED$_b$ ($\Phi<2\degr$) based on the lower limit on the Doppler factor. The viewing angle is probably closer to 0, which means extreme beaming of the light. Though there is a numerical tension between the Doppler factors from the kinematical and the SED modeling, we can infer from the analyses that the jet was pointing very close to the observer's line of sight at that time. We also note that the errors on the nutation angle (i.e. half-angle of the jet) and precession angle (i.e. half-angle of the precession cone), combined with the short nutation period (1.6 years) still allow to compare the Doppler factors. We thus infer that the viewing angle changes are able to explain the new SED state in the framework of the precession$+$nutation model presented in this paper.

\section{Generalized approach to modelling radio light curves}
\label{lightcurves}
Given our analysis that can explain the (quasi)periodic radio outbursts by jet precession and nutation, radio light curves could in general be modelled as a superposition of periodic and non-periodic, stochastic processes. In the power-density spectrum $S(f)$ or periodograms, periodic processes show up as peaks at specific frequencies $f_{\rm per}$, while the non-periodic, stochastic processes are represented by a power-law function of the frequency $S(f)\propto f^{\gamma}$ \citep{timmer}, with $\gamma=-1$ corresponding to the flickering or ``red noise" process, $\gamma=0$ stands for the white noise, and $\gamma=-2$ stands for the random-walk noise. In our model of the radio flux variability, based on the bulk jet precession and nutation, the power-law noise is complemented by the periodic signal that consists of a precession with a frequency $\omega_{\rm p}$ and a nutation with a generally higher frequency $\omega_{\rm n}=2(\omega_{\rm orb}-\omega_{\rm p})$, where $\omega_{\rm orb}$ is the orbital frequency of the BH binary.

\begin{figure*}[h!]
    \centering
    \includegraphics[width=0.28\textwidth]{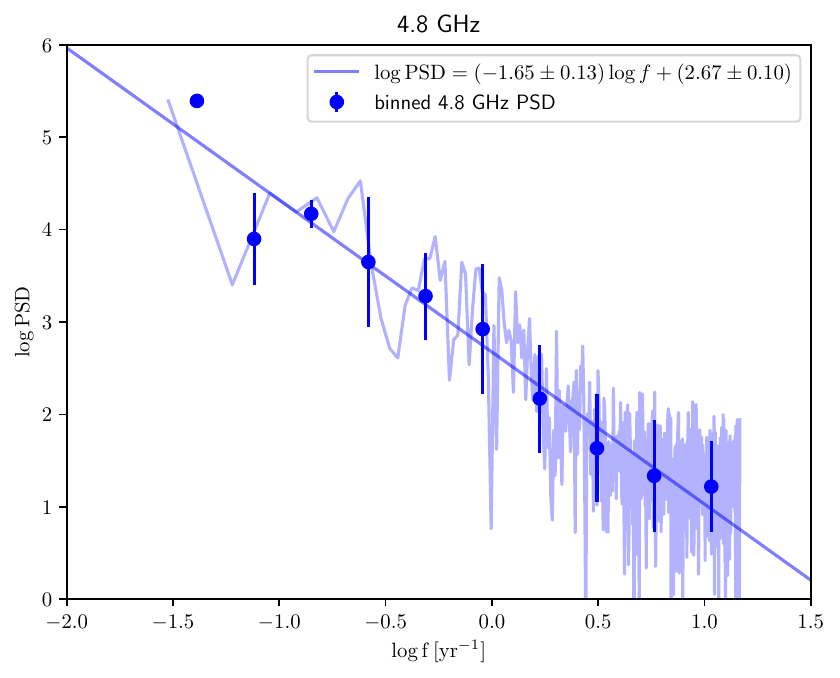}
     \includegraphics[width=0.28\textwidth]{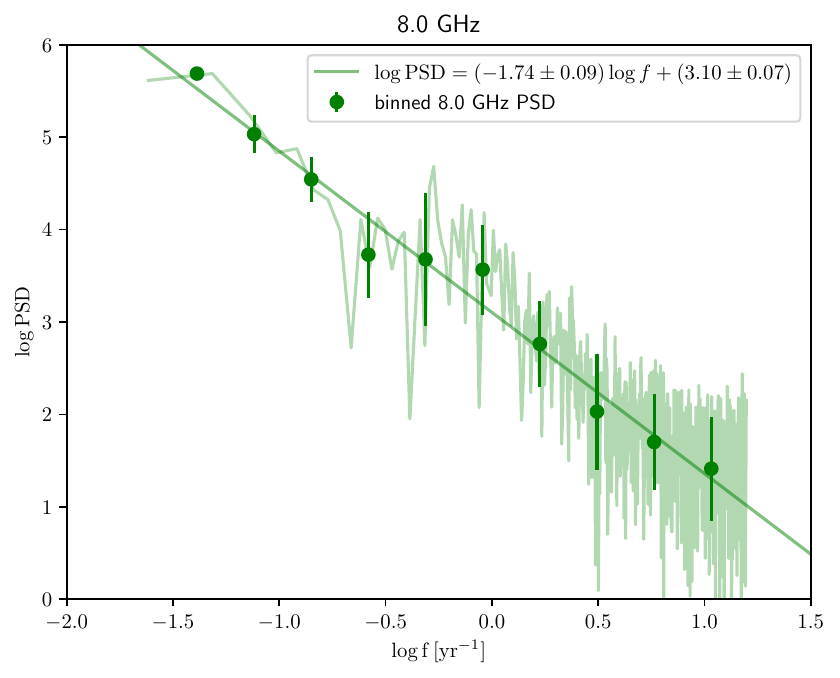}
      \includegraphics[width=0.28\textwidth]{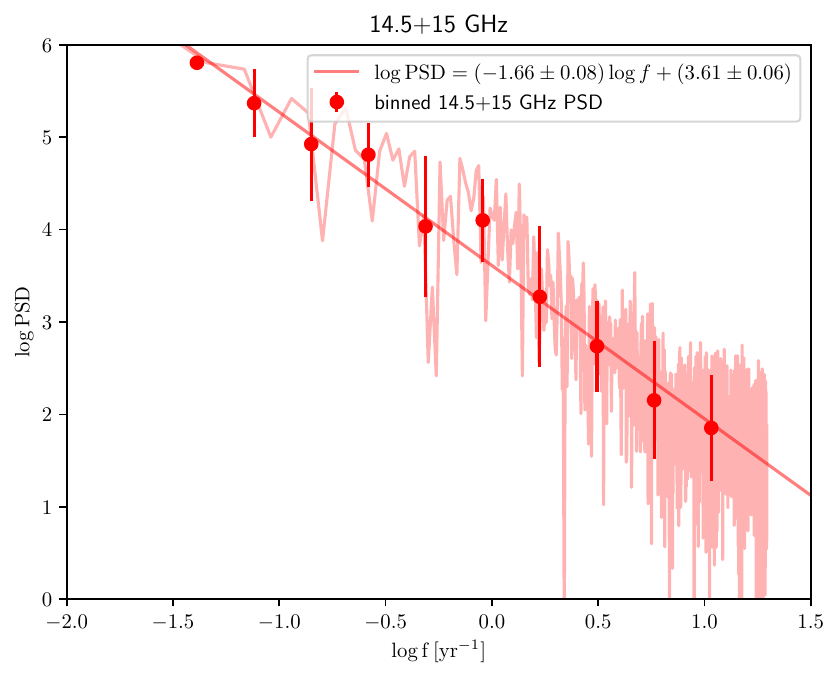}
    \caption{ The power spectral densities (PSDs) calculated for the continuum light curves at 4.8 (left panel, blue line), 8.0 (middle panel, green line), and 14.5$+$15 GHz (right panel, red line). The mean values for 10 bins are shown as points with the corresponding errorbars. The binned PSDs were fitted with simple power-law functions, $\log{{\rm PSD}}= \gamma \log{f}+\beta$ with the mean slopes of $\gamma=-1.65$, $-1.74$, and $-1.66$ and the mean intercepts of $\beta=2.67$, $3.10$, and $3.61$ for $4.8$, $8.0$, and $14.5+15$ GHz, respectively.}
    \label{fig_psd}
\end{figure*}

\begin{figure*}[h!]
    \centering
    \includegraphics[width=\textwidth]{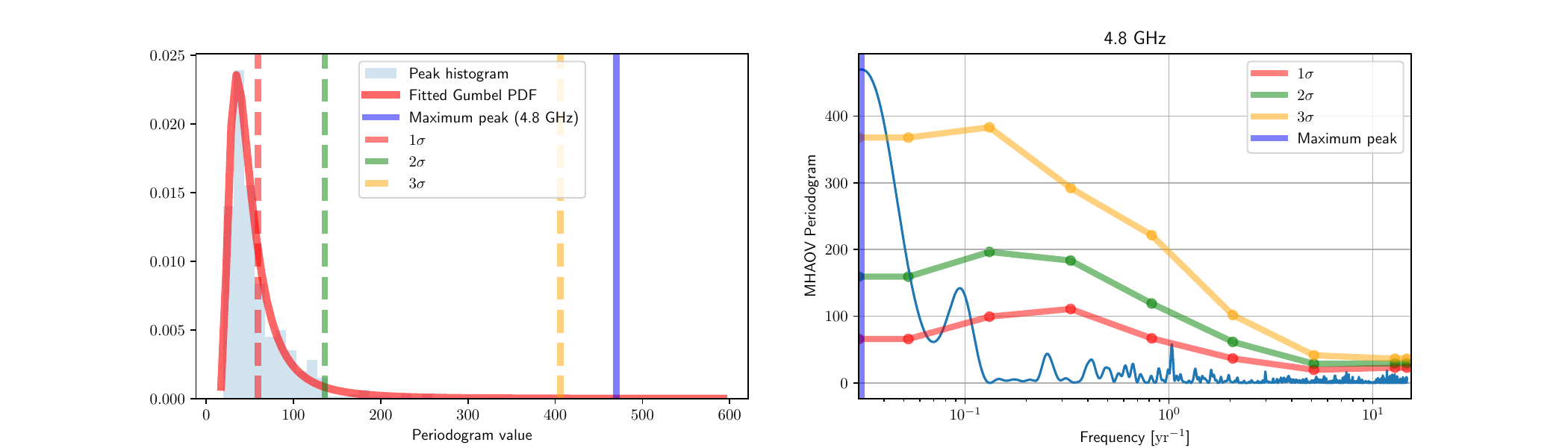}
      \includegraphics[width=\textwidth]{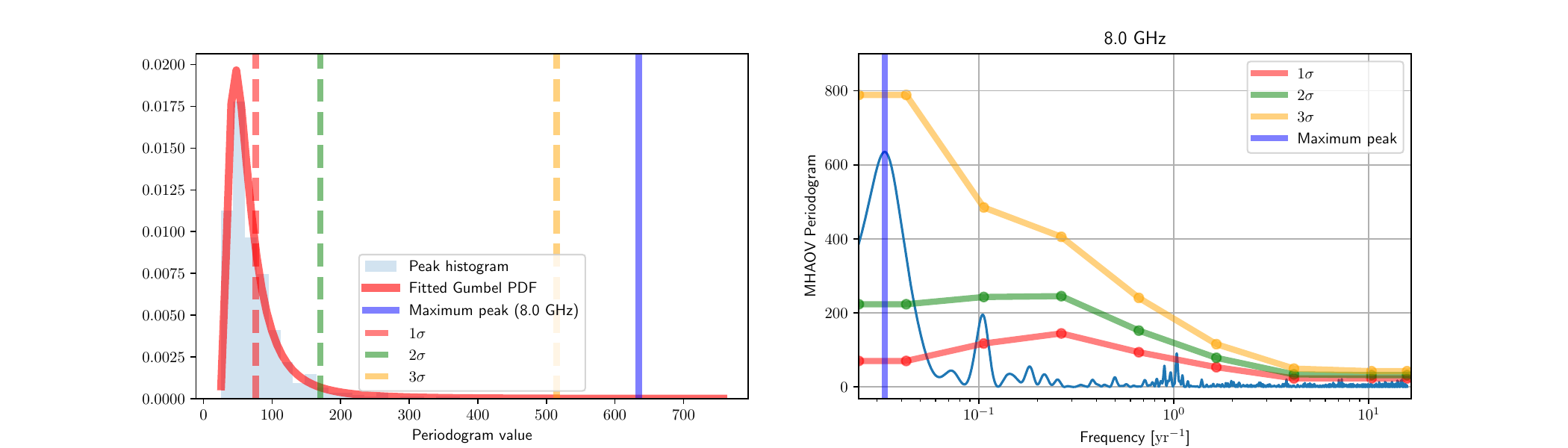}
       \includegraphics[width=\textwidth]{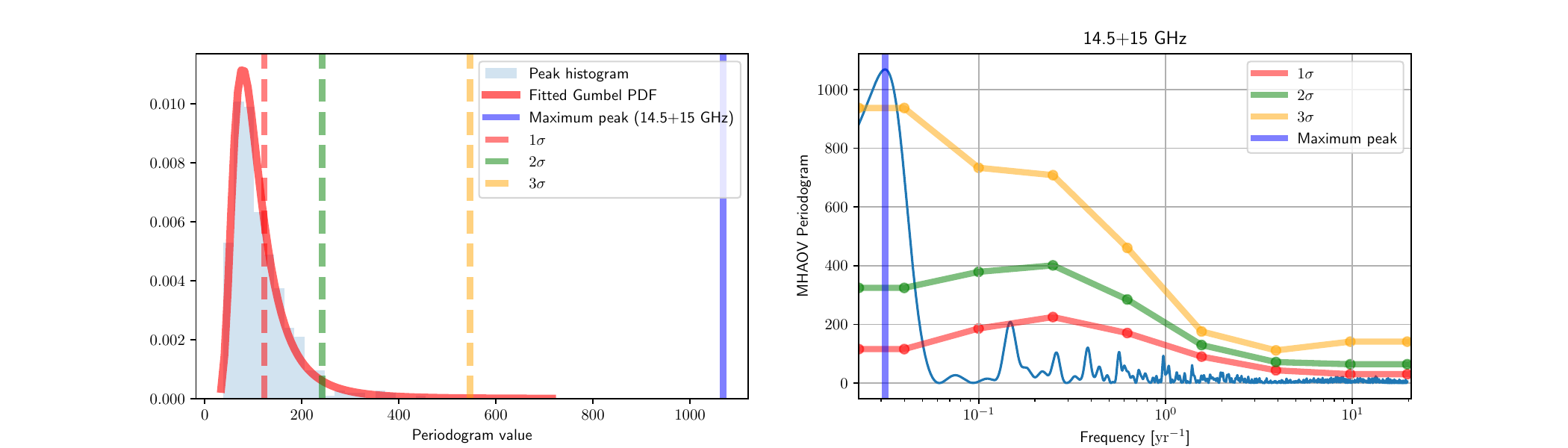}  
    \caption{The periodograms calculated based on the multiharmonic analysis of variance (MHAOV) for the continuum light curves at 4.8, 8.0, and 14.5$+$15 GHz (from the top to the bottom). In the left panel, we show the histogram of MHAOV periodogram peaks constructed from 50 generated boostrapped light curves with 10 best identified peaks. From the fitted Gumbel probability density function, we inferred 1$\sigma$ (0.3173, red), 2$\sigma$ (0.0455, green), and 3$\sigma$ (0.0027, orange) confidence levels for the whole frequency interval where the smallest frequency is given by the light curve duration, while the highest frequency is set to $(2\times \text{mean sampling separation})^{-1}$, i.e. the Nyquist frequency. In the right panel, the confidence levels depicted in the MHAOV periodogram are calculated for specific frequency bins whose central frequencies are marked by points and they exhibit a decreasing tendency towards higher frequencies.}
    \label{fig_mhaov}
\end{figure*}

To demonstrate this specifically, we take the radio continuum light curves at 4.8, 8.0, and the combined 14.5+15 GHz light curve and calculate their power spectral densities (PSDs) using the fast Fourier transform algorithm. The PSDs are shown in Fig.~\ref{fig_psd}. It is seen that the PSDs decrease approximately as simple power-law functions. To analyze this, we binned the Fourier powers into 10 equal-sized bins between the minimum and the maximum frequencies. The Fourier power per bin was calculated as an average value of the powers that fall into the bin and the uncertainty was determined as the standard deviation, see ten points in Fig.~\ref{fig_psd}. The binned power density spectra were then fitted with simple power-law functions using the form $\log{{\rm PSD}}=\gamma \log{{\rm f}}+\beta$. The best fit power-law functions are shown as lines in Fig.~\ref{fig_psd}. For all the three radio frequencies, the slope is the same within uncertainties, $\gamma\sim -1.7$, while the intercept increases from $\beta=2.67$ to $\beta=3.61$ from 4.8 to $14.5+15$ GHz, respectively. This demonstrates the presence of the stochastic process with the red-noise and the damped random-walk nature \citep{timmer,2009ApJ...698..895K}. The value of $\gamma$ falls into the interval $(-2,-1)$ inferred by \citet{2020ApJS..250....1T} for a limited blazar sample, though in the $\gamma$-ray domain.  The presence of periodic processes is not apparent in PSDs.

The potential periodicities appear when one expresses PSDs considering the red-noise power-law process, such as in the periodogram of the orthogonal multi-harmonic analysis of variance \citep[MHAOV; ][]{1996ApJ...460L.107S}, which is suitable for unevenly sampled light curves. In Fig.~\ref{fig_mhaov}, we show calculated MHAOV periodograms for the three frequencies -- 4.8, 8.0, and the combined 14.5+15 GHz -- from the top to the bottom panel, respectively. In the periodograms, we show $1\sigma$, $2\sigma$, and $3\sigma$ significance levels that are inferred from fitting the generalized extreme value distribution functions to the histograms of periodogram peak distributions. The periodogram peak distributions are constructed from a few hundred bootstrap realizations. In the left panel of Fig.~\ref{fig_mhaov}, we infer confidence levels for the whole frequency range, while in the right panel, we calculate 1$\sigma$, 2$\sigma$, and 3$\sigma$ levels for different frequency bins with the logarithmic increment of $\log{2.5}\simeq 0.4$ whose central frequencies are represented by points. For 4.8 GHz light curve, we find the periodogram peaks at 0.97, 10.62, and 32.47 years that are at least at 1$\sigma$ level. For 8.0 GHz light curve, we find similar periodicity peaks at 0.97, 9.56, and 30.45 years with the significance of $\gtrsim 1\sigma$. For the combined 14.5+15.0 GHz light curve, the periodicity peak distribution is more complex with the peaks at 1.03, 1.78, 2.59, 3.84, 6.82, and 31.43 years that are close to at least 1$\sigma$ significance level, see Fig.~\ref{fig_mhaov}. For all the three light curves, the most significant peak is at $\sim 30-32$ years. The periodicity analysis demonstrates the stochastic nature of the radio variability for OJ~287 in conjunction with periodic, deterministic processes at different periods. The periodic process close to one year (though at only 1$\sigma$ significance) may be associated with the jet nutation, while the longer periods ($\sim$10 and $\sim$30 years) that have $\sim 2-3\sigma$ significance could be associated with the precession in the framework of our geometric precession-nutation model. The intermediate candidate periods at $14.5+15.0$ GHz, i.e. $1.78$, $2.59$, $3.84$, and $6.82$ years, could be related to the orbital motion of the putative SMBH binary.

\begin{figure*}[h!]
    \centering
   \includegraphics[width=0.28\textwidth]{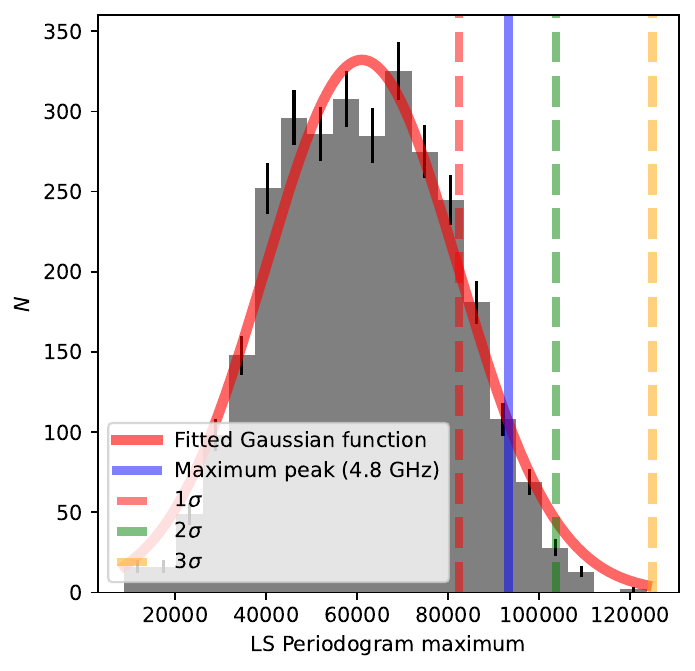} \includegraphics[width=0.6\textwidth]{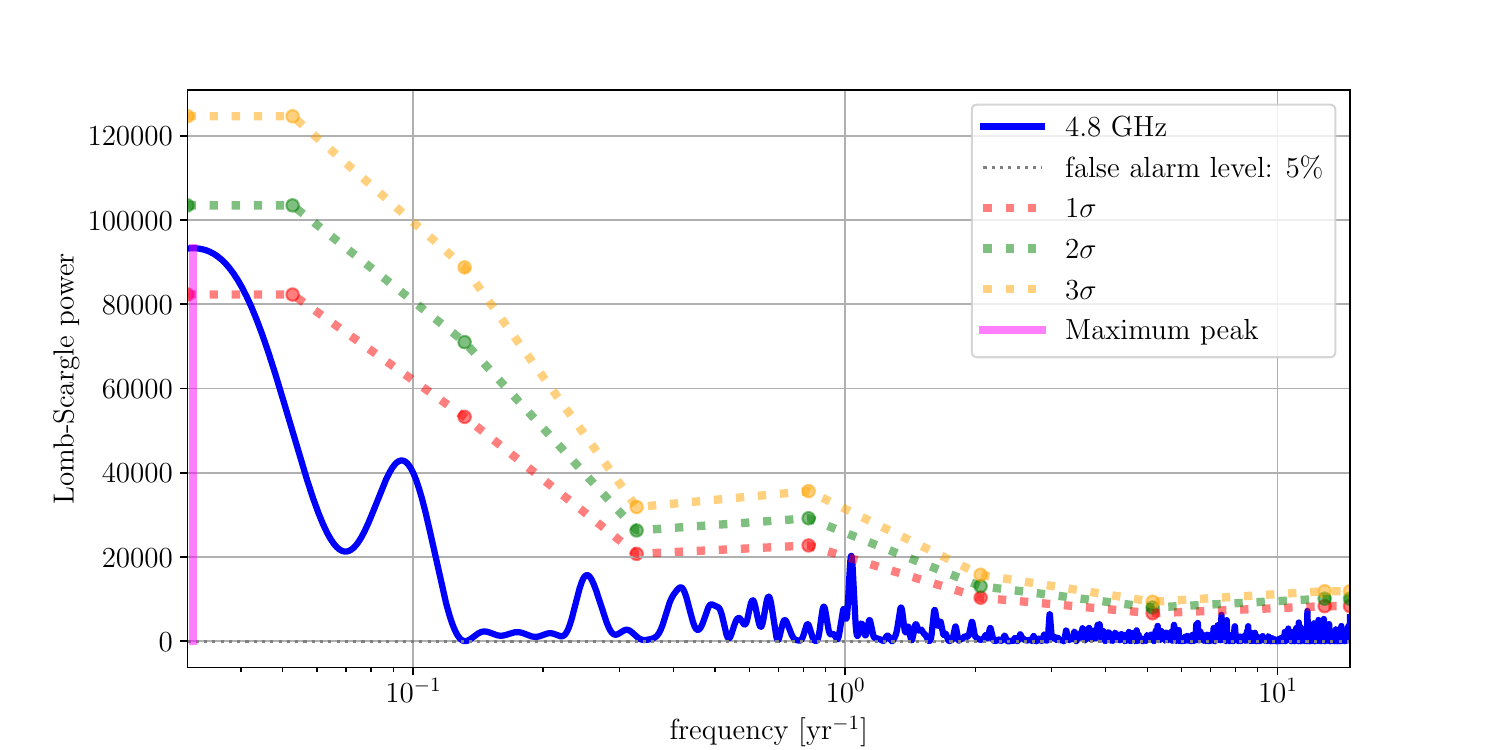}
   \includegraphics[width=0.28\textwidth]{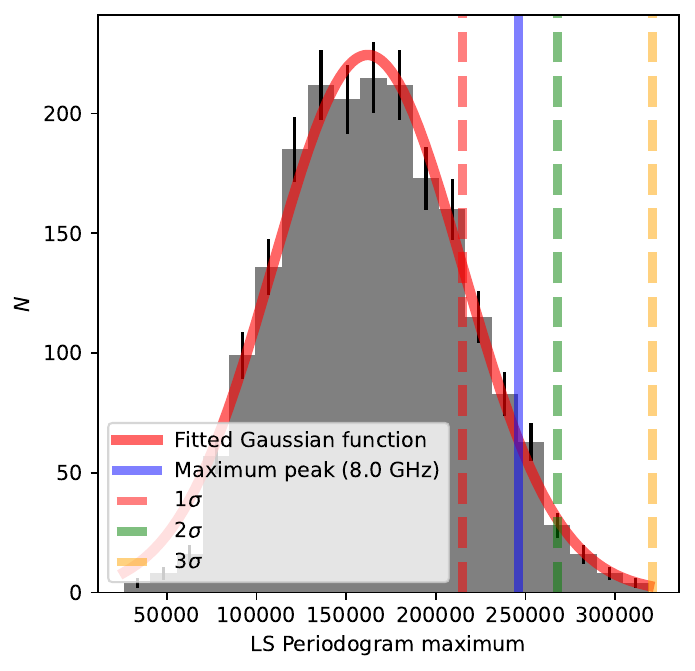} \includegraphics[width=0.6\textwidth]{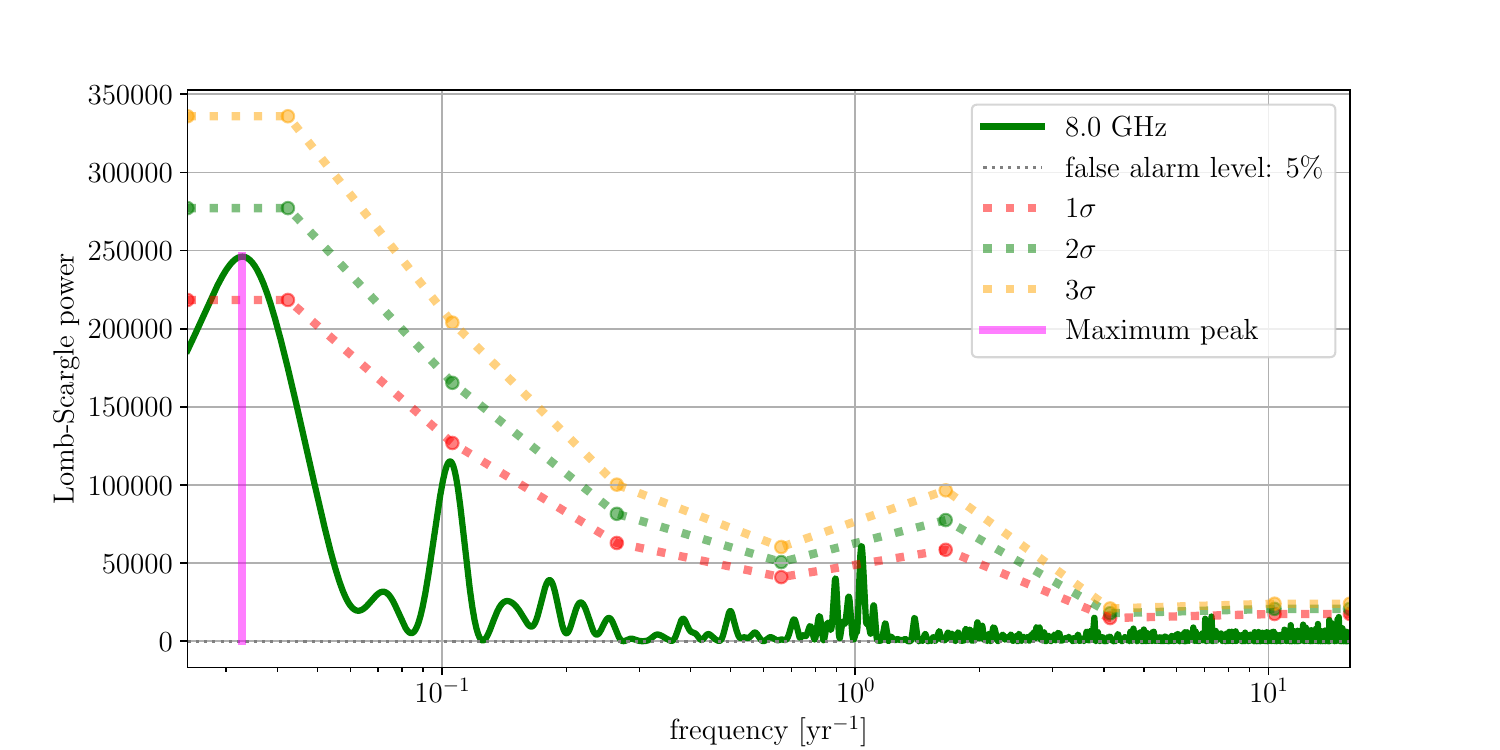}
    \includegraphics[width=0.28\textwidth]{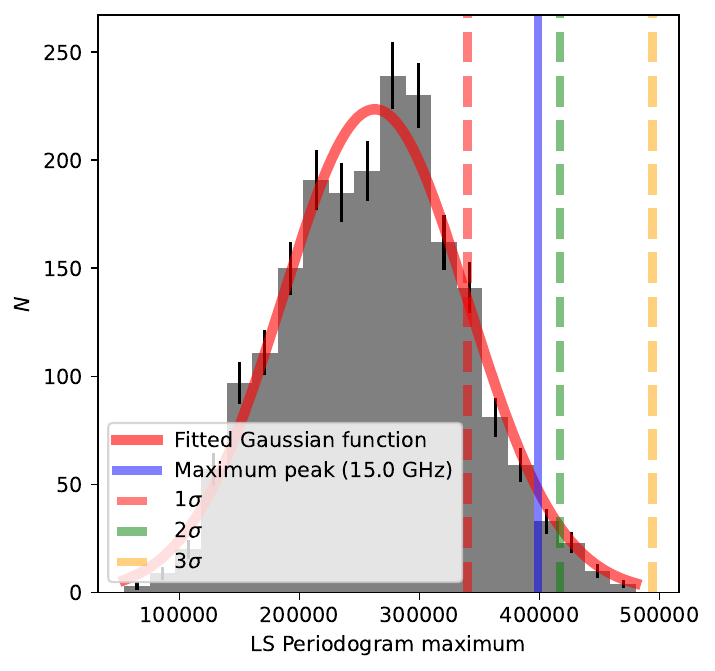} \includegraphics[width=0.6\textwidth]{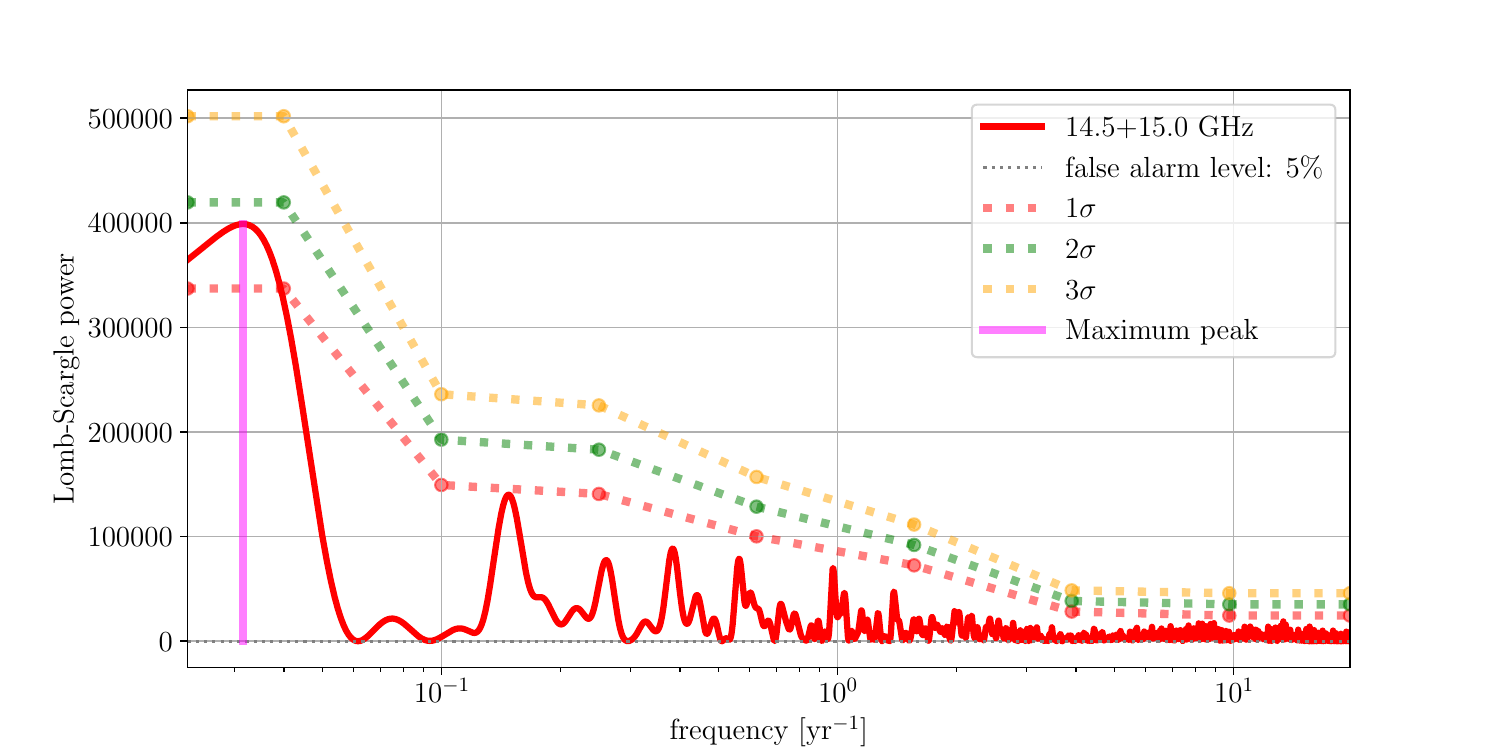}
        \caption{From the top to the bottom panels, Lomb-Scargle (LS) periodograms for the flux densities at $4.8$ (blue line), $8.0$ (green line), and $14.5+15.0$ GHz (red line), respectively. In the left panels, we show the distributions of LS periodogram peaks inferred from 2000 bootstrap realizations of the radio light curves. The vertical lines mark 1, 2, and 3$\sigma$ confidence intervals calculated for the whole frequency range. In the right panels, we show LS periodograms with marked 1, 2, and 3$\sigma$ confidence levels that were inferred for frequency bins whose central points are represented by points. The dotted horizontal gray line marks 5\% false alarm probability level.}
    \label{fig_lomb_scargle}
\end{figure*}

\begin{figure}[h!]
    \centering
    \includegraphics[width=\columnwidth]{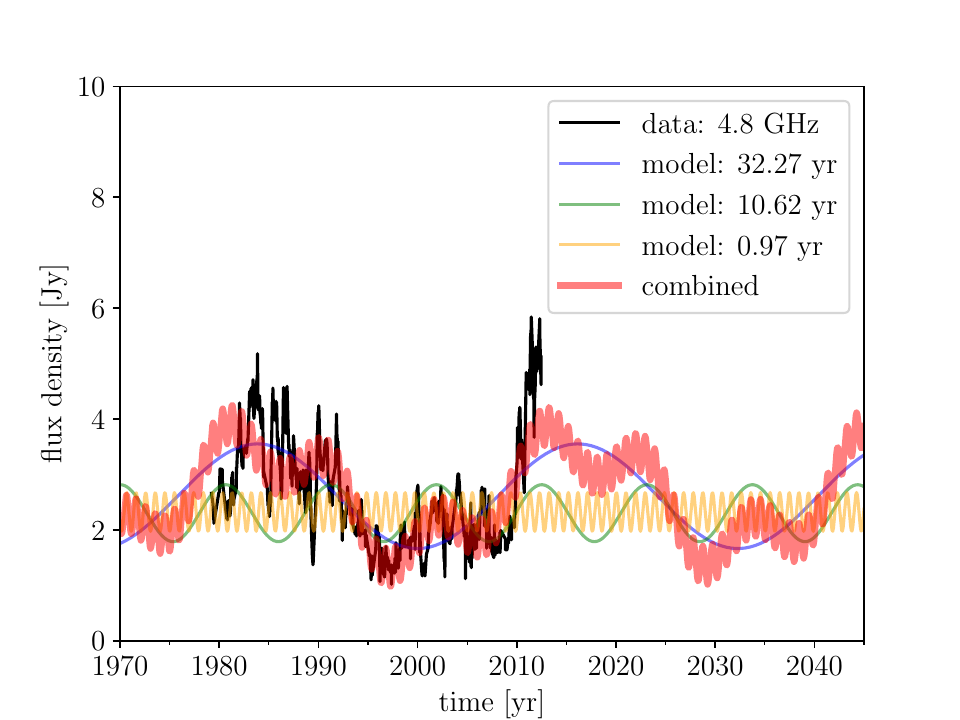}
      \includegraphics[width=\columnwidth]{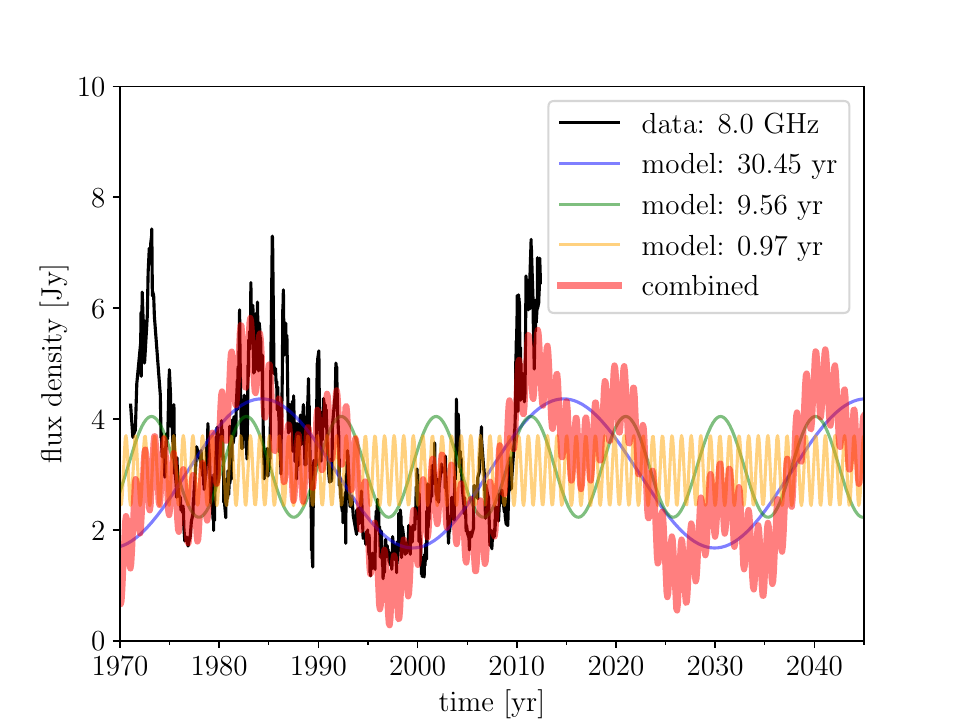}
        \includegraphics[width=\columnwidth]{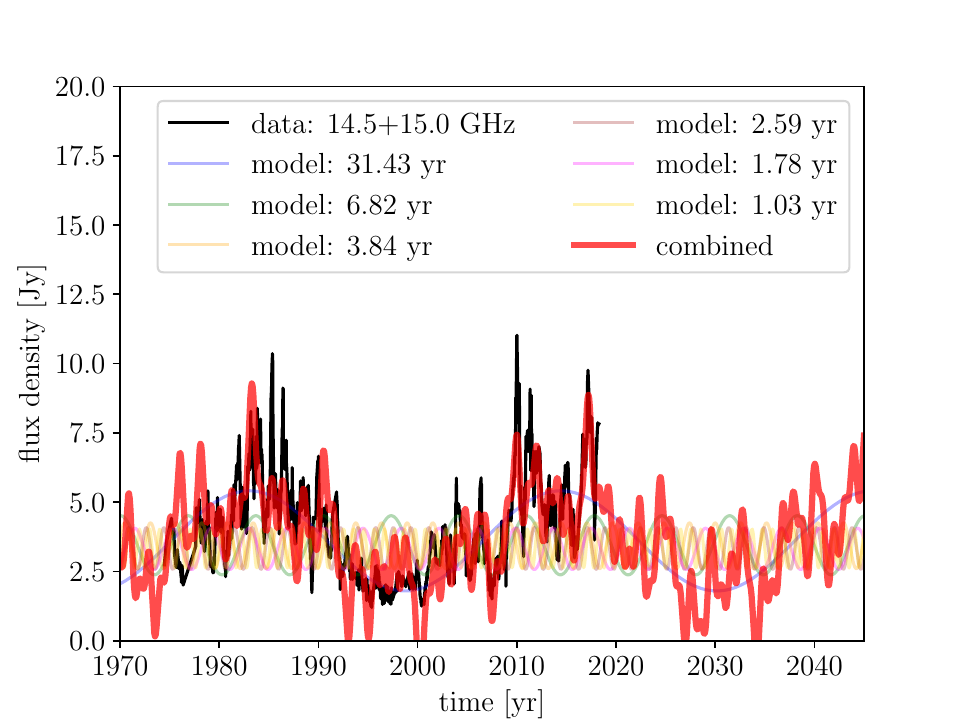}
        \caption{From the top to the bottom panels: modelled radio light curves versus observed continuum flux densities at 4.8 GHz, 8.0 GHz, and 14.5+15.0 GHz. For 4.8 and 8.0 GHz light curves, the model light curves are constructed by summing three signals with periods of $\sim 30$, $10$, and $1$ years, while at 14.5+15.0 GHz, the final model light curve is more complex with $30$- and $1$-yr periods present, and additional four periodic signals present.}
    \label{fig_lomb_scargle_model}
\end{figure}

Interestingly, the same periodicities are also reproduced using the Lomb-Scargle (LS) periodogram, see Fig.~\ref{fig_lomb_scargle}. In the left panels of Fig~\ref{fig_lomb_scargle}, we show the histograms of LS periodogram maximum constructed from 2000 bootstrap realizations. From the fitted Gaussian function, we inferred 1, 2, and 3$\sigma$ confidence levels for the whole frequency range. In the right panel, we show LS periodograms for 4.8, 8.0, and 14.5$+$15.0 GHz light curves, where we also depict 1, 2, and 3$\sigma$ confidence levels calculated for different frequency bins in a similar way as for the MHAOV periodograms (using 1000 bootstrap realizations for each frequency bin). In this case, the periodocity peak significance is generally lower, with the peaks at $\sim 30$ years and $\sim 1$ year being above the 1$\sigma$ level (see the 8.0 GHz light curve), while the other peaks that are close to and below $\sim 10$ years are at or mostly below 1$\sigma$ level.

The combined model light curves versus the observed light-curves at 4.8, 8.0, and 14.5+15.0 GHz are shown in Fig.~\ref{fig_lomb_scargle_model}. The model light curves are constructed as the sum of periodic components and can reproduce well the main features of the observed radio light curve, with the potential to predict future radio outbursts of OJ~287, which can be utilized for testing the model. For 4.8 and 8.0 GHz radio light curves, the main features can be reproduced by summing three signals with periods of $\sim 30$, $\sim 10$, and $\sim 1$ year, which is roughly consistent with the longer precession period, short nutation timescale, and an additional periodic signal that could be associated with the orbital period of the binary black hole. For the higher frequency at $14.5+15$ GHz, the periodogram is more complex, with the $\sim 30$- and $\sim 1$-year signals present, but additional four periodic signals (6.82, 3.84, 2.59, 1.78 years) are needed to reproduce the full variability. We note that at 14.5$+$15 GHz, the model has also predicted correctly the radio flare around 2017.4. A smaller radio flare is expected around 2024.8-2024.9.
In the following subsection, we approach the light curve modelling from a different perspective, making use of the autoregressive intergrated moving average (ARIMA) models.

\subsection{Light curve modelling using ARIMA model: determining timescale of stochastic processes}
\label{subsec_arima}

We test if the three radio light curves could be modelled using the autoregressive integrated moving average model (ARIMA), which can be applied in astronomical time series analysis \citep{2018FrP.....6...80F,2019AJ....158...58C} to describe time series using stationary stochastic processes \citep[see e.g.][and references therein for the application on blazar light curves]{2020ApJS..250....1T}. Training the ARIMA model on the observed light curve or its part, one can make a prediction for the future evolution of the system and compare with the actual observations to see whether the source behaves in a stationary stochastic way as described by ARIMA on shorter or longer timescales. Further motivation for such an analysis is given by \citet{2016MNRAS.461.3145V} who show that stochastically driven flux variability may mimic quasi-periodic behaviour in AGN light curves, especially when only few cycles are captured ($\leq 2$; see also \citeauthor{2020A&A...634A.120A}, \citeyear{2020A&A...634A.120A} for the periodicity detection in $\gamma$-ray detected \textit{Fermi}-LAT blazars). 

ARIMA$(p,d,q)$ process contains $p$ autoregressive terms, $d$ denotes the order of time series differencing, and $q$ stands for the number of moving-average terms. The autoregressive part AR$(p)$ calculates the flux density at time $t$ as the linear combination of $p$ weighted past values of the flux density. The moving average MA$(q)$ part calculates the current value as the sum of the current white noise and $q$ weighted past white-noise values $\epsilon$. The integrated order of differencing $d$ ensures that the studied light curve is stationary. The final ARIMA$(p,d,q)$ model equation for the once-differenced time series, i.e. light curve transformed by $f_{t}'=f_{t}-f_{t-1}$ , can generally be expressed as,
\begin{align}
    f_{t}'=&c+\alpha_1 f_{t-1}'+\alpha_2 f_{t-2}'+\ldots+\alpha_{p} f_{t-p}'+\notag\\
    &+\epsilon_{t}+\theta_1 \epsilon_{t-1}+\theta_2 \epsilon_{t-2} \ldots \theta_{q}\epsilon_{t-q}\,,
    \label{eq_ARIMA_eq}
\end{align}
where the parameters $c$, $\alpha_i$, and $\theta_{j}$ can be determined using the Box-Jenkins method as implemented e.g. in \texttt{python} \texttt{statsmodels}.

For each radio light curve $(4.8, 8.0, 14.5+15.0){\rm GHz}$, we first performed a linear interpolation to regular time intervals separated by 0.1 years. Subsequently, with the help of the \texttt{python} module \texttt{statsmodels}, we performed the analysis to determine the order of $p$, $d$, and $q$ parameters. We provide details about the ARIMA model fitting in Appendix Subsection~\ref{appendix_arima}, in particular Table~\ref{tab_ARIMA_results1} and Fig.~\ref{fig_ARIMA_pdq_scan1}. Using (partial) autocorrelation functions as well as AIC and BIC values obtained from fitting different ARIMA models ($p=[0,\ldots,15]$, $d=[0,1]$, $q=[0,\ldots,15]$), we determined the preferred ARIMA model with $p=2$, $d=1$, and $q=1$ for the lower frequency light curves at 4.8 and 8.0 GHz. This class of ARIMA models has the smallest BIC, while the AIC indicates the higher $p$ order of $\geq 5$. For the 14.5+15.0 GHz light curve, the preferred model is ARIMA(1,1,2) corresponding to the smallest BIC. The Augmented Dickey-Fuller (ADF) test indicates that its ADF $p$-value drops below the threshold value of 0.05 after the first differenciation of all the three light curves, which supports setting $d=1$ for reaching stationarity. The best-fit coefficients for the corresponding AR and MA terms are listed in Table~\ref{tab_ARIMA_results1} for each light curve. In Fig.~\ref{fig_ARIMA_fit}, we show the best-fit to the light curves for the corresponding ARIMA$(p, d, q)$ model.

We also perform additional tests using the light curves that were regularly sampled to $0.05$-year time steps. For cases $d=0$ and $d=1$, the number of AR terms $p$ is in the range between 1 and 3, while the number of MA terms $q$ is in the range between 0 and 4, see Table~\ref{tab_ARIMA_results2} and Fig.~\ref{fig_ARIMA_pdq_scan2}. The small increase in $p$ and $q$ order thus merely reflects the time-step decrease by a factor of two, while the timescale of the stationary stochastic process remains in the range of $0.05-0.2$ years.

Furthermore, we apply fractional differencing to all the three light curves using the \texttt{python} library \texttt{fracdiff} \citep{de2018advances}. This ensures that light curves can be described by a stationary stochastic process, while still keeping the largest possible signal. For a larger time step of 0.1 years, the term $d$ reaches values intermediate between 0 and 1 and less than 0.5. For a smaller time-step of 0.05 years, $d$ decreases to close to zero for 4.8 and 14.5$+$15 GHz light curves, while it is $d=0.41$ for the 8.0 GHz light curve. The $p$ and $q$ values for fractionally differenced light curves are displayed in Fig.~\ref{fig_ARIMA_pdq_scan3}. For the time step of 0.1 years and the model with the smallest BIC, $p$ and $q$ are consistently equal to one, while for the time step of 0.05 years, $p$ is in the range if $1-3$ and $q=0$. Hence, again the characteristic timescale for the stationary stochastic process is $\sim 0.05-0.15$ years.  

We also fit the ARIMA model to the temporal evolution of the quasi-stationary component a, which was interpolated with the time-step of 0.15 years. Using fractional differencing and the ADF test, we determined the difference degree of $d=0.555$ to reach the stationarity of the temporal evolution, see Fig.~\ref{fig_compa_AIC_BIC}. Based on the minimum AIC and BIC values, we found that the optimal ARIMA models have a low order of autoregressive and moving average terms with $p\leq 3$ and $q\leq 2$, which indicates the timescale of $\sim 0.15-0.45$ years for stationary stochastic processes. In Fig.~\ref{fig_compa_ARIMA_fit_forecast}, we show the best-fit ARIMA(1,0.555,0) model (left panel) and the forecast testing (right panel). 

The fitted ARIMA models and the ADF test suggest that without differencing, radio light curves are non-stationary, i.e. they exhibit a long-term trend, which is apparent by eye. The long-term jet precession and nutation are viable candidate mechanisms to address this trend as we showed in Section~\ref{section_modelling_precession_nutation}. Forecast mean ARIMA models, which were trained on approximately the half of the corresponding radio light curves, tend to converge towards a constant value, which reflects the short-term memory of the preferred ARIMA models, i.e. $p$ and $q$ are at most 2 for the time step of 0.1 years, hence they depend on at most two lagged flux density and white noise values going back in time by $\lesssim 0.2$ years. The $p$ order of $\leq 2$ was also found by \citet{2020ApJS..250....1T} while fitting blazar $\gamma$-ray light curves using the ARIMA model. The low order implies the presence of a damping timescale of the propagating disturbances in the compact disc-jet system, beyond which they do not affect significantly the emission processes.

Within 2$\sigma$ confidence interval, the forecast values are consistent with the observed radio outburst around $\sim 2010$-$2011$, see Fig.~\ref{fig_ARIMA_forecast}. However, the sum of three main deterministic periodic processes indicated by the periodograms, in contrast to ARIMA stationary stochastic processes, can reproduce the main long-term light curves trends, see Fig.~\ref{fig_lomb_scargle_model}, with a clear trend forecasting in comparison to the ARIMA model.

Overall, the preferred ARIMA models with the weighted terms lagged by $<0.2$ year imply that the OJ~287 system is dominated by stochastic processes on short timescales, which is also consistent with the simple power-law shape of PSDs, see Fig.~\ref{fig_psd}. In contrast, on the timescales $\geq 1$ year the system is dominated by deterministic periodic processes related to precession and nutation jet motions, which is indicated by precession-nutation model fits and periodogram peaks, see Figs.~\ref{fig_mhaov} and \ref{fig_lomb_scargle}. 


\subsection{Comparing precession parameters for a sample of Blazars}
\label{sec_comparison}
In this section, we extend our analysis to a larger number of systems in order to identify sources where jet precession/nutation seems to be present.

In order to explain the structural changes on parsec scales and the concurrent flux-density changes observed for individual blazars, several authors have employed bulk precession of the jet. 
In Tables~\ref{precession} and \ref{precession1} we list sources for which the entire set of precession parameters has been successfully fitted or estimated. 
The tabulated parameter values have been obtained from the literature and go back to studies obtained in various frequency bands. The information is based on data sets widely differing in quality and quantity, hence the uncertainties in the precession parameters also vary substantially.

\begin{figure}[h!]
    \centering
\begin{minipage}{\columnwidth}
\includegraphics[width=\columnwidth]{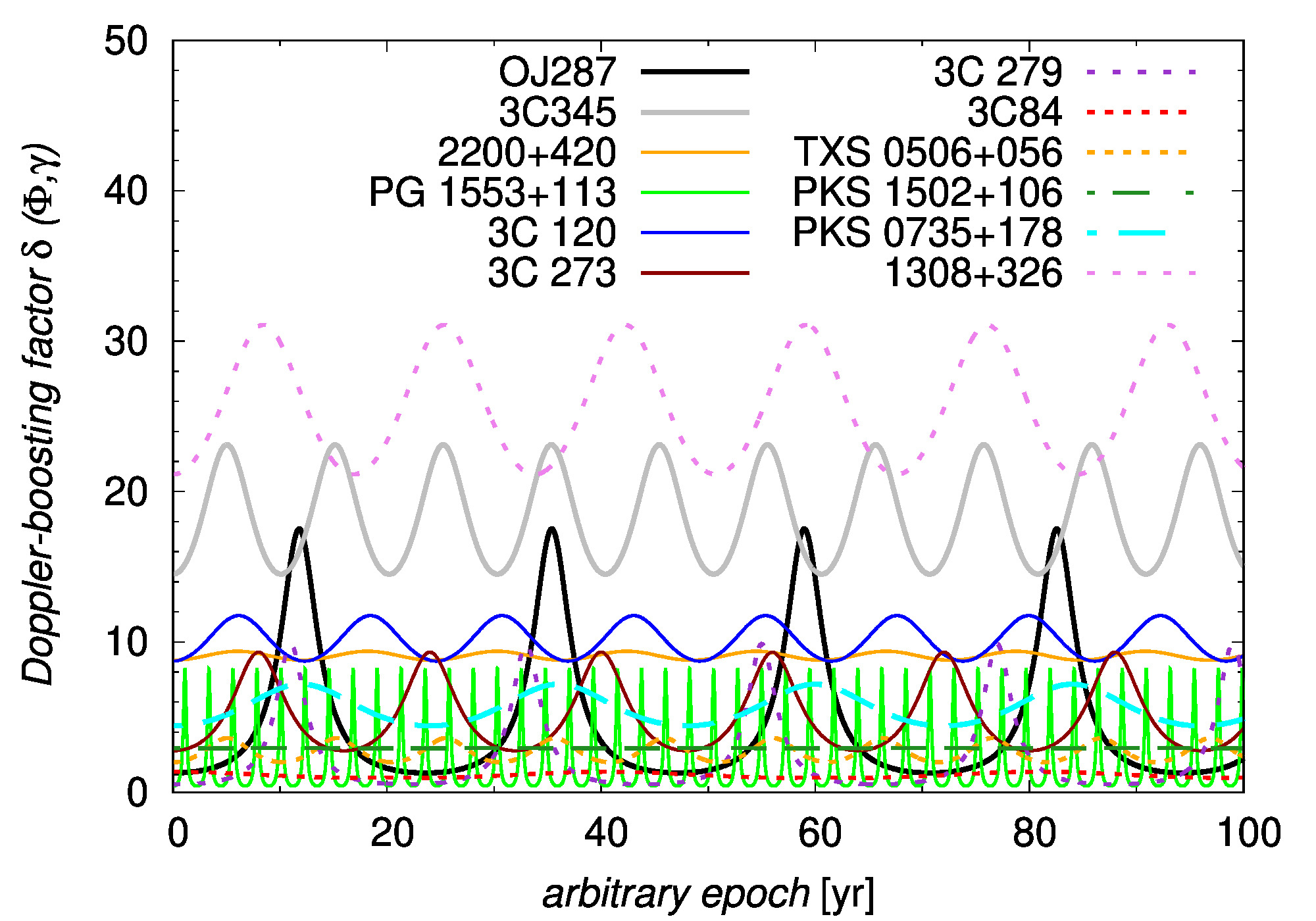}
[a]
\end{minipage}
\begin{minipage}{\columnwidth}
\includegraphics[width=\columnwidth]{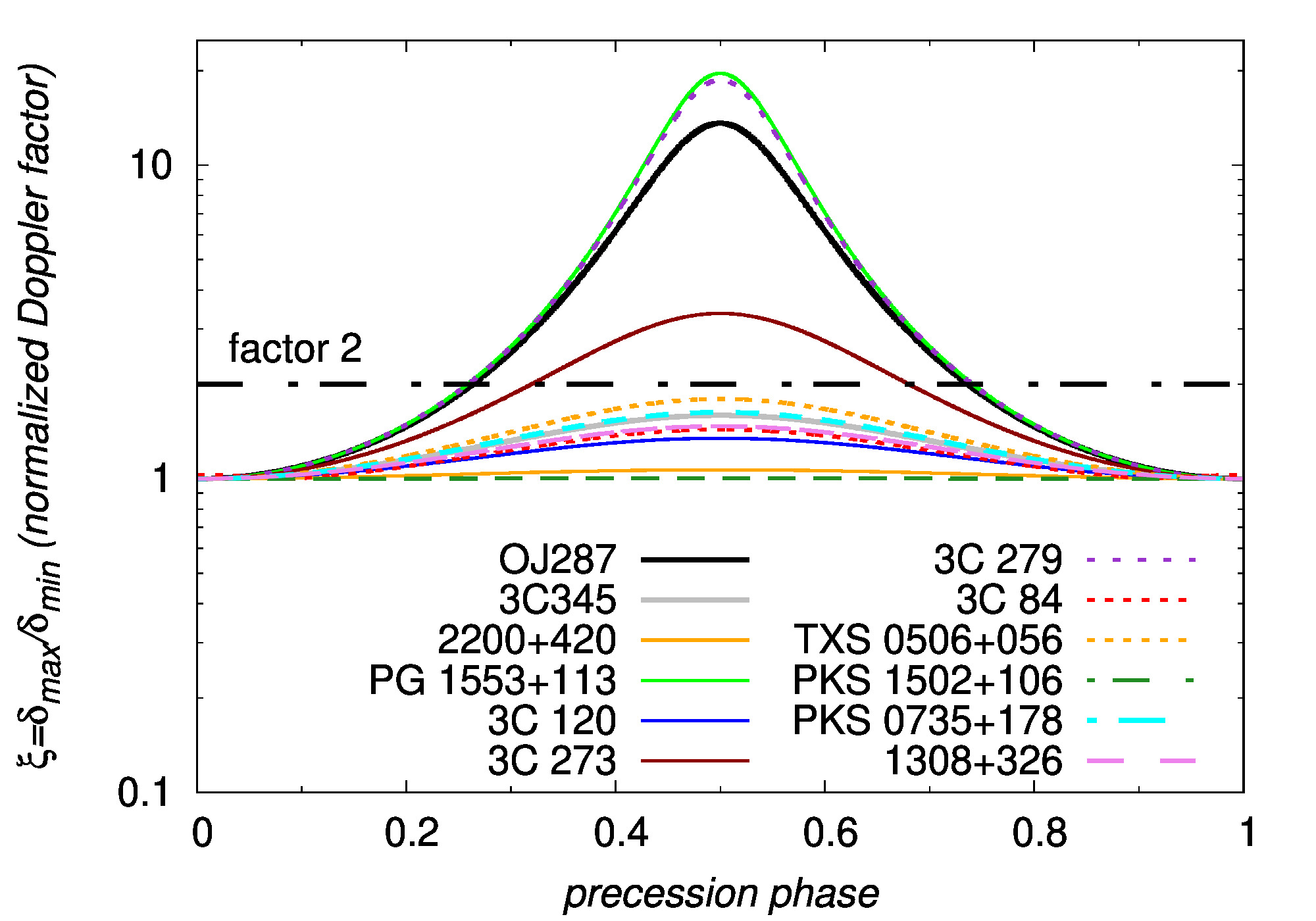}
[b]
\end{minipage}
        \caption{[a] shows the time-dependent Doppler-boosting factor $\delta(t, \gamma, \Phi)$ 
        for the radio galaxies listed in Table~\ref{precession}. 
        In [b] the ratio of maximum to minimum Doppler factor 
        $\xi=\delta_{\rm max}/\delta_{\rm min}$ 
        derived for one precession phase is displayed. The dot-dashed line denotes a ratio $\xi=2$; most of the sources lie below this limit (8/12), while four lie above.}
    \label{fig_jet_precession}
\end{figure}

As noted earlier, jet precession is intimately connected to the radio flux variability, via the Doppler factor.
For favourable configurations the Doppler beaming can significantly boost the non-thermal jet emission. 
For comparison, we calculate and plot in Fig.~\ref{fig_jet_precession}[a] the time-variable (and periodic) Doppler boosting factor $\delta$ for most of the sources in Tables~\ref{precession} and \ref{precession1}, using the tabulated values of the model parameters. 
These AGN, which are classified as BL Lac object (0735+178, 2200$+$420, PG1553$+$113, OJ\,287, TXS 0506+056), 
Quasar (3C\,273, 3C\,279, 3C\,345, 1308+326, PKS 1502$+$106), Seyfert 1 galaxy (3C\,120), and radio galaxy (3C\,84), have estimated precession periods of the order of 1 to 40 years (as deduced from the literature). Recently, jet precession for the nearby low-luminosity radio galaxy M81 was confirmed with the period of $\sim 7$ years \citep{2023arXiv230300603V}. However, another solution is also discussed, specifically a precession-nutation model with the 7-year period corresponding to nutation, while precession with the period $\sim 800$ years can account for the observed long-term trend of the position angle. Hence, because of the ambiguity, this source is not included in our sample of precessing jets.

Observationally, perhaps a more meaningful parameter than the Doppler factor is the ratio of the maximum to the minimum Doppler factor, $\xi\equiv \delta_{\rm max}/\delta_{\rm min}=(S_{\rm max}/S_{\rm min})^{1/(p+\alpha)}$, which are computed from the temporal profile of the calculated Doppler factor inferred from the observational radio VLBI data, see Eq.~\eqref{eq_doppler_boosting}. 
This ratio $\xi$ is a more realistic measure of variability, being proportional to the ratio of the maximum to the minimum flux density.
However, since the values of the Doppler factor ratio depend on the ratio between the maximum and minimum flux densities in the light curves, the observationally inferred values of $\xi$ would strongly depend on the considered epochs of monitoring. If an outburst is included, the ratio would increase, while epochs of nearly quiescence would lead to a low ratio. We therefore included information on the considered epochs in Tables~\ref{precession} and \ref{precession1}.
Figure~\ref{fig_jet_precession} [b] displays the values of $\xi$ for the sources listed in Table~\ref{precession}. 

We list the computed values of the maximum Doppler-factor ratio $\xi_{\rm max}$ (the minimum is unity that implies zero or a negligible variability due to the Doppler boosting of a precessing jet) in Table~\ref{table_Doppler_ratio}.
For several sources listed in this table, the precession model fits the light-curves well. In \citet{britzen3c}, the correlation between the radio flux-density evolution and jet precession is shown. For 3C~84, the radio flux-density is modeled and nicely fits the OVRO and UMRAO data (15 GHz). A superposition of the observed optical magnitude in the B band and the Doppler factor of the precession model is shown for 3C~279 \citep{AbrahamCarrara}. For 3C~345 and 3C~120 the authors show the contribution of the precessing jet to the flux-density in the $B$-band (3C~345: \citeauthor{CaproniAbrahama2004}, \citeyear{CaproniAbrahama2004};  3C~120: \citeauthor{CaproniAbrahamb2004}, \citeyear{CaproniAbrahamb2004}). For PG 1553+113, the Doppler boosting factor is superposed to the \textit{Fermi} $\gamma$-ray light curve \citep[see Fig. 2, bottom panels in][]{Caproni2017}.

\begin{table*}[h!]

  \caption{A selection of AGN with variability and/or jet morphological changes that can be modelled
   applying a jet-precession model. 
   References are: 1) \citet{AbrahamCarrara}; 2) \citet{AbrahamRomero}; 3) \citet{Britzen2010b}; 4) \citet{Caproni2013}; 5)  \citet{Caproni2017};  6) \citet{CaproniAbrahama2004}; 7) \citet{CaproniAbrahamb2004}; 8) \citet{Britzen2018}; 9)  \citet{Britzen2017}. 
   Besides for OJ~287, the tabulated parameters of the sources have been derived from the cited papers. The parameters are explained in Table~\ref{tab_parameters_precession_nutation}. The ``considered period'' is the time period for which the listed parameters have been derived.}
    \centering{
    \resizebox{1.0\textwidth}{!}{
    \begin{tabular}{c|c|c|c|c|c|c|c|c|c}
    \hline
    \hline
    AGN   & 3C\,279 &  3C\,273 & PKS 0735$+$178 &  2200$+$420 & PG 1553$+$113  &  3C\,345 & 3C\,120 & OJ\,287 &1308$+$326 \\
    \hline
    Type  & OVV quasar & Quasar & BL Lac & BL Lac & BL Lac & Quasar & Seyfert 1 & BL Lac  & Quasar \\
    z &$0.538$ &$0.158$ &$0.45 \pm 0.06$ & 0.0686 &$0.49 \pm 0.04$ &  $0.5928$ & $0.033$ & $0.306$ & $0.997$\\
    $P_{\rm p, obs}$ [yr] & $22.0$ &$16.0$ &$\simeq 24$ & $12.11\pm0.65$ &  $2.245 \pm 0.011$ &$10.1 \pm 0.8$ &$12.3 \pm 0.3$ &$24 \pm 2$ &$16.9 \pm 0.3$ \\
    $\gamma$ & $9.1 \pm 0.6$ &$10.8 \pm 0.6$ & $\simeq 8$ &$5.35 \pm 1.31$ &$24.629 \pm 3.853$ & $12.5 \pm 0.5$ & $6.8 \pm 0.5$ & $10 \pm 1$ &$16.2 \pm 4.4$ \\
    $\eta_0$ [deg] & $37 \pm 1$ &$-120 \pm 3$ &$225.6$ & $-165.86\pm 0.16$ & $48.154 \pm 0.885$ &$-96 \pm 5$ & $-108 \pm 4$ &  $-149 \pm 40$ & $-42.8$ \\
    $\Phi_0$ [deg] &   $21.5 \pm 2$ &$10.0 \pm 0.8$ & $9.8$ & $4.43 \pm 2.16$ &$15.254 \pm 0.171$ & $2.6 \pm 0.5$ & $4.8 \pm 0.5$ &$12 \pm 3$ & $1.65 \pm 0.45$ \\
    $\Omega_{\rm p}$ [deg] &  $15.7 \pm 1.5$ &$3.9 \pm 0.5$ &1.85 & $0.51\pm 0.24$ &$10.059 \pm 0.156$ &  $1.3 \pm 0.5$ &$1.5 \pm 0.3$ & $10 \pm 3$ & $0.93 \pm 0.24$ \\
    considered period& 1964-1991& 1960-1990& 1995-2009& 1980.93-2007.682& 2009-2016& 1977-1995& 1976-1999& 1995-2017&  1995-2014\\
    Model & SMBBH & SMBBH & non-ballistic & SMBBH, nutat. &  SMBBH, $\gamma$-flares & SMBBH &   SMBBH & SMBBH or LT & Prec. jet (SMBBH)  \\
    \hline
    References & 1) & 2) & 3)  & 4)  & 5) & 6) & 7) & 8) & 9) \\
    \hline
    \end{tabular}}}
    \label{precession}
\end{table*}

\begin{table}[h!]
  \caption{Previous table continued. References are: 10)  \citet{britzentxs}, 11) \citet{britzen3c}, and 12) \citet{britzen_1502}.}
    \centering{
    \resizebox{0.49\textwidth}{!}{
    \begin{tabular}{c|c|c|c}
    \hline
    \hline
    AGN   & TXS 0506+056 & 3C 84& PKS 1502+106 \\
    \hline
    Type  & BL Lac & radio galaxy&Quasar\\
    z &$0.3365$ & $0.0176$&1.838\\
    $P_{\rm p, obs}$ [yr] & $10.2 \pm 1.1$ &$40.5 \pm 0.2$  & $20.2 \pm 2.7$\\
    $\gamma$ & $2.1^{+1.2}_{-1.1}$& $1.05$& $1.6 \pm 0.4$\\
    $\eta_0$ [deg] & $-93.7 \pm 94.3$ &$30$ & $113.6 \pm 11.4$ \\
    $\Phi_0$ [deg] &   $10.3 \pm 24.3$ &$130$ & $1.90  \pm 1.4$ \\
    $\Omega_{\rm p}$ [deg] &  $19.5 \pm 31.9$ &$45$ & $0.7 \pm 0.4$\\
    considered period&2009-2018& 2000-2018& 1997-2020\\
    Model & SMBBH or LT precession & SMBBH& SMBBH or LT precession  \\
    \hline
    References & 10) & 11)& 12) \\
    \hline
    \end{tabular}}}
    \label{precession1}
\end{table}


\begin{table}[h!]
    \caption{Derived maximum Doppler-factor ratio $\xi_{\rm max}$ for the AGN compared in this paper.}
    \centering
    \begin{tabular}{c|c}
    \hline
    \hline
    Source & $\xi_{\rm max}$\\
    \hline
    3C\,84 & $1.44$\\
    3C\,120 & $1.35$\\
    TXS 0506$+$056 & $1.80$\\
    PKS 0735+178& $1.63$\\
     OJ\,287 & $13.61$ \\
     3C\,273 & $3.37$\\
    3C\,279 & $18.69$\\
    1308+326 & $1.47$\\
     PKS 1502+106& $1.003$\\
     PG1553$+$113 & $19.63$ \\
    3C\,345 & $1.59$ \\
    2200$+$420 & $1.07$ \\
    \hline
    \end{tabular}
    \label{table_Doppler_ratio}
\end{table}

\subsection{Comparison of Doppler factors derived via precession model with those derived from shock-in-jet scenarios}
\label{sec_doppler_fac_prec_shock}
\citet{homan2021} derive Doppler factors for the MOJAVE survey sources based on the assumption that all sources in their sample have the same intrinsic median brightness temperature Tb$_{int}$. The Doppler factor for each source is then derived by dividing the sources median observed brightness temperature by Tb$_{int}$. The majority of the derived Doppler factors for the MOJAVE sources lie in the range between 0 and 40 (Fig. 7a in their paper). \citet{hovatta} determined the variability brightness temperatures of the fastest flares in 22 and 37 GHz light-curves. By assuming the same intrinsic brightness temperature for each source, they calculated the Doppler boosting factors. We compare the values obtained by \citet{homan2021} and \citet{hovatta} with the range of values for the Doppler factor (and the mean value) obtained by applying our precession model in Table \ref{Doppler}.

The values for 3C84, 3C~120, TXS 0506+056, PKS~0735+178, OJ~287, PKS~1502+106, PG~1553+113 and 2200+420 calculated based on the precession model fall in a similar range as those obtained by \citet{hovatta} and \citet{homan2021}. The largest discrepancy between the precession derived Doppler factors and the other estimates are found for 3C~273, 3C~279, 1308+326, and 3C~345. In addition to 3C~273, the values obtained by \citet{homan2021} also disagree with those obtained by \citet{hovatta} for the three other sources.  

Since Doppler factors can not be measured directly, the derived values rely on model assumptions. For the Doppler factors obtained by \citet{homan2021}, the apparent speeds of components and the brightness temperature based on these speeds are used to derive a Doppler factor. To our knowledge, the MOJAVE team assumes the shock-in-jet scenario and does not account for any geometric effect (e.g. precession) which can lead to different apparent speeds. In the case of the values obtained by \citet{hovatta}, the brightness temperatures of the fastest flares are used to determine the Doppler factor. The fastest flare might be due to a shock, an instability or magnetic reconnection and does thus not necessarily serve as a indicator of the boosting. Both approaches (apparent speed, fastest flare) thus rely on model assumptions. 

Both teams provide one single value for the Doppler factor. The Doppler factor based on the observed precession takes decades of measurements into account and can predict the evolution of the Doppler factor. This predicted Doppler factor evolution can be confronted with future measurements.
\begin{table*}[h!]
\centering
    \caption{The range of Doppler factors $(\delta_{\rm min},\delta_{\rm max})$ derived via the precession model (this paper). In addition, we list the mean Doppler factor calculated as $\overline{\delta}=0.5(\delta_{\rm min}+\delta_{\rm max})$. We list Doppler factors derived from the literature (determined without assuming precession) for the AGN compared in this paper. No Doppler factor has been determined for "-" entries. }
    \centering
    \begin{tabular}{c|rcccc}
    \hline
    \hline
    Source & $\overline{\delta}_{\rm prec}$\,$(\delta_{\rm min},\delta_{\rm max})$&$\delta_{T_{B}}$\citep{homan2021}&$\delta_{\rm var}$\citep{hovatta}&&\\
    \hline
    3C\,84 & 1.16 (0.95, 1.37) &4.3&0.3&\\
    3C\,120 & 10.24 (8.73, 11.75) &5.7&5.9&\\
    TXS 0506$+$056 & 2.81 (2.01, 3.61) &1.8&-&\\
    PKS 0735+178& 5.81 (4.42, 7.19) &5.1&3.8&\\
    OJ\,287 & 9.40 (1.29, 17.51) &46.3&17.0& \\
    3C\,273 & 6.04 (2.77, 9.32) &21.8&17.0&\\
    3C\,279 & 5.19 (0.53, 9.86) &140.2&24.0&\\
    1308+326& 26.12 (21.15, 31.08) &5.1&15.4&\\
    PKS 1502+106& 2.93 (2.93, 2.94)  &5.1&12.0&\\
    PG~1553$+$113 & 4.33 (0.42, 8.24) &1.4&-&\\
    3C\,345 & 18.81 (14.51, 23.11) &47.7&7.8& \\
    2200$+$420 & 9.08 (8.79, 9.38) &18.3&7.3& \\
    \hline
    \end{tabular}
    \label{Doppler}
\end{table*}



\section{Discussion}
\label{sec_discussion}
Many observational studies of AGN radio variability have been reported in the literature \citep[e.g.,][]{Hughes1992, Aller1999, fuhrmann}.

The kinematics of AGN jets on the parsec scale of AGN is usually probed using VLBI observations. 
Brighter parts in the jet -- so-called components -- can be traced and are seen to move at apparent speeds typically of 
the order of 5-10 times the speed of light (see e.g. CJF -- Caltech-Jodrell Bank Flat-Spectrum sample: \citeauthor{Britzen2008}, \citeyear{Britzen2008}; for MOJAVE: \citeauthor{Lister}, \citeyear{Lister}). 
The physical nature of these apparently moving jet components is still unclear, 
in particular the question whether the apparent motion merely reflects a pattern speed, or a real material object like a blob of synchrotron plasma is still open.
The most often commonly invoked scenario to explain the component motion in jets is that of a shock travelling within the jet stream -- the so-called shock-in-jet model.
This directly relates to the most common explanation for variability as outlined in the Introduction.

Despite the prominence of the shock-in-jet scenario in the literature, several studies have indicated that the jet kinematics may in fact be largely dictated by geometrical processes. 
In particular, jet precession has been invoked to explain the observed phenomena \citep[e.g.,][]{AbrahamRomero, nadia, Abraham2018}.

Precession as a potential cause of AGN flux variability in AGN has been discussed by a number of authors in the past \citep[see e.g.,][]{Caproni2013,Abraham2018}. 

Precession is not a property exclusively associated with blazars only; Seyfert 1 galaxies as well as radio galaxies exhibit signs for precession -- 3C\,84 being a prominent example \citep{britzen3c}.

\subsection{Origin of jet precession and nutation}
\label{precession_origin}
Precession and nutation of a jet basically reflects the motion of its base.
Depending on the dominant mechanism for jet launching, the base can be close to the central BH (Blandford-Znajek jet, \citealt{znajek})
or to the surrounding accretion disk (Blandford-Payne jet, \citealt{payne}).
Essentially, both types of outflow can be set into precession, so long as requisite non-axisymmetric forces are present.
In general terms, this requires a multi-component system with misaligned rotation axes.
These components could be a BH-disk combine, or a SMBBH.
Precession could then arise from
\begin{itemize}
 \item[(a)] torques on the accretion disk induced by a secondary BH, or
 \item[(b)] torques on the jet launching  primary BH induced by a secondary BH,
 \item[(c)] the misalignment of the BH spin axis from the angular momentum vector of the accretion inflow, leading to LT precession of the disk that may lead to jet precession. 
\end{itemize}
For the case of OJ\,287  (for details see \citealt{Britzen2018}), all three scenarios have been considered for the periodic motion inferred for the disk-jet system; 
however, considering timescales of the order of $10$ years, option (b) could be excluded,
leaving for the probable cause of precession an efficient disk-jet coupling. 
We further refer to, e.g. \citet{CaproniAbrahamb2004} for the application of the SMBBH model to the jet kinematics, and to \citet{Caproni2004} for the observational analysis of the presence of LT precession in several AGN.


Both processes, (a) and (b), can be expected to occur in the aftermath of galaxy mergers, and are thus a natural outcome
of hierarchical galaxy formation scenario.
The relevance of SMBBH formation has been highlighted in numerous studies (e.g., \citealt{begelman, wilson, mayer}). 
One expects a SMBBH to form during the final merger phase 
(e.g. \citealt{mayer}, and references therein), or a misalignment of disk and BH spin axes as a direct result of two 
supermassive BHs merging. 
Process (a) and (b) requires two massive BHs with sub-parsec separation, thus in the final merger stage. 
Process (c) arises due to the frame-dragging in the vicinity of a spinning BH and requires the offset of 
the accretion flow angular momentum vector from the BH spin. 
Such an offset can arise in the case of either:
\begin{itemize}
  \item[(i)] post-merger systems due to their random orientations of the original orbital plane of the merging BHs with respect to the accretion flow around a (primary) BH,
  \item[(ii)] accretion flows that are fed by winds from a disk-like stellar system as expected in common models 
             for the hot accretion flow for Sgr~A* \citep{Cuadra, Shcherbakov, Yalinewich}. For this example the misalignment between the accretion disk axis and the BH spin is 
             specifically discussed in \citet{2013MNRAS.432.2252D}.
\end{itemize}
Note that in scenario (c) the BH-driven jet is launched without precession as the BH axis remains 
stable in space, while it is the accretion disk which is precessing. However, disk winds and jets launched from the disk will follow the disk precession and thus set their jet axis in precession. As a result, the environment of the central spine jet will change in time, and also the spine jet will structurally follow the precession of the disk jet and disk (see e.g. \citealt{2013Sci...339...49M}).

Thus, there is an overall consistency between the precessing jet model for AGN, presented in our work, and the large-scale 
picture of galaxy formation.

\subsection{Numerical simulations of jet precession}
\label{simulations}
Mounting evidence for jet precession has been found for an increasing number of AGN, as discussed above. 
Additional support to this scenario comes from observations of micro-quasars, i.e. stellar binary systems in which a compact object accretes matter from a donor star and a jet is launched \citep{Mirabel1999}.

As a typical example, the GRAVITY collaboration recently confirmed evidence for precession in the microquasar SS433 \citep{gravity}, as already reported in several studies (see e.g. \citealt{blundell}). 
SS433 is known to eject a two-sided radio jet precessing with a period of 162.5 days \citep{margon}. 
Recent hydrodynamical simulations are capable of reproducing the observed precession signatures of SS433 very well
(see e.g. \citealt{monceau2014, monceau2015, monceau2017}).
Another such example is the micro-quasar GX\,339-4, an X-ray detected BH binary. It also shows variability in the IR and X bands, which could be modeled via jet precession
driven by the LT mechanism \citep{malzac}.
Yet another case of jet precession is the micro-quasar V404 Cygni \citep{miller}. Also the jet precession in LS I +61$^\circ$303 \citep[][and references therein]{Wu2018} is likely the result of LT precession \citep{Massi2010}.
 
Since 3-dimensional (3D) general relativistic magnetohydrodynamic simulations are very CPU expensive,
only recently the evolving precession (that is essentially 3D) of such systems could be numerically investigated
(see reviews by \citealt{hawley, marti}). 
Recent progress has been reported by \citet{liska} who investigated the dynamical evolution of a tilted thick accretion disk
around a rapidly spinning BH.
In fact, the authors find that the disk-jet system as a whole undergoes a LT precession.
These simulations may serve as clear evidence that the jets can be used as probes of disk precession. 
The precession time scales obtained within their simulations are consistent with the timescales that were derived for precession and nutation in OJ\,287 \citep{Britzen2018}.
However, we caution that despite the clear precession seen in these simulations, the application of these simulations to the observationally inferred precession of the AGN central engine is somewhat limited.
The reason is that the precession amplitude found in these simulations decreases on longer time scales (after several
precession periods) and the system gradually realigns.
Furthermore, \citet{liska19} have shown by highly resolved GR-MHD simulations of the BH-disk-jet system that any misalignments may 
disappear due to the Bardeen-Petterson effect. 
Nonetheless, the onset of disk warping and subsequent wind launching directed along the disk rotation axis seems confirmed \citep{white}.

The evolution of jet launching in binary systems has also been investigated recently \citep{sheikhnezami2015, sheikhnezami2018}, 
albeit not in the relativistic regime.
The authors investigate using non-relativistic 3D MHD simulations the evolution of jets launched from the disk around a primary compact object.
The disk starts warping and with that the direction of the disk jet changes.
The jet axis starts tracing a circle, indicating strongly the  formation of a precession
cone.
In particular, it could be shown how the spiral structure of the accretion disk subsequently affects the mass loading and the magnetic field of the disk outflow, 
namely leading to a spiral magnetohydrodynamic structure of the jet
\citep{sheikhnezami2022}.
Also here, CPU constraints as well as the physical constraints like the mass loading of the disk limit simulations time scales to less than
an orbital time scale of the secondary (however 1000s of disk orbital periods at the jet foot-point).

\subsection{Implications of jet precession for multi-wavelength flaring}
\label{multi-wavelength}
In this paper we focus on radio variability of AGN which seems to arise from jet precession. 
However, several authors have presented evidence that jet precession may also be responsible for the variability observed in other bands of the electromagnetic spectrum. 
A deeper examination of the multi-wavelength aspect of jet precession is planned for a forthcoming paper, but some examples will be mentioned here. 
Since precession is a geometric phenomenon, it would affect the synchrotron flux density at a wide range of frequencies. 
For OJ\,287 as a representative of BL Lac sources, the optical emission is dominated by the synchrotron radiation of the jet and its observed optical periodicity is expected to be at least partly related to the jet precession (e.g., \citealt{Abraham, Britzen2018}). In the case of OJ\,287 we found a correspondence between the radio periodicity and the periodicity seen in the optical \citep{Britzen2018}. 
The time interval of the optical flaring is roughly half of the radio flaring.  
The authors suggest that most likely both periodic phenomena originate from jet precession. 

The precession phase in OJ\,287 relates to the SED variability, as discussed in the following subsection.
\citet{stamerra} use multi-wavelength observations of PG1553+113 to investigate if the observed modulation 
(gamma-rays, optical) can be explained with geometrical variations in the jet, possibly pointing to jet precession. They discuss PG1553+113 as a probe for a geometrical periodic modulation.
Independent evidence for jet precession being responsible for the periodic multi-wavelength flaring in PG1553+113 has been presented by \citet{Caproni2017}. 
\citet{tavani} also discuss the possible gamma-ray (from \textit{Fermi}-LAT observations) signatures of a SMBBH in PG 1553+113.
In the case of 3C\,84, \citet{britzen3c} find evidence for precession of the central radio structure and a correlation between the radio and gamma-ray light-curves with the higher energy emission lagging the low-energy emission. This delayed correlation is difficult to explain in terms of the shock-in-jet scenario. A plausible scenario, however, is that the correlation arises from jet precession and the time delay due to the emissions originating in different parts of the radio structure.
\citet{bhatta} studied the deterministic aspect of the $\gamma$-ray variability of 20 $\gamma$-ray bright blazars dependent on the source class. They find that the dominant physical processes in FSRQs are more of deterministic nature.
The following four sources from our paper are in common with \citet{bhatta}: 3C\,273, 3C\,279, PKS 1502+106, and 2200+420.
In PKS 1502+106, they find the strongest deterministic case.

\subsection{Precession phase related to time variable SED}
Since precession results in a time-variable Doppler factor, this should also influence the temporal evoluton of the SED, affecting the synchrotron as well as the Inverse-Compton emission in tandem. Indeed, we show in this paper that the new spectral features seen in the 2015-2017 multi-wavelength high activity state of OJ\,287 \citep{kushwaha} could correspond to a special precession phase.
In this paper, we demonstrate that the Doppler factors derived from SED-fitting and precession-fitting peak at the same time. We conclude that the viewing angle changes caused by precession also induce variations of the SED state.

\subsection{Precessing jets as potential neutrino emitters}
\label{neutrinos}
\citet{britzentxs} have reported evidence that the neutrino emission observed from TXS 0506+056 might be related to a collision of jetted material. 
In their model, a special viewing angle of the precessing inner jet plays a major role in obtaining the right circumstances for neutrino generation. 
The timescale inferred from the observed precession signatures is of the order of 10 yr. 
The observational clue presented in that work appears promising for addressing the issue of neutrino production, intricacy of which is highlighted, e.g., by \citet{rodrigues}. 
Most recently, further neutrino emission has been observed from the direction of TXS~0506+056 on April 18, 2021 by Baikal-GVD \citep{baikal}.
On September 18, 2022, the IceCube Collaboration indicated that another neutrino
arrived from the direction of TXS 0506+056 \citep{tjus}. Both events support the prediction for repeated neutrino-emission based on the precession model for this source and a supermassive binary black hole model \citep{britzentxs, debruijn}.
For another IceCube candidate emitter, the AGN PKS 1502+106, \citet{britzen_1502} find evidence for precession (and nutation). 

Four independent neutrino detectors - including IceCube - reported neutrino emission in December 2021 \citep[see e.g.,][]{baikal, sahakyan}. PKS 0735+178 is located just outside the 90\% point-spread function containment of the IceCube event, within the expected systematic uncertainty of the determination of the arrival direction.
For PKS~0735+178, evidence for precession has been presented in \citet{Britzen2010b}.



\subsection{Lighthouse effect versus Precession}
\label{lighthouse}
In case of well observed and sufficiently long monitored VLBI jets there is a simple phenomenological way to discriminate between geometric
precession and internal helicity. 
In case of geometric precession the pitch of the jet helix (which is the height of one complete helix turn, measured parallel 
to the axis of the helix) is nearly constant. 
The helix we see is only a pattern, the components move essentially along straight or ballistic trajectories. 
Whether the cause of the periodic jet launching direction is the spin-orbit precession, Newtonian binary motion \citep[e.g.][]{Kun2014}, LT precession \citep[e.g.][]{britzentxs}, etc., the result is the same, a namely constant pitch. 

In case of internal helicity the pitch is not constant, and the jet components move along the helix \citep[e.g.][]{steffen,hardee}. 
This is the lighthouse model (see below), and the Kelvin-Helmholtz (or any magnetic field/instability related) model.
While Kelvin-Helmholtz instabilities can be expected from almost any hydrodynamic shear flow, 
it remains unclear whether the flow pattern can actually develop into a large-scale (meaning pc-long) structure. 
Numerical simulations show that these instabilities develop rather quickly, resulting more in a turbulent cocoon between the
jet beam and the ambient gas than in a wave-like, bent jet morphology \citep[see e.g.][]{1997ApJ...479..151M}.

If one is able to measure and de-project the pitch, one can discriminate between external (changing jet launching direction) or internal helical motion (the jet-components follow MHD-determined paths).

As an alternative explanation for the observed (intra-day) variability of blazars, \citet{Camenzind} proposed 
the so-called {\em lighthouse effect}.
In this model, radiating plasma blobs are injected along the magnetized relativistic jet and then follow the overall jet kinematics.
The radiation from the relativistically moving blobs is beamed in the direction of motion and thus when the blob follows a helical trajectory, its light beam sweeps across the observer repeatedly, giving rise to a quasi-periodic emission flares.
Due to the overall acceleration and collimation of the bulk outflow, the geometry of the helical trajectory probably changes along the jet.
We note that the lighthouse model is a kinematic model which neglects the dynamics of the jet flow.
In fact, the nature and the dynamics of the postulated jet blobs is not explained.
Naturally, any MHD jet is rotating, however, the poloidal motion dominates rotation for super-Alfv\'enic flows.
Also, due to the rotation, MHD jets naturally imply the existence of a jet helical magnetic field.
This is different from jet precession where the motion of the ejected blobs can be purely poloidal - being ejected ballistically into different directions over time. The nozzle itself is rotating (i.e. precessing), but the jet velocity is expected to exceed the precession speed substantially, so in principle even a jet with purely poloidal speed and magnetic field will change its viewing angle. The latter does not happen in the light-house model.

From an observational point of view, the periodicity time scale and the number of events detected in a given time 
may vary from source to source and also from time to time for a given source.
\citet{Britzen2000} showed that the simultaneous detection of radio/optical/gamma-ray flaring in the blazar PKS 0420-014 is in 
agreement with the predictions of the lighthouse model.
Similarly for 3C\,454.3, \citet{Qian2007} found periodic variations in the radio flux that are consistent with the lighthouse scenario.

%

\subsection{OJ~287: Unsolved questions regarding the ``Impact-Model"}
\label{impactmodel}
There seems to be a need to explore an alternative to the currently popular model of the double-peaked optical outbursts appearing every 12 years \citep{aimo, valtonen2009} which invokes periodic piercing of the orbiting secondary BH through the accretion disk of the primary BH. The disk-piercing model has scored {\bf well} in predicting the times of the giant optical flares (with $\sim$ 12 year periodicity), by suitably updating of the input dynamical parameters. 
However, \citet{komossa} did not detect the outburst predicted for October 2022 by the ``Impact-model". Neither was the predicted thermal bremsstrahlung spectrum observed (at any epoch). The Impact-Model also predicts a very large mass of the primary black hole ($\sim10^{10} M_\odot$). \citet{komossa} estimate the mass of the primary black hole in OJ 287 to be $\sim10^8 M_\odot$ and thus confirm the mass value proposed by \citet{Britzen2018}, based on the precession model. \\

The Impact-Model scenario is a hybrid in the sense that it identifies the periodic giant optical flares to thermal radiation arising from the disturbed/ejected gas in the disk as the latter is impacted by the accretion disk of the primary BH by the orbiting secondary BH. An expected corollary of this is that the optical flaring event would be followed by an enhancement of jet activity (i.e., synchrotron radiation) on account of gradual accretion of the disturbed disk gas on to the primary BH. Thus, conceivably, jet precession and resulting variability of its synchrotron and IC flux could take place even in this scenario as well. However, as the immediate major outcome of the impact (identified with the giant optical flare), the model invokes free-free radiation from the impact-heated gas belonging to the accretion disk. One would then also expect to see bright optical/UV emission lines from the same cloud of thermal plasma, although the required cooling of the gas cloud to an appropriate temperature might delay the onset of bright recombination line emission. Such emission features, to our knowledge, have never been reported for this source and therefore it has steadfastly retained its BL Lac classification. Oddly, this has happened despite its high brightness (m$_{\rm V}\sim$ 12 mag) as well as the fact that the free-free cooling rate is very rapid in the temperature regime of 10$^5$-10$^6$ K. An intensive optical spectral monitoring would play a critical role in placing this model on a firm footing (or, otherwise).
A few observational pointers relevant to this issue are:\\
\begin{itemize}
\item
We are NOT dealing with a thin accretion disk (BL Lac). Hence the putative ``impact'' is a nebulous concept \citep[see ][]{villforth2010}.
\item
For all the flares, the primary peak lacks a radio counterpart, while some secondary peaks have shown radio counterparts.
\item
Optical polarisation can be very high even for a primary peak \citep{villforth2010}. This is a problem for the thermal interpretation \citep{valtaoja2000, villforth2010}.
\item
The amplitude of the optical flares has been steadily declining from flare to flare. This holds for both primary and secondary peaks in a flare.
\end{itemize}

At present, the key arguments in support of the thermal optical interpretation for the giant optical flares comes from the observed drop in the fractional linear polarisation during the giant optical flares, a possible indicator of enhanced thermal contamination of the optical synchrotron radiation \citep{valtonen2017, kushwaha2018}. While such a contamination may well be occurring, we would like to recall that the drop in a fractional polarisation with a rise in brightness may also occur in case of a mildly swinging synchrotron jet \citep{gopalwiita1992}. Thus, we believe that a clinching evidence in favour of the thermal interpretation of the giant optical flare awaits an unambiguous detection of optical/UV emission lines following the flares. Given the high brightness of this blazar, even a 1-metre class optical telescope should suffice. 

Recently, \citet{2021ApJ...917...43S} performed GRMHD simulations of an orbiting perturbed (secondary black hole) punching through the hot flow surrounding the primary SMBH. Thus, they correctly accounted for the ADAF typical for BL Lac sources. The model can address the 12-year optical periodicity by the enhanced accretion that occurs every time the secondary component passes through the accretion flow (twice per orbit), while the 24-year periodicity in the radio light curves could be attributed to the ejected plasmoids due to the viewing angle close to the jet axis (the counter-jet and its components are deboosted). However, the model does not provide a prediction for the light curves in different bands yet. In addition, the duty cycle of the modelled flares appears rather short in their model, while the large-amplitude radio flares seen in OJ287 have a long duty cycle of $\sim 10$ years more consistent with the smooth viewing angle changes due to jet precession.

Finally, we may recall that in the literature, there are claims that the giant optical flares are accompanied by a Seyfert-like behaviour, e.g., a soft X-ray excess \citep{pal2020}. However, the central engine in these objects is known to accrete at a high Eddington rate (e.g. \citet{tortosa}, unlike BL Lacs. Moreover, in order to justify the analogy, one expects to observe optical emission lines which are ubiquitous in case of Seyferts.
A more detailed analysis of these issues is planned for a future publication.

\subsection{Some observational challenges to the shock-in-jet scenario}
\label{challenges}
While a coincidence is often seen between flaring at millimetre wavelengths and the injection of a radio knot into the base of the parsec-scale VLBI jet \citep[see e.g.][]{savolainen}, this is not always the case. As an example, \citet{Bach} studied the connection using the flares in the radio and optical light-curves of BL\,Lac, and found that only some of the new VLBI components had an associated event in the light curves. Several AGN, where no component ejections could be detected in VLBI observations coinciding or following radio outbursts have been reported (e.g., PKS 0735+178, \citeauthor{Britzen2010b}, \citeyear{Britzen2010b}; 3C~454.3 \citeauthor{Britzen2013}, \citeyear{Britzen2013}).\\

Stationary radio components, which maintain a constant separation from the core, have been observed in the VLBI images of 1803+784 (\citet{britzen_1803_2005}) and several other AGN. For Gamma-ray bright blazars, an exhaustive list of stationary components (which do not move along the jet) can be found in \citet{jorstad}.

Such stationary radio knots have been identified with ``recollimation shocks" envisioned to form within the relativistic jet flows (e.g., \citealt{stawarz}). 
However, \citep{britzen1803} shed new light on them, by demonstrating that while the ``stationary" knots do not appear to move along the jet, they do move (rather rapidly) orthogonal to the main jet ridge line in 1803+784. 
For OJ\,287 this kind of motion is shown to be consistent with jet nutation \citep{Britzen2018}. Throughout this paper we call these components quasi-stationary components. They need to be distinguished from those stationary components, where no motion at all is observed over a longer time span. This is the case for, e.g., Mrk~501 \citep{edwards, britzen_mrk501}.


AGN feedback models struggle to steadily quench star formation. Kinetic jets are less efficient in quenching star formation in massive galaxies unless they have wide opening or precession angles \citep{su}. According to this study, AGN feedback models require a jet with an optimal energy flux and a sufficiently wide jet cocoon with long enough cooling time at the cooling radius. 

In contrast to straight jets, precessing jets sweep through a larger volume of the interstellar medium (ISM). Therefore they have the potential for enhanced feedback into the ISM as well as the intergalactic medium, i.e. quenching or sometimes triggering star formation \citep[e.g., ][for a molecular jet]{aalto} or providing radio-mechanical heating feedback that prevents hot X-ray-emitting atmospheres in galaxy clusters from runaway cooling and thus stabilizes them \citep[see e.g.][]{2022NatAs...6.1008Z}.

On a different scale, in Henize 2-10, a dwarf starburst galaxy with a potential central massive black hole, a precessing bipolar outflow connecting the region of the black hole with a star-forming region has been reported \citep{schutte}. 

\subsection{Why the shock-in-jet model is not sufficient to explain variability}
\label{shock-in-jet}

The past decades have seen a predominant interpretation of variability and jet morphology in terms of the shock-in-jet scenario. While the shock-in-jet scenario might explain radio knots, this scenario is insufficient to explain the observed flux-density and structural changes of many AGN jets. The dominance of the shock-in-jet scenario over the past decades is no argument against exploring mechanisms that might unravel more basic details about the way the jet and the central engine of AGN work. 
As noted above and in other studies, the precession model can explain certain observational aspects of the radio jets and of AGN variability that are difficult to comprehend relying solely within the framework of the shock-in-jet scenario. We think that precession therefore modulates the appearance of the jet.

Identifying deterministic patterns within variability data will enable us to discriminate between geometric (precession) phenomena and stochastic (e.g., flaring) events. This will provide a deeper and more fundamental understanding of AGN physics. It will also allow us to predict times of distinct states, e.g., flaring states due to predictable precession events (reversal points, see Fig.~4[b] in \citeauthor{Britzen2018}, \citeyear{Britzen2018}). This is progress compared to multiple multi-wavelength campaigns which, in the past, had to be planned and conducted ``blindly" due to the lack of knowledge of predictable, correlated and inter-connected processes. Applying this previous knowledge to predict the deterministic behaviour will allow us to much more carefully explore also those processes which are not deterministic but of stochastic nature. 
Magnetic reconnection might play an important role in the launching of jets as well as the injection of components \citep{Vourellis2019, Ripperda2020, Nathanail2020}.




\section{Conclusions}
\label{sec_conclusions}
In this paper we have revisited the key problem of flux variability/flaring of AGN at radio frequencies. 
Conventionally, it is interpreted in terms of the "shock-in-jet" scenario and its variants. 
Recently this paradigm has encountered problems in explaining certain instances of radio knot ejection unaccompanied by a flare and vice versa. 
A similar difficulty is posed by the correlated flux variability of 3C\,84 at gamma-ray and radio frequencies, albeit the gamma-ray profile 
exhibiting a very large time lag of $\simeq$ 300-400 days.

We have therefore proposed an alternative (deterministic/geometric) interpretation according to which the observed structure and flux density of 
the parsec scale radio jet predominantly reflects the precession (and nutation) of the relativistic jet, leading to a time variable Doppler beaming
of its emission. 
We note that the detection of the signatures of nutation (a second-order precession effect) in the light curves of certain blazars provides 
further support to this interpretation. 
Hence, the monitoring of radio flux and parsec-scale structure of blazars is likely to facilitate extraction of more information about the jet
ejection process as compared to its subsequent evolution via shocks, etc. 
The recent confirmation of repeated neutrino emission from TXS 0506+056 fosters the jet precession model.
We also find that the jet precession scenario is in accord with the concept of hierarchical evolution of the universe wherein mergers of galaxies 
(along with their central BH) are supposed to be a key ingredient.\\
As a caveat, the applicability of our model in connecting the jet kinematics to features in the light-curve is unclear for M87 (e.g. \citealt{britzenm87}) and Sgr A$^\star$ \citep{gravitys}. 
For M87, while indication for jet precession has been found \citep{britzenm87} and a tilted disk/jet system 
has been considered \citep{chatterjee},
its link to radio flux variability needs to be explored further. In the case of Sgr A$^\star$, the presence of a jet is still to be demonstrated (e.g., \citealt{falcke1999, li2013, zadeh2020, fragione2020, cecil}). For Sgr~A*, asymmetric X-ray flares seem to be associated with nearly edge-on orbits \citep{2015A&A...573A..46M,2017FoPh...47..553E,2017MNRAS.472.4422K}, which could be the sign of the accretion-flow precession or the disk warping due to the Bardeen-Petterson effect. \citet{2013MNRAS.432.2252D} modelled tilted thick accretion flows around Sgr~A* and obtained a good agreement with the near-infrared and the mm-domain observations in terms of the variability and the spectral evolution.\\

In the following we recapulate our main arguments and the results obtained.

(1)
Precession of AGN jets can generally be triggered in a binary black hole system or by the Lense-Thirring effect on the accretion disk. 
Simultaneously, both effects can also lead to nutation of the disk-jet system. 
In this paper, we have applied a kinematic precession model that is of general nature, and can, thus, be applied independently of the origin 
of the precession. 
We have provided a concise mathematical description of our modeling of jet precession and 
nutation, extending the approach of Britzen et al. (2018). 

(2)
We apply the kinematic precession model to twelve AGN. 
We model their radio light curves and determine the long-term evolution of the Doppler factor.
From our comparison of the derived Doppler factors with those derived by the well-known shock-in-jet scenario,
we conclude that precession is likely to play a dominant role in causing the (radio) variability, and the apparent speeds of the jet components.  

(3)
A corrollary of the above is that a time-varying Doppler factor should also be reflected in the temporal evolution of the SED. 
Indeed, we show
that the Doppler factor derived from SED-fitting and the one derived independently from precession-fitting (based on the derived VLBI jet kinematics) 
peak at the same time taking OJ~287 as an example. 
We thus propose that the change of viewing angle over time, as envisioned in our precession model, can explain the variation in the SED state.

(4)
The point that needs to be emphasized is that precession and nutation of a jet basically reflect the varying orientation of its base and at the same time, they can be detected by studying the pc-scale motion of the VLBI knots.
Signatures of precession are also frequently observed on large-scales in Karl G. Jansky Very Large Array (VLA) and MERLIN radio maps of powerful jet sources \citep[e.g.,][]{krause}.
For a complete sample of 33 3CR radio sources the authors find strong evidence for jet precession in 24 cases (73\%).
We argue in case of a collimated disk wind or jet, the torques needed for causing the precession and nutation 
can be induced on the accretion disk by a secondary BH, or by Lense-Thirring precession by an inclined BH axis.
In case of a precessing spine jet launched by the Blandford-Znajek process, torques on the jet launching primary BH may be induced 
by a secondary BH. 
State-of-the art numerical simulations confirm our scenario.

(5)
Fitting radio light curves using the ARIMA model implies that the OJ~287 system is dominated by stochastic processes on rather short timescales ($\lesssim $0.2 years), while the clearly present long-term trends are dominated by deterministic periodic processes related to precession and nutation jet motions on timescales  $\geq1$ year.

(6)
While we mainly focus on radio variability in this paper, we briefly discuss other bands of the electromagnetic spectrum as well,
finding evidence for variability caused by jet precession at these higher frequencies as well. We further argue that jet precession and the subsequent strong dynamical changes in the jet kinematics may eventually have also led to neutrino emission in three AGN. 
The three known radio sources (TXS 0506+056, PKS 1502+106, PKS 0735+178) have been identified by IceCube as neutrino emitters.

(7)
Earlier, we had pointed out problems with the popular model of the quasi-periodically flaring blazar OJ 287, which invoked a secondary black hole repeatedly impacting the accretion disk surrounding the primary black hole \citep[see][]{Britzen2018}. Here we have noted additional difficulties with that scenario, accentuated due to the lack of evidence for optical/UV recombination line emission associated with the giant optical flares which are purported to arise from free-free emission \citep[e.g., ][] {valtonen2009}.

(8) We hope that the various observational evidences for the AGN jet precession, highlighted and modelled in this work, would encourage long-term VLBI and SED monitoring of bright blazars.

\begin{acknowledgments}
The authors appreciate the significant and important help provided by the anonymous referee. The insightful suggestions enhanced the presentation of the manuscript's results considerably.
The authors thank N. MacDonald for carefully reviewing the manuscript and for providing many valuable comments and thoughts that significantly improved the paper. 
We thank A. Witzel, S.-J. Qian, J.-P. Breuer, C. Raiteri, M. Villata, Z. Abraham, and M. Krause for very helpful discussions. 
Special thanks go to P. Kushwaha for providing data and a plot modified for this manuscript. M.Z. acknowledges the financial support by the GA\v{C}R EXPRO grant No. 21-13491X ``Exploring the Hot Universe and Understanding Cosmic Feedback". G-K. acknowledges a Senior Scientist fellowship from the Indian National Science Academy. E.K. thanks the Hungarian Academy of Sciences for its Premium Postdoctoral Scholarship. This research has made use of data from the OVRO 40-m monitoring program \citep{Richards} which 
is supported in part by NASA grants NNX08AW31G, NNX11A043G, and NNX14AQ89G and NSF grants AST-0808050 and AST-1109911. This work was supported in part by the Deutsche
Forschungsgemeinschaft (DFG) via the Cologne Bonn Graduate School (BCGS), the Max Planck Society through the International Max Planck Research School (IMPRS) for Astronomy and Astrophysics as well as special funds through the University of Cologne. Conditions and Impact of Star Formation is carried out within the Collaborative Research Centre 956, sub-project [A02], funded by the Deutsche Forschungsgemeinschaft (DFG) – project ID 184018867. This research was funded in part by the Austrian Science Fund (FWF) [P31625].
This research has made use of data from the University of Michigan Radio Astronomy Observatory which has been supported by the University of Michigan and by a series of grants from the National Science Foundation, most recently AST-0607523. This research has made use of data from the
MOJAVE database, which is maintained by the MOJAVE team \citep{Lister2018}. 
\end{acknowledgments}



\appendix

Here we provide additional statistical details related to the parameter inference of the precession-nutation model and the alternative interpretation of the light-curve variability using autoregression and moving-average models.

\section{Markov Chain Monte Carlo precession-nutation parameter inference}
\label{appendix_mcmc}

\subsection{Precession parameters based on the position-angle evolution}

\begin{figure*}[h!]
    \includegraphics[width=\textwidth]{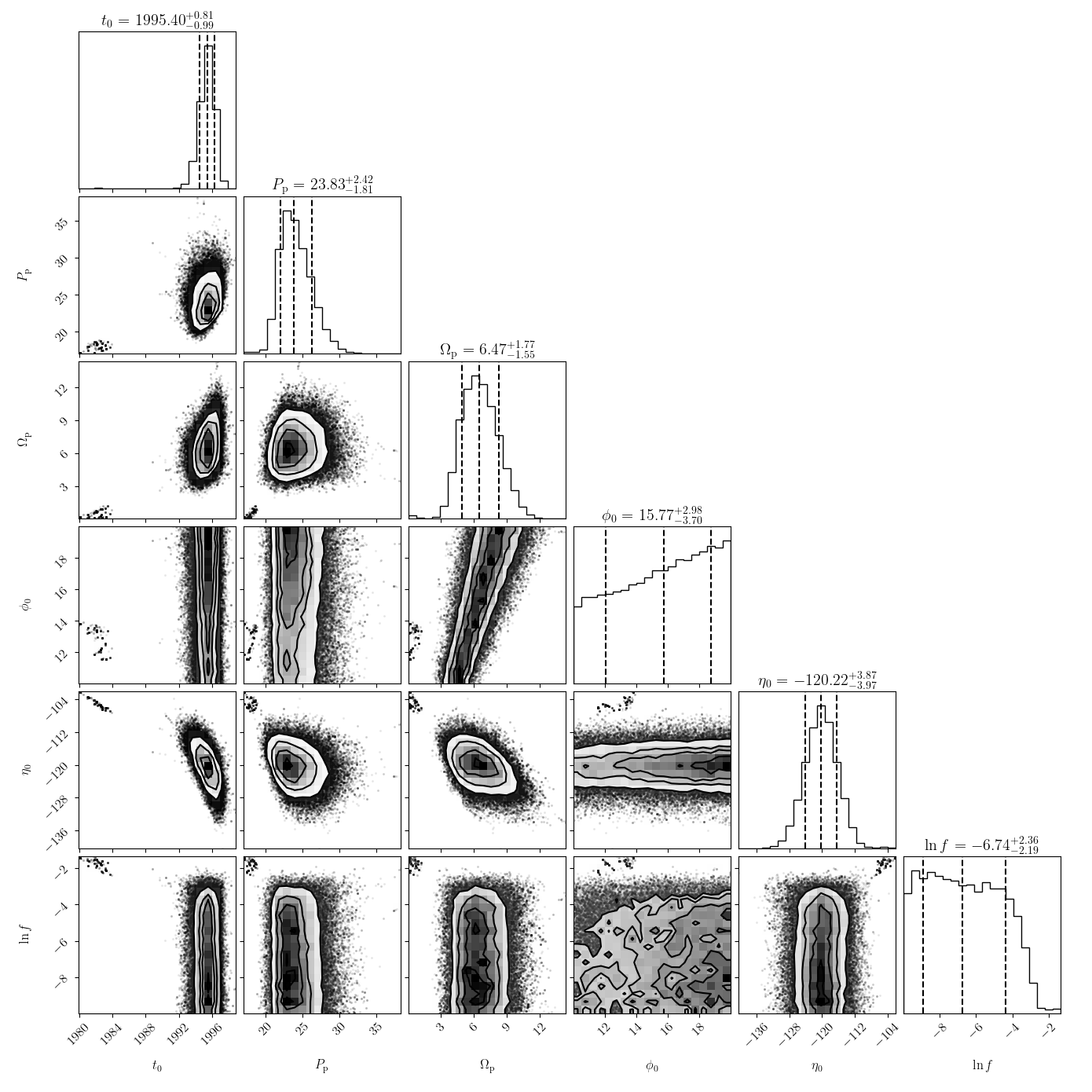}
    \caption{Marginalized one-dimensional likelihood distributions and two-dimensional likelihood contours at 1$\sigma$, 2$\sigma$, and 3$\sigma$ confidence levels for the parameters related to the position-angle temporal evolution due to the bulk precession of the jet. The dashed vertical lines in diagonal panels stand for 16\%, 50\%, and 84\% percentiles of the inferred posterior parameter likelihood distributions. }
    \label{fig_mcmc_eta}
\end{figure*}

\begin{figure}[h!]
    \centering
    \includegraphics[width=\columnwidth]{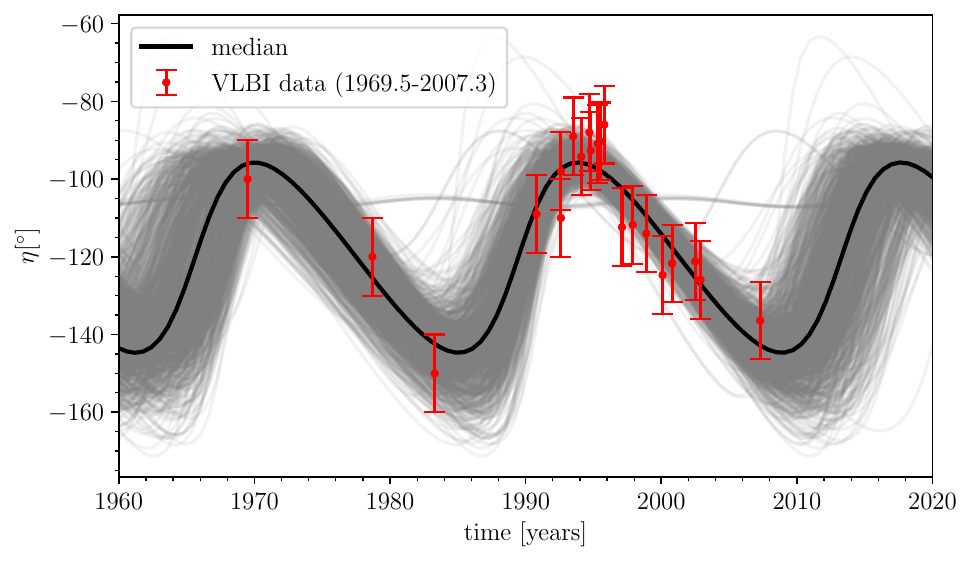}
    \caption{Median model of the position angle temporal evolution (black solid line) based on the posterior parameter distributions. Gray lines are based on 1000 samples of the parameter chains. Red points with errorbars are based on VLBI observations.}
    \label{fig_mcmc_model_eta}
\end{figure}

In Fig.~\ref{fig_mcmc_eta}, we show the corner plot with marginalized one-dimensional likelihood distributions and two-dimensional likelihood contours for the precession parameters $t_0$, $P_{\rm p}$, $\Omega_{\rm p}$, $\phi_0$, and $\eta_0$ inferred from the temporal evolution of the ejected component position angles $\eta$. We also display the posterior distribution for the variance underestimation factor $\log f$, which is included in the definition of the likelihood function. Hence we deal with 6 free parameters. In Fig.~\ref{fig_mcmc_model_eta}, we compare the posterior median model with the VLBI data, including 1000 random samples of parameter chains. This way the precession model uncertainties around the model median are apparent. We summarize the non-zero flat prior parameter ranges in Table~\ref{tab_priors_eta}.

\begin{table}[h!]
    \centering 
    \label{tab_priors_eta}
    \caption{Summary of non-zero flat prior parameter ranges for the precession parameters inferred from the temporal evolution of the jet position angle.}
    \begin{tabular}{c|c}
    \hline
    \hline
     Parameter  & Prior range  \\
     \hline
      $t_0$ [yr]   & [1970, 2010] \\
      $P_{\rm p}$ [yr] & [10,40]  \\
      $\Omega_{\rm p}\,[^{\circ}]$ & [0,90] \\
      $\phi_0\,[^{\circ}]$ & [10,20] \\
      $\eta_0\,[^{\circ}]$ & [-180,0] \\
      $\log{f}$ & [-10,1] \\
      \hline
    \end{tabular}   
\end{table}

\subsection{Precession parameters based on the apparent-velocity evolution}

\begin{figure*}[h!]
    \includegraphics[width=\textwidth]{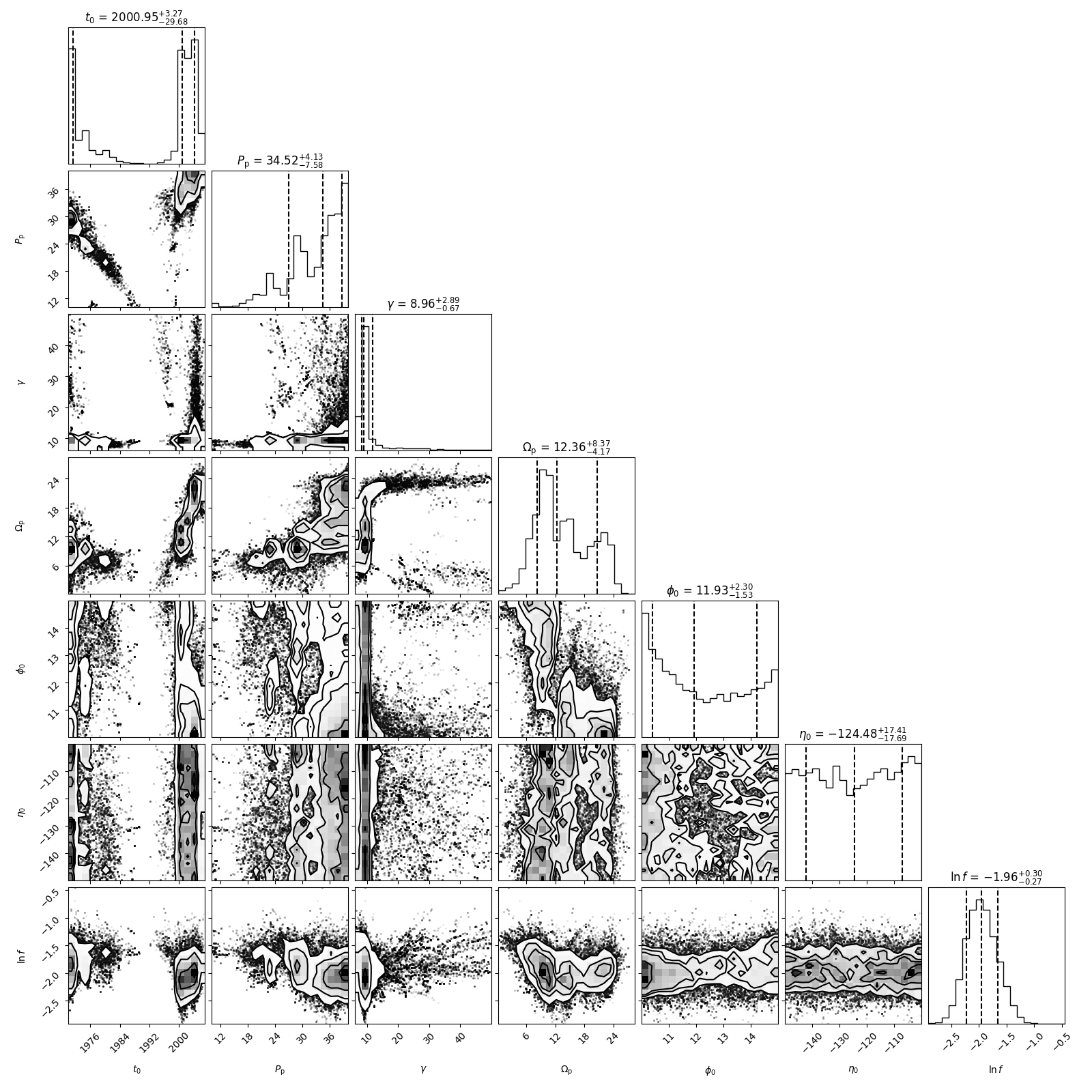}
    \caption{Marginalized one-dimensional likelihood distributions and two-dimensional likelihood contours at 1$\sigma$, 2$\sigma$, and 3$\sigma$ confidence levels for the parameters related to the apparent-velocity temporal evolution due to the bulk precession of the jet. The dashed vertical lines in diagonal panels stand for 16\%, 50\%, and 84\% percentiles of the inferred posterior parameter likelihood distributions. }
    \label{fig_mcmc_bapp}
\end{figure*}

\begin{figure}[h!]
    \centering
    \includegraphics[width=\columnwidth]{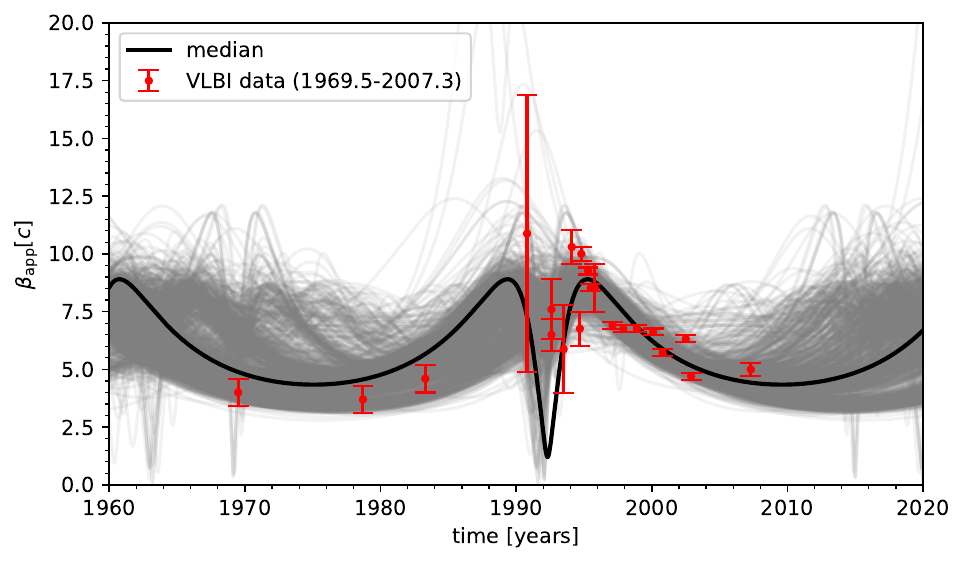}
    \caption{Median model of the apparent-velocity temporal evolution (black solid line) based on the posterior parameter distributions. Gray lines are based on 1000 samples of the parameter chains. Red points with errorbars are based on VLBI observations.}
    \label{fig_mcmc_model_bapp}
\end{figure}

In Fig.~\ref{fig_mcmc_bapp}, we show the corner plot with marginalized one-dimensional likelihood distributions and two-dimensional likelihood contours for the precession parameters $t_0$, $P_{\rm p}$, $\gamma$, $\Omega_{\rm p}$, $\phi_0$, and $\eta_0$ inferred from the temporal evolution of the ejected component apparent velocities $\beta_{\rm app}$. We also display the posterior distribution for the variance underestimation factor $\log f$, which is included in the definition of the likelihood function. Hence we deal with 7 free parameters. In Fig.~\ref{fig_mcmc_model_bapp}, we compare the posterior median model with the VLBI data, including 1000 random samples of parameter chains. This way the precession model uncertainties around the model median are apparent. We summarize the non-zero flat prior parameter ranges in Table~\ref{tab_priors_bapp}.

\begin{table}[h!]
    \centering 
    \label{tab_priors_bapp}
    \caption{Summary of non-zero flat prior parameter ranges for the precession parameters inferred from the temporal evolution of the jet apparent velocity.}
    \begin{tabular}{c|c}
    \hline
    \hline
     Parameter  & Prior range  \\
     \hline
      $t_0$ [yr]   & [1970, 2007] \\
      $P_{\rm p}$ [yr] & [10,40]  \\
      $\gamma$ & [1, 50]\\
      $\Omega_{\rm p}\,[^{\circ}]$ & [0,50] \\
      $\phi_0\,[^{\circ}]$ & [10,15] \\
      $\eta_0\,[^{\circ}]$ & [-150,-100] \\
      $\log{f}$ & [-10,1] \\
      \hline
    \end{tabular}   
\end{table}

\subsection{Precession-nutation parameters based on the stationary-component evolution}

\begin{figure*}[h!]
    \includegraphics[width=\textwidth]{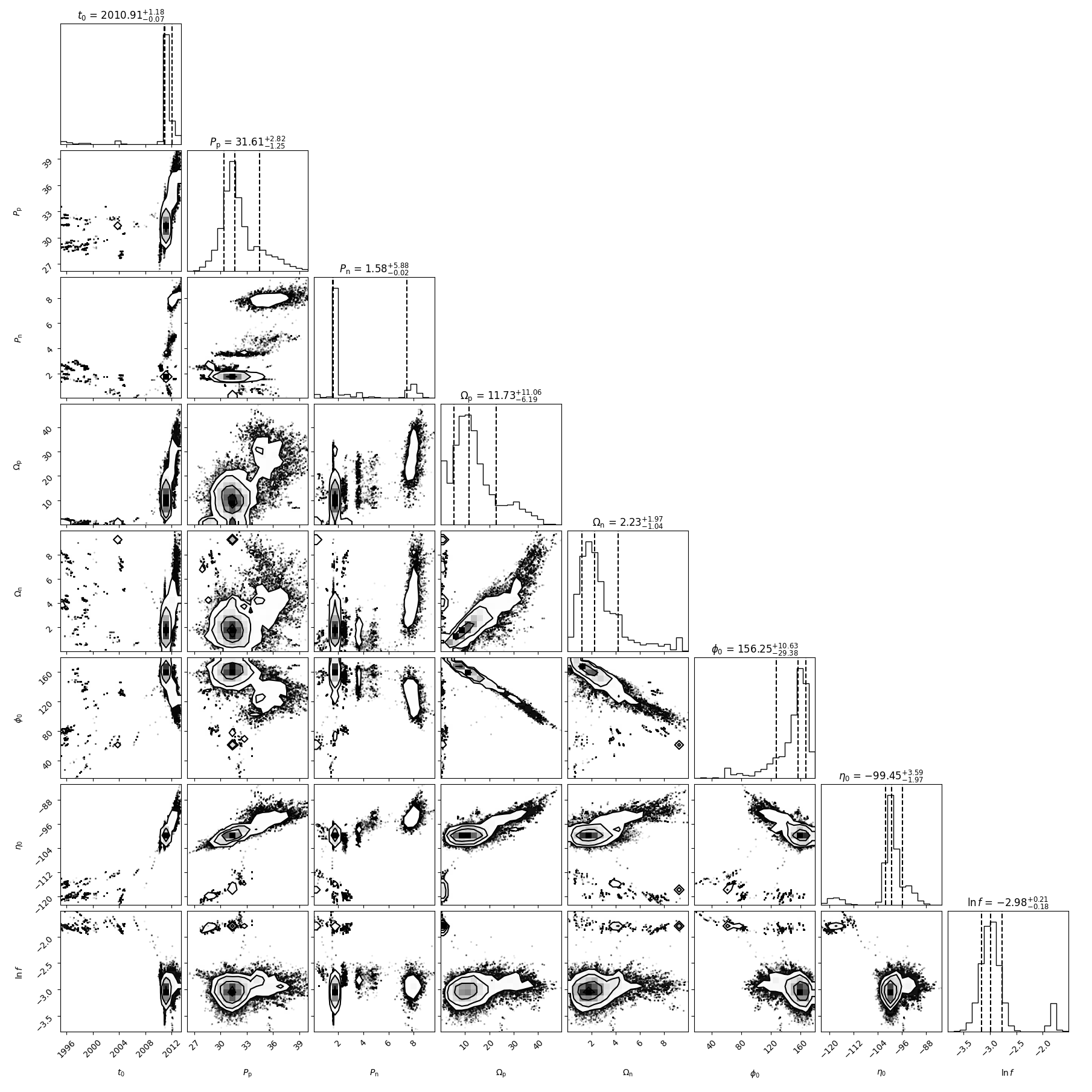}
    \caption{Marginalized one-dimensional likelihood distributions and two-dimensional likelihood contours at 1$\sigma$, 2$\sigma$, and 3$\sigma$ confidence levels for the parameters related to the temporal evolution of the stationary component due to the precession-nutation motion of the jet. The dashed vertical lines in diagonal panels stand for 16\%, 50\%, and 84\% percentiles of the inferred posterior parameter likelihood distributions. }
    \label{fig_mcmc_compa}
\end{figure*}

\begin{figure}[h!]
    \centering
    \includegraphics[width=\columnwidth]{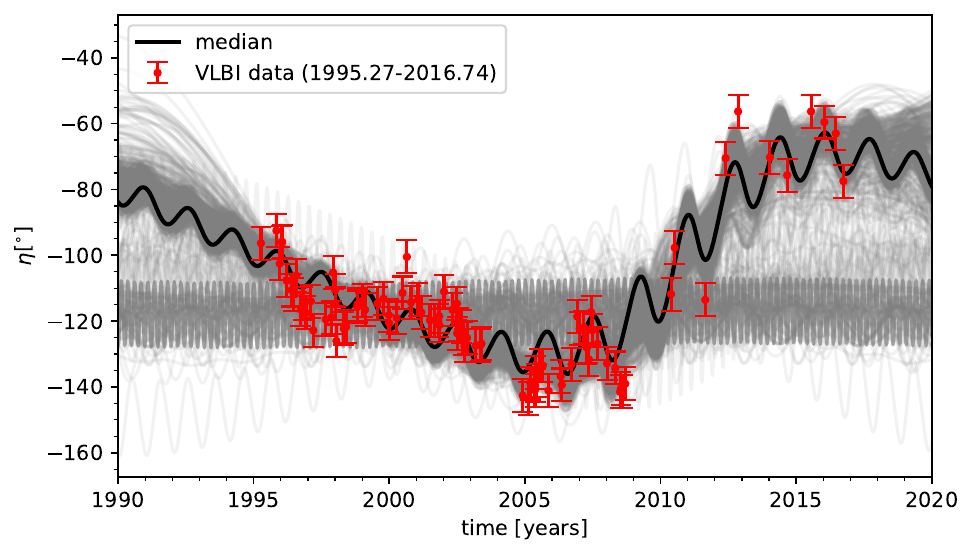}
    \caption{Median model of the temporal evolution of the position angle of the stationary component (black solid line) based on the posterior parameter distributions. Gray lines are based on 1000 samples of the parameter chains. Red points with errorbars are based on VLBI observations.}
    \label{fig_mcmc_model_compa}
\end{figure}

In Fig.~\ref{fig_mcmc_compa}, we show the corner plot with marginalized one-dimensional likelihood distributions and two-dimensional likelihood contours for the precession-nutation parameters $t_0$, $P_{\rm p}$, $P_{\rm n}$, $\Omega_{\rm p}$, $\Omega_{\rm n}$, $\phi_0$, and $\eta_0$ inferred from the temporal evolution of the position angle of the stationary component \textbf{a}. We also display the posterior distribution for the variance underestimation factor $\log f$, which is included in the definition of the likelihood function. Hence we deal with 8 free parameters. In Fig.~\ref{fig_mcmc_model_compa}, we compare the posterior median model with the VLBI data, including 1000 random samples of parameter chains. This way the precession-nutation model uncertainties around the model median are apparent. We summarize the non-zero flat prior parameter ranges in Table~\ref{tab_priors_compa}.

\begin{table}[h!]
    \centering 
    \label{tab_priors_compa}
    \caption{Summary of non-zero flat prior parameter ranges for the precession-nutation parameters inferred from the temporal evolution of the stationary component \textbf{a}.}
    \begin{tabular}{c|c}
    \hline
    \hline
     Parameter  & Prior range  \\
     \hline
      $t_0$ [yr]   & [1995, 2016] \\
      $P_{\rm p}$ [yr] & [10,40]  \\
      $P_{\rm n}$ [yr] & [0,10]  \\
      $\Omega_{\rm p}\,[^{\circ}]$ & [0,90] \\
       $\Omega_{\rm n}\,[^{\circ}]$ & [0,10] \\
      $\phi_0\,[^{\circ}]$ & [0,180] \\
      $\eta_0\,[^{\circ}]$ & [-180,-50] \\
      $\log{f}$ & [-10,1] \\
      \hline
    \end{tabular}   
\end{table}

\section{Light curve modelling using autoregressive integrated moving average (ARIMA) models}
\label{appendix_arima}

Here we provide additional plots and supporting statistical verification related to the radio light curve modelling using stationary stochastic processes represented by ARIMA$(p,d,q)$ models described in Subsection~\ref{subsec_arima}.

We apply the \texttt{python} function \texttt{ARIMA} in \texttt{statsmodels} that requires a regular binning of the light curves. We interpolate the data into two regular time grids -- one with the time step of 0.1 years and the other with the time step of 0.05 years. The mean time sampling of the light curves is $\sim 0.03$ years, hence we do not apply a smaller time step than $0.05$ years.

\begin{figure}[h!]
    \centering
    \begin{tabular}{ccc}
    \includegraphics[width=0.33\textwidth]{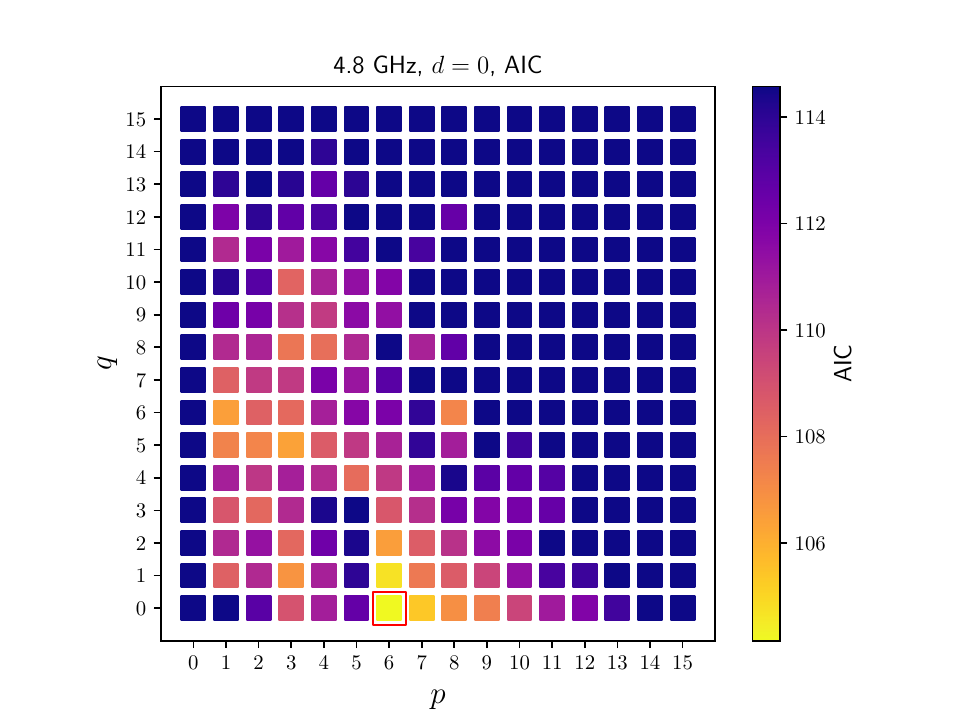} &  \includegraphics[width=0.33\textwidth]{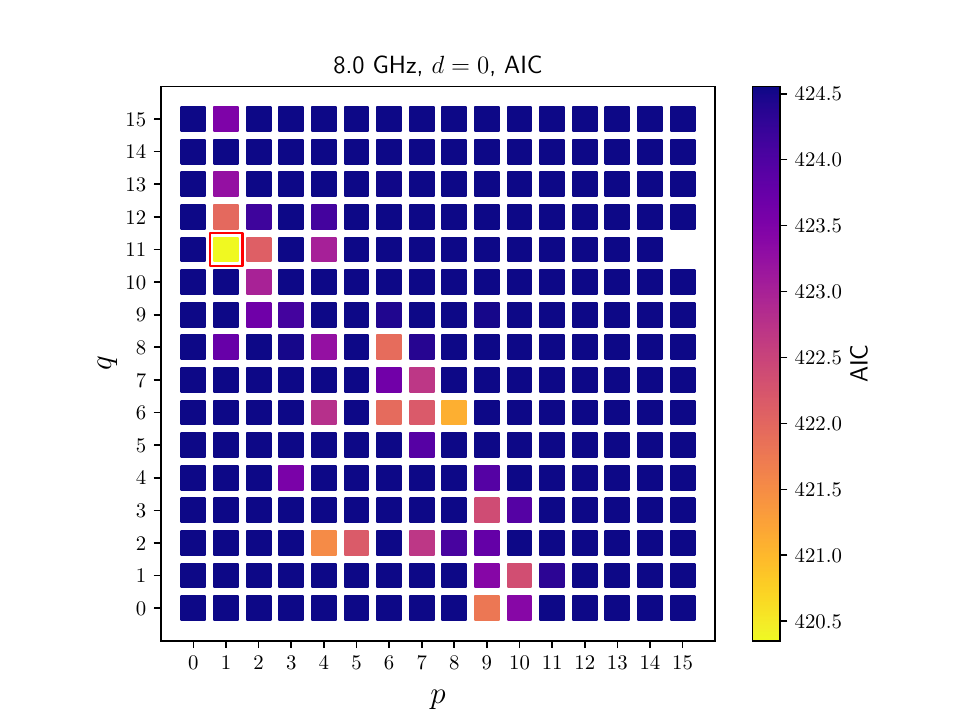} & \includegraphics[width=0.33\textwidth]{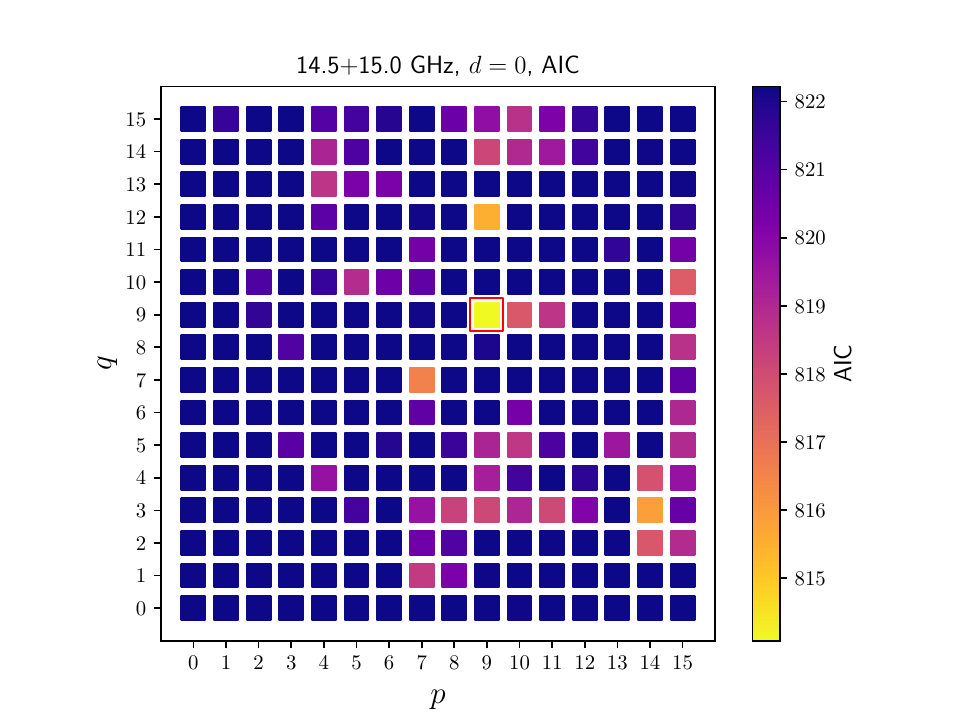}\\
     \includegraphics[width=0.33\textwidth]{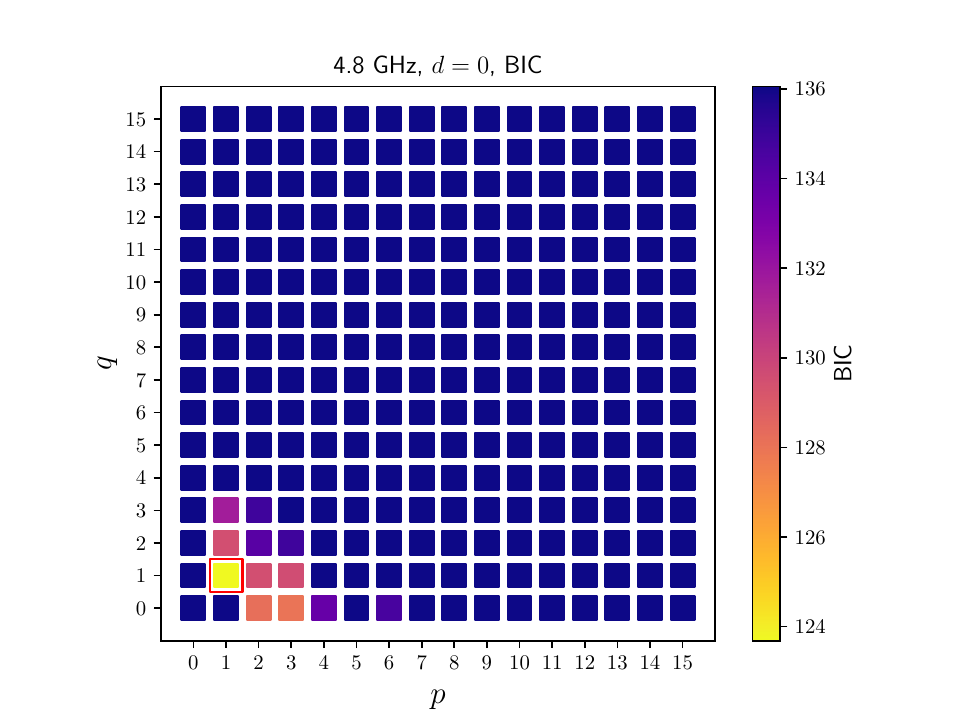} &  \includegraphics[width=0.33\textwidth]{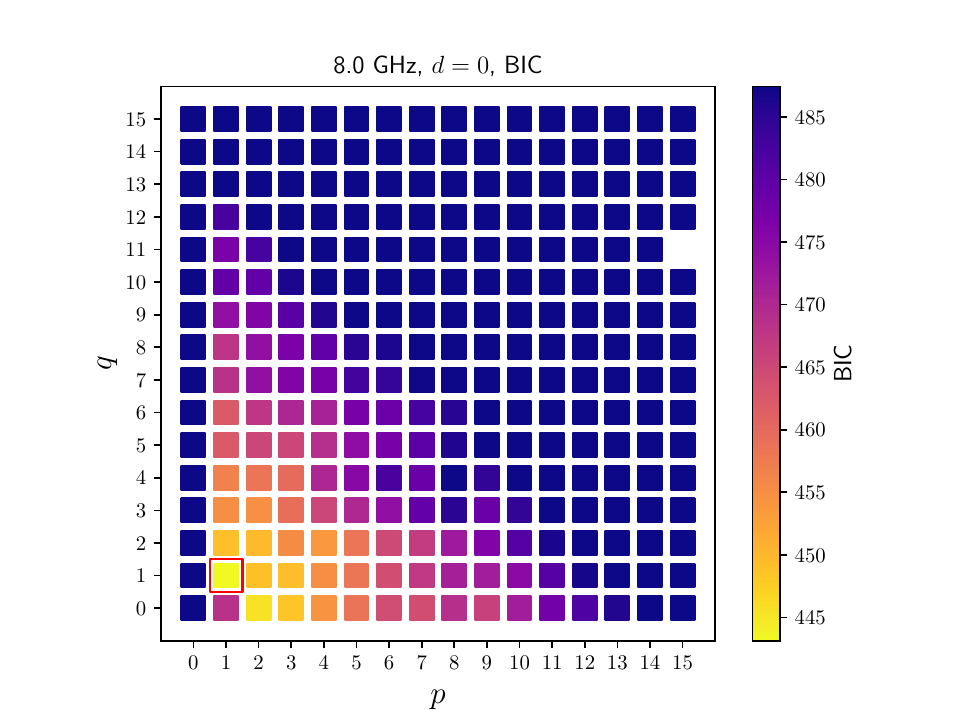} & \includegraphics[width=0.33\textwidth]{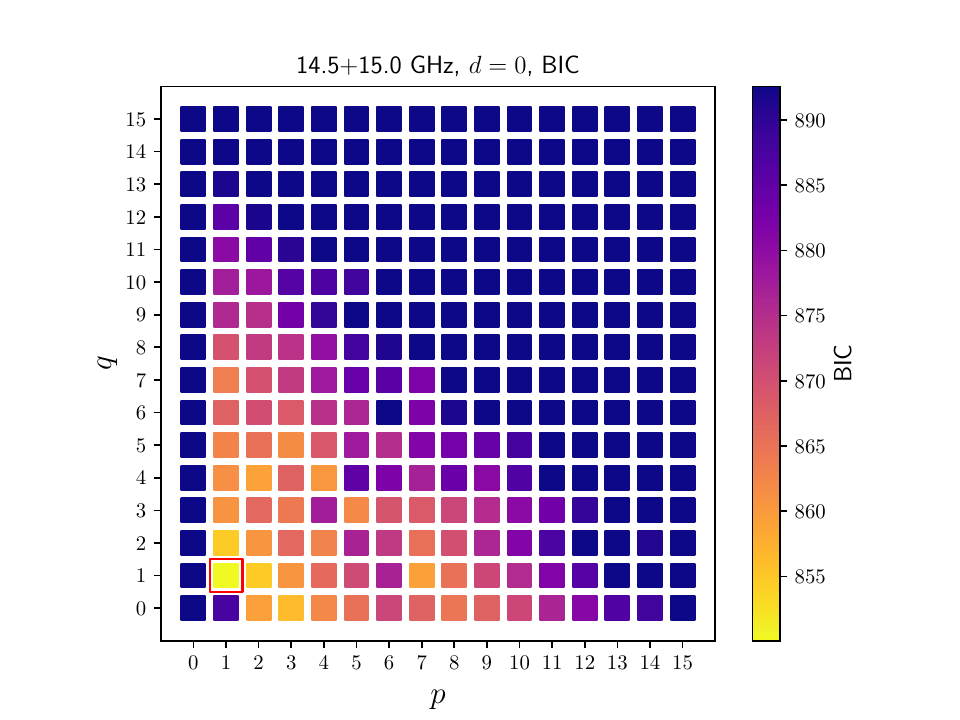}\\
      \includegraphics[width=0.33\textwidth]{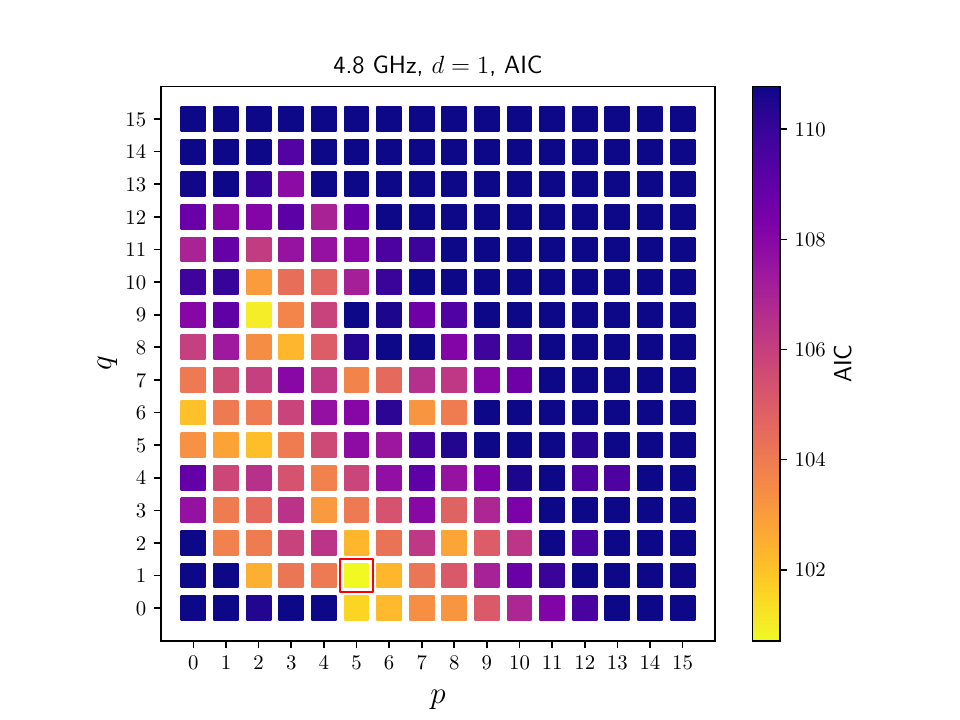} & \includegraphics[width=0.33\textwidth]{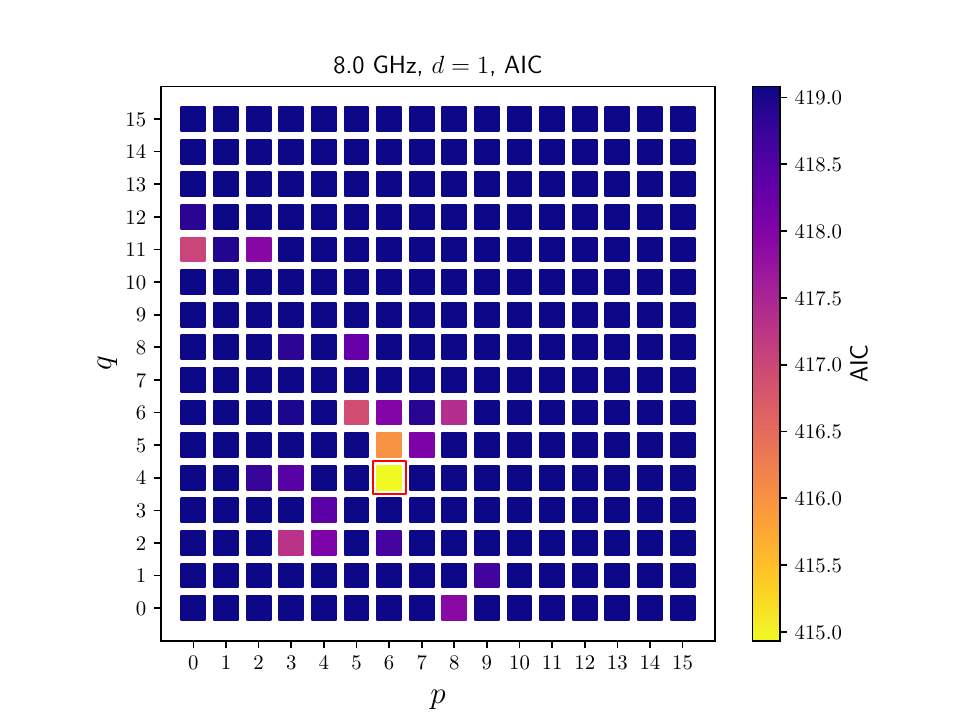} &\includegraphics[width=0.33\textwidth]{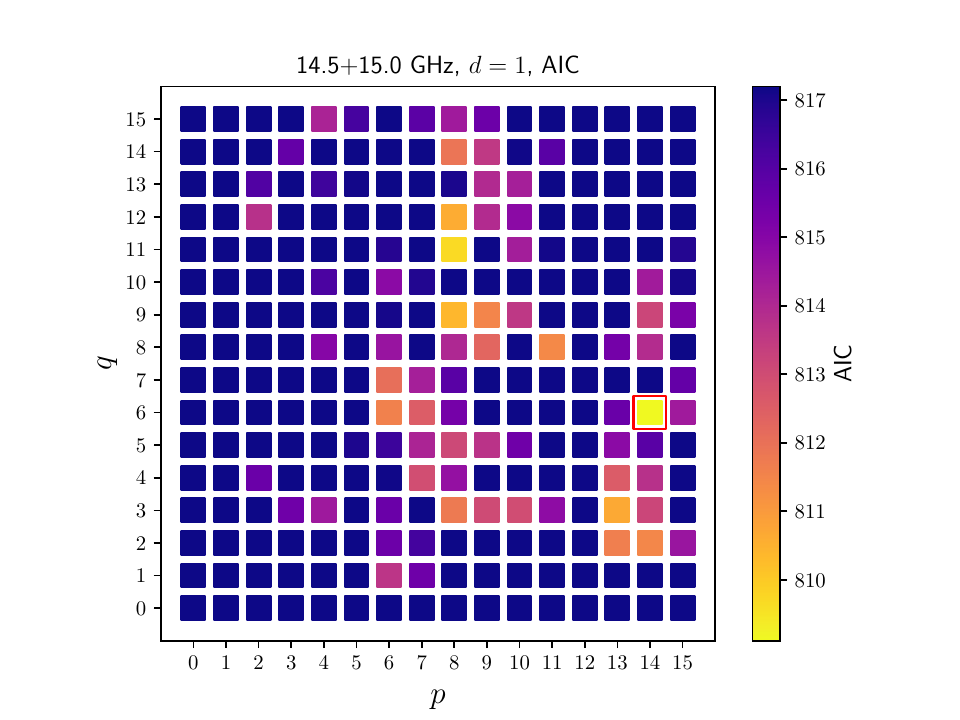}\\
       \includegraphics[width=0.33\textwidth]{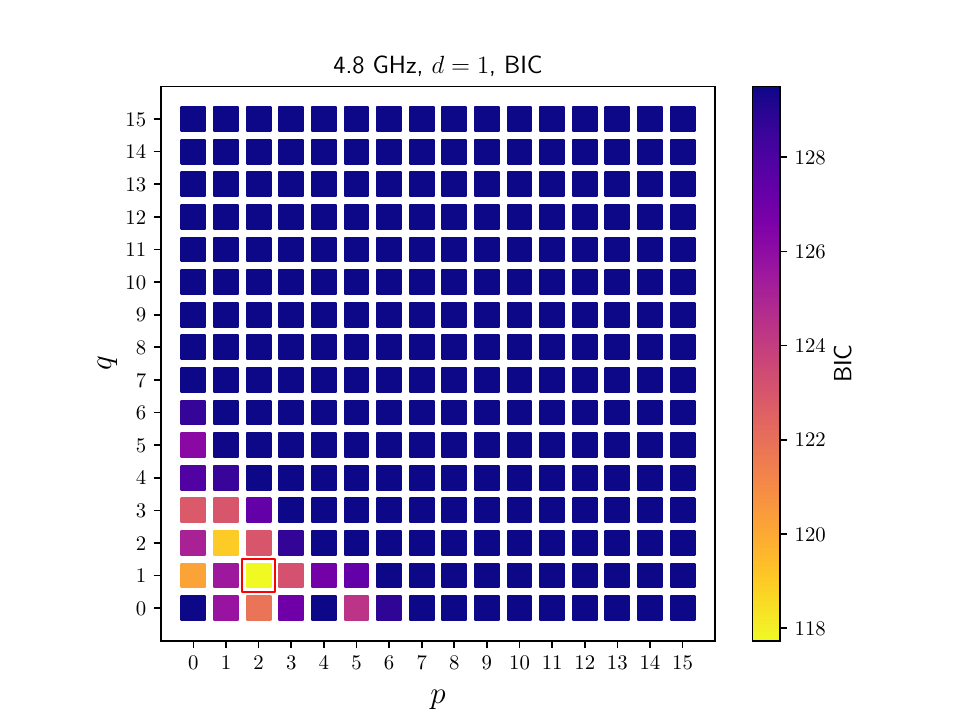} & \includegraphics[width=0.33\textwidth]{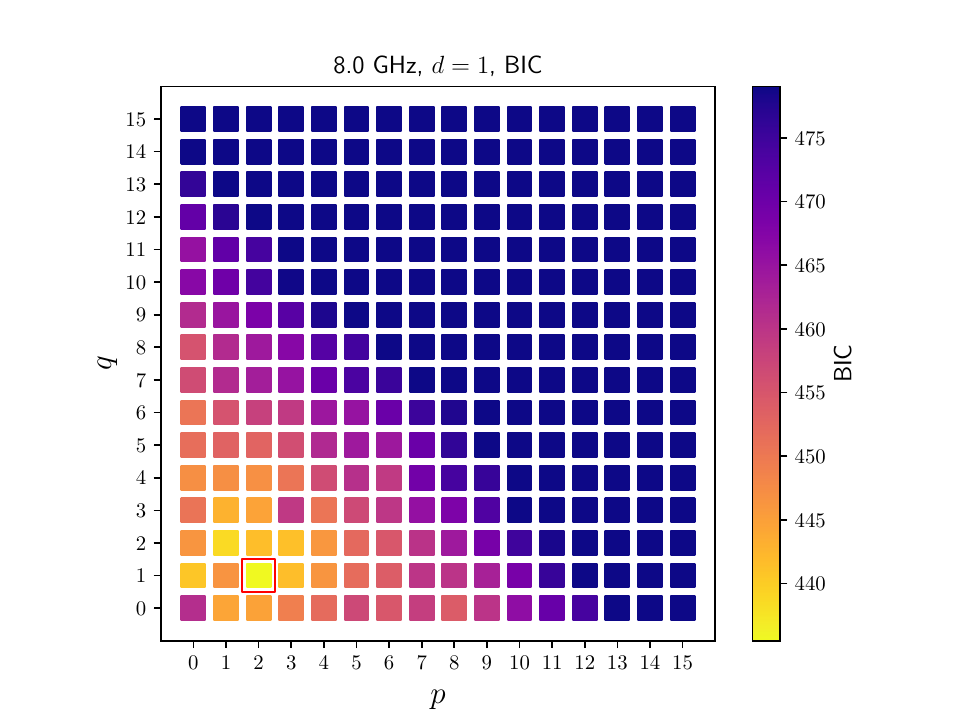} &\includegraphics[width=0.33\textwidth]{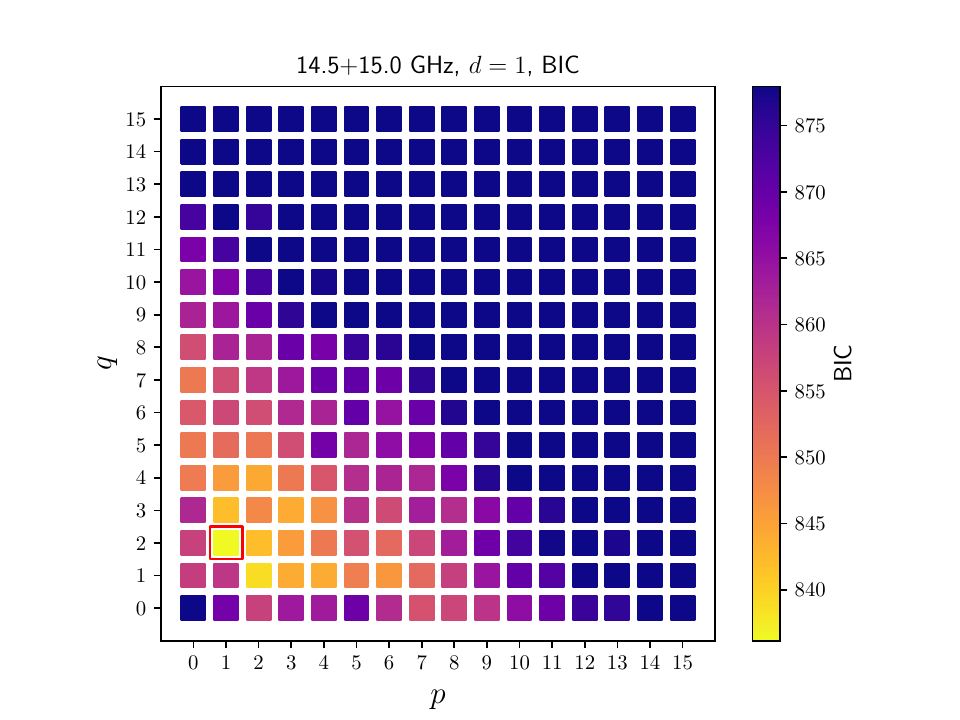}\\
    \end{tabular}
    \caption{Distribution of Akaike and Bayesian Information Criteria (AICs and BICs) for ARIMA(p,0,q) (upper two rows) and ARIMA(p,1,q) (lower two rows) models fitted to 4.8, 8.0, and 14.5+15.0 GHz light curves (from the left to the right columns). AIC and BIC minima are denoted by red-framed squares. The radio light curves were interpolated to the regular time-step of 0.1 years. The white squares in the middle upper two panels represent $p$ and $q$ parameters, for which the ARIMA model fit did not converge.} 
    \label{fig_ARIMA_pdq_scan1}
\end{figure}

In Table~\ref{tab_ARIMA_results1}, we describe basic results of fitting different ARIMA$(p,d,q)$ models to light curves at 4.8, 8.0, and 14.5$+$15 GHz that were interpolated to regular 0.1-year time steps. Here $f_{t}'$ denotes the light curve that is differenced once, i.e. $f_{t}'=f_{t}-f_{t-1}$. The table lists Augmented Dickey-Fuller (ADF) test statistic values and the corresponding $p$-values for no differencing and the first-order differencing. Then we include the preferred ARIMA model for the first-order differencing including its Akaike and Bayesian information criterium (AIC and BIC) values as well as its best-fit parameters. For completeness, we also list minimum AIC and BIC values for $d=0$ and $d=1$ cases as well as corresponding ARIMA models, which we inferred from AIC and BIC values as functions of different $p$ and $q$ parameters, see Fig.~\ref{fig_ARIMA_pdq_scan1}. For completeness, we evaluate the degree of fractional differencing using the \texttt{python} library \texttt{fracdiff} \citep{de2018advances}, for which the light curves become stationary, i.e. they pass the Augmented Dickey-Fuller test at the 95\% confidence level, and at the same time they keep the maximum possible signal for the stationary case.  

\begin{figure}[h!]
    \centering
    \begin{tabular}{ccc}
    \includegraphics[width=0.33\textwidth]{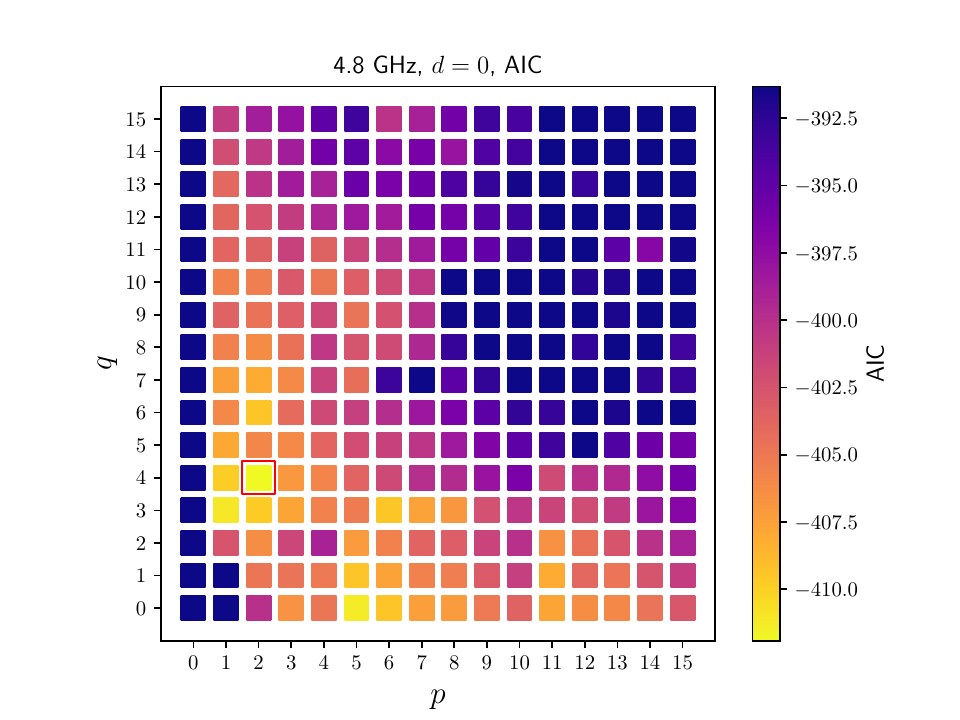} &  \includegraphics[width=0.33\textwidth]{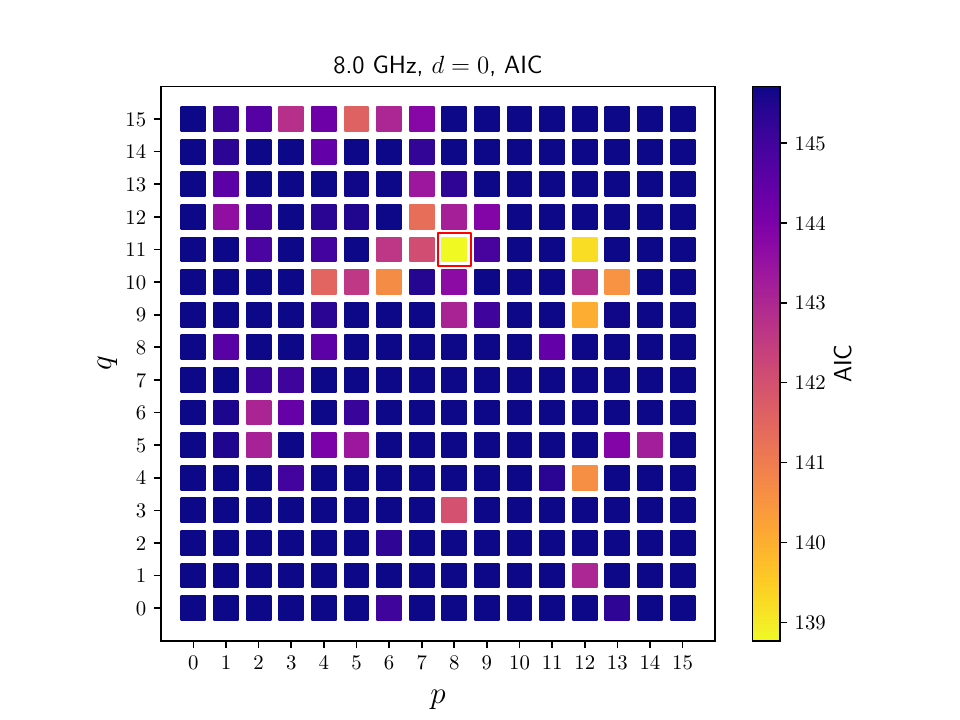} & \includegraphics[width=0.33\textwidth]{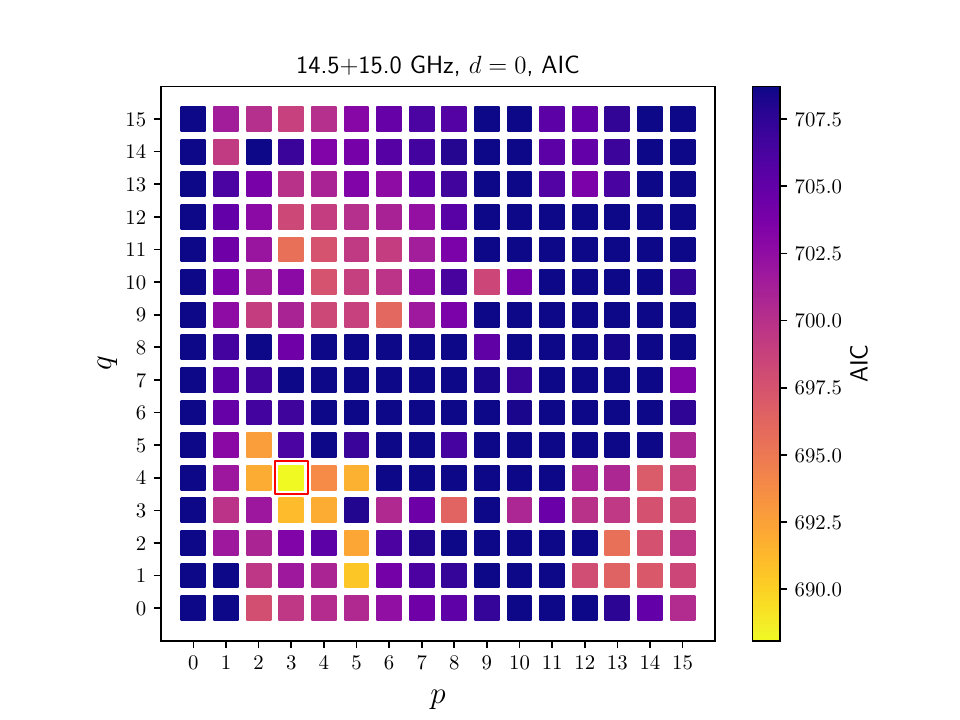} \\
     \includegraphics[width=0.33\textwidth]{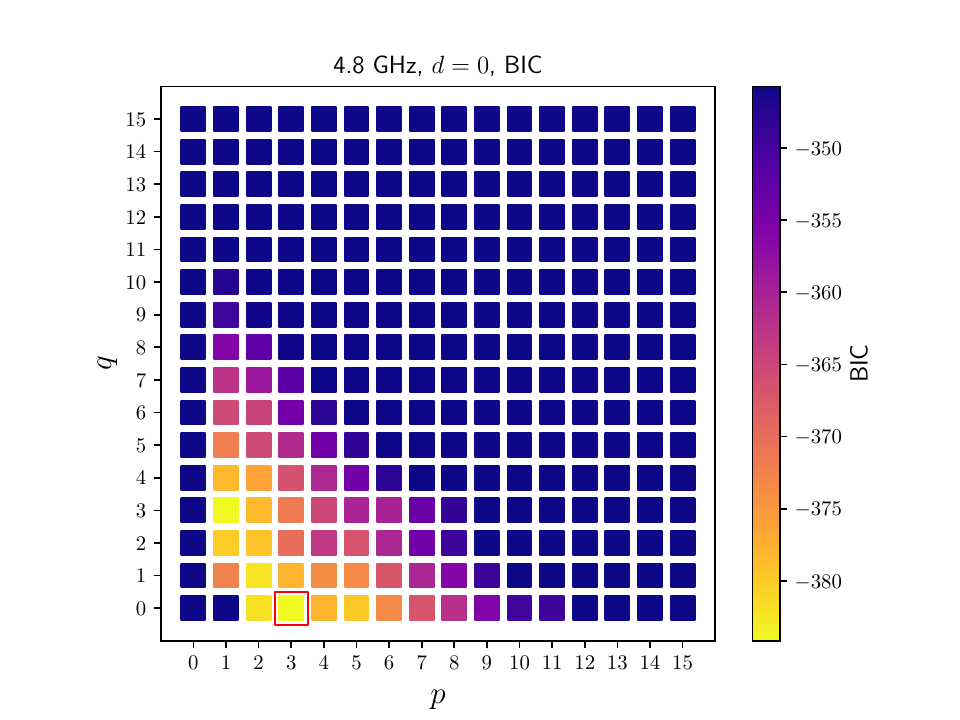} &  \includegraphics[width=0.33\textwidth]{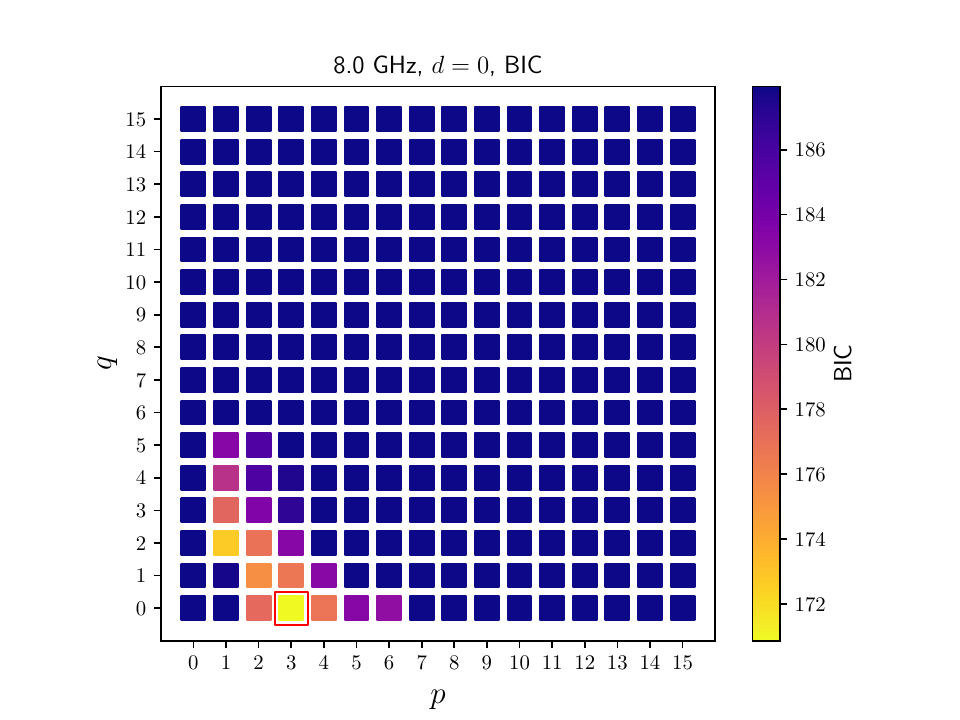} & \includegraphics[width=0.33\textwidth]{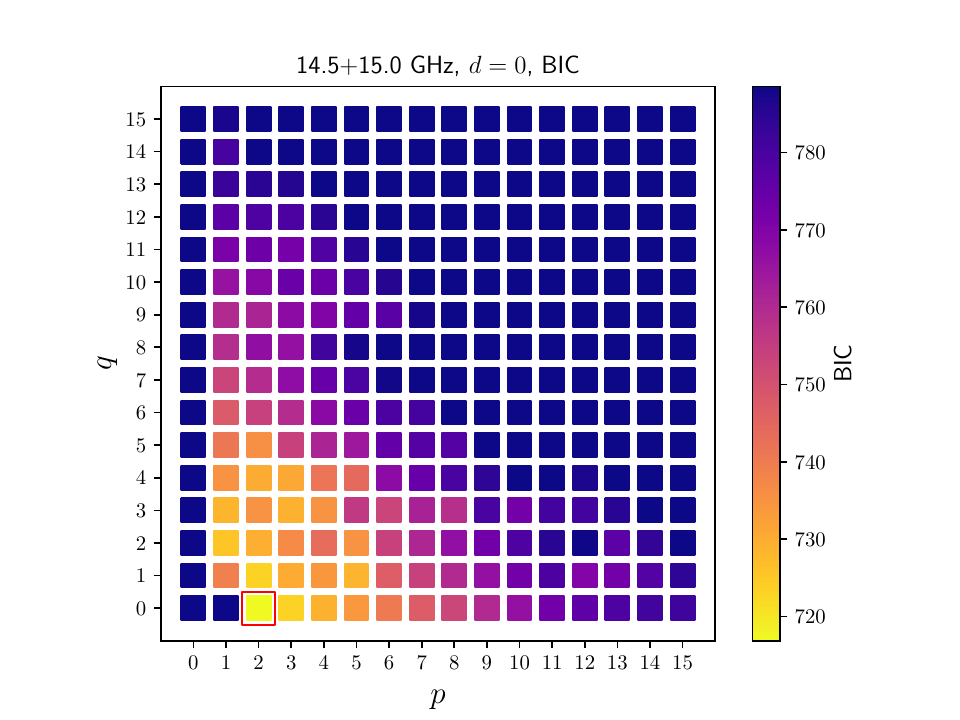} \\
      \includegraphics[width=0.33\textwidth]{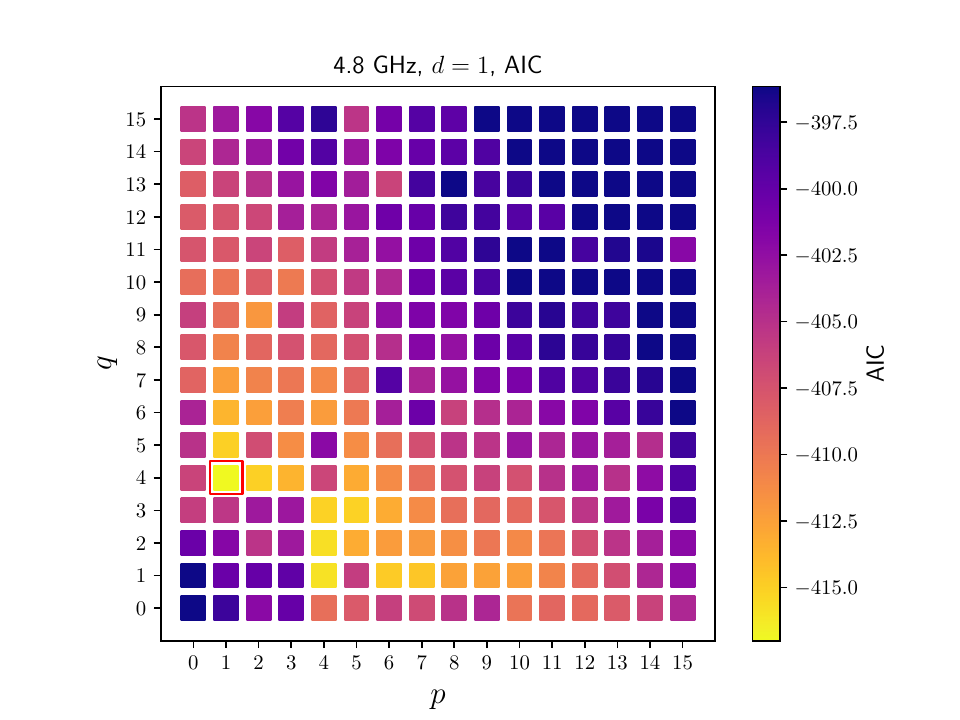} & \includegraphics[width=0.33\textwidth]{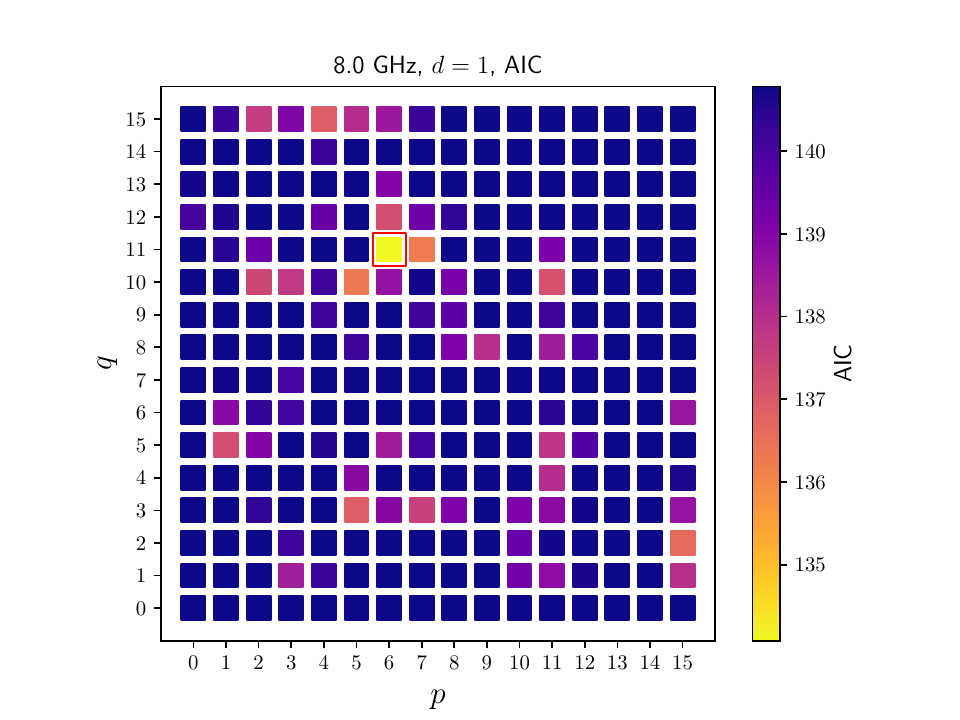} & 
     \includegraphics[width=0.33\textwidth]{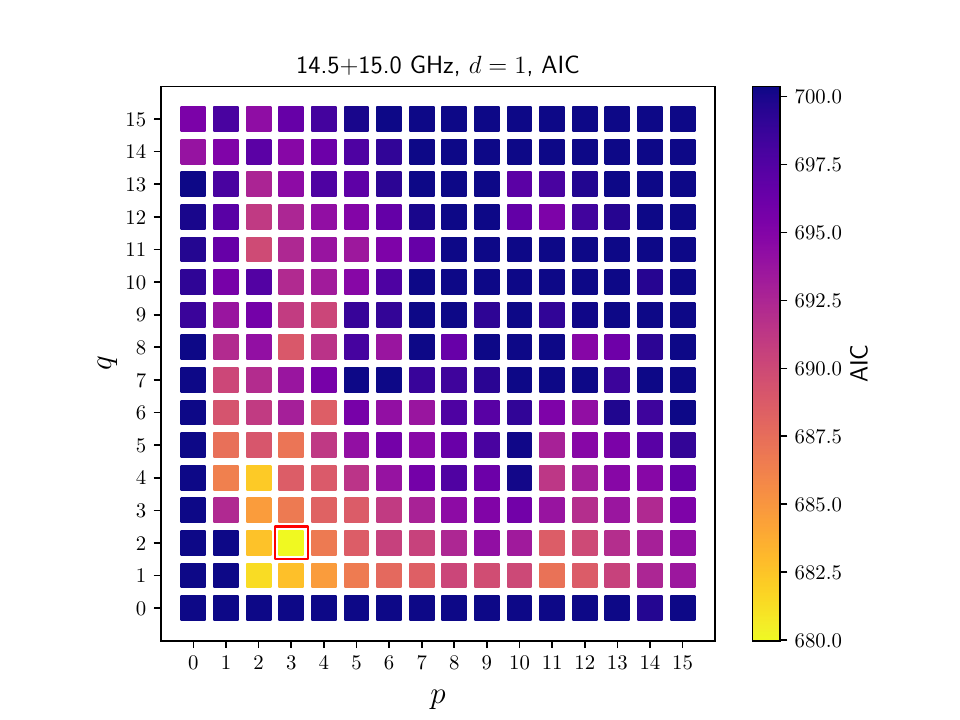} \\
       \includegraphics[width=0.33\textwidth]{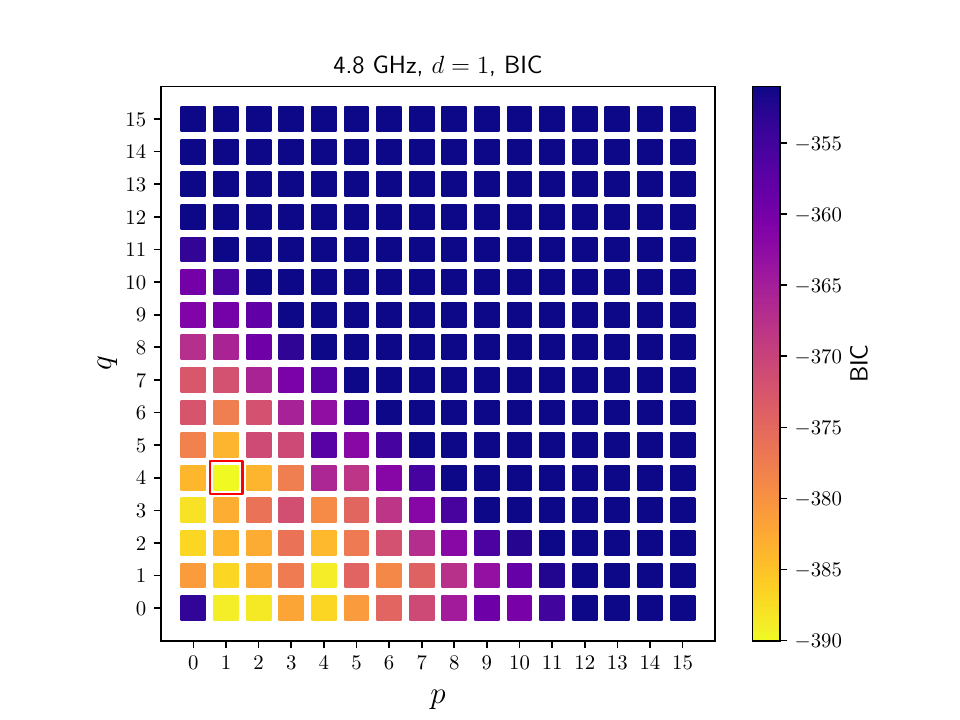} & \includegraphics[width=0.33\textwidth]{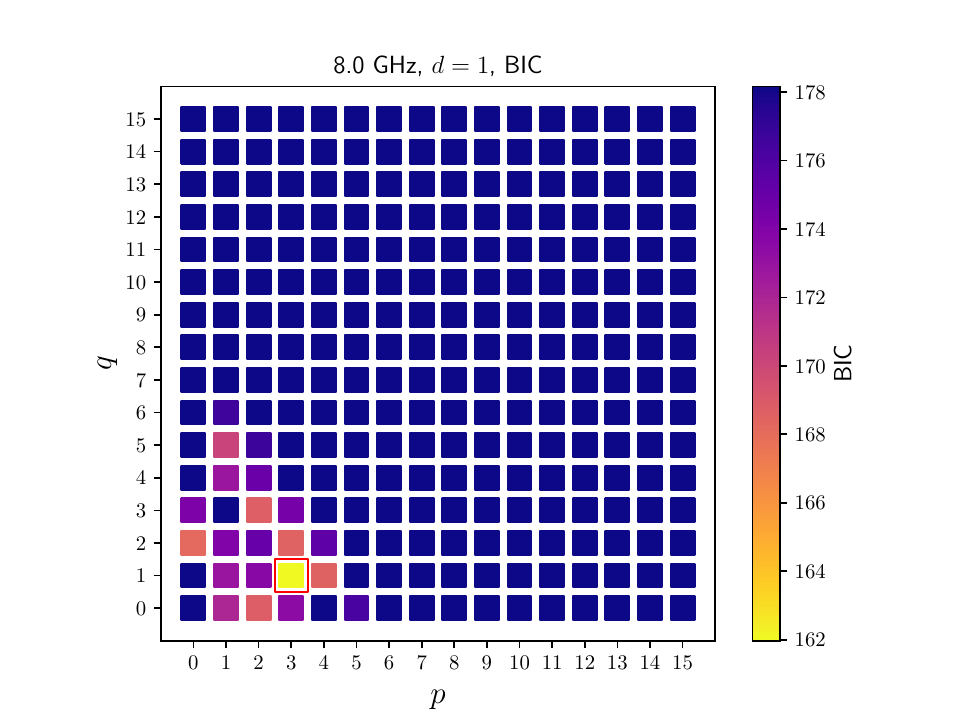} & \includegraphics[width=0.33\textwidth]{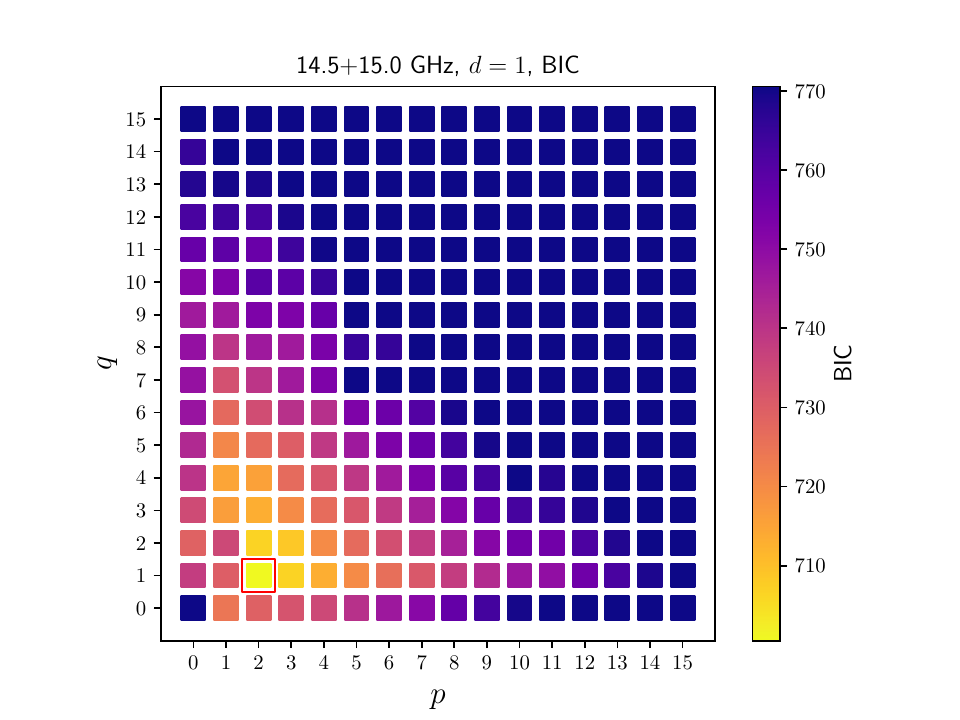}\\
    \end{tabular}
    \caption{Distribution of Akaike and Bayesian Information Criteria (AICs and BICs) for ARIMA(p,0,q) (upper two rows) and ARIMA(p,1,q) (lower two rows) models fitted to 4.8, 8.0, and 14.5+15.0 GHz light curves (from the left to the right columns). AIC and BIC minima are denoted by red-framed squares. The radio light curves were interpolated to the regular time-step of 0.05 years.} 
    \label{fig_ARIMA_pdq_scan2}
\end{figure}

In Table~\ref{tab_ARIMA_results2}, we list the same quantities and parameters for the case when the radio light curves are interpolated to regular 0.05-year time steps. We may notice small differences of the $p$ and $q$ order for the ARIMA models corresponding to the smallest BIC values. However, considering a smaller time step by a factor of two, $p$ and $q$ values increased at most by a factor of two, which implies that stochastic stationary processes are relevant on the timescales of $\sim 0.05-0.2$ years, i.e. significantly smaller timescales in comparison with the precession and nutation timescales. The AIC and BIC values for different $p$ and $q$ parameters are graphically shown in Fig.~\ref{fig_ARIMA_pdq_scan2} for the case of $0.05$-year time step.

\begin{figure}[h!]
    \centering
    \begin{tabular}{ccc}
    \includegraphics[width=0.33\textwidth]{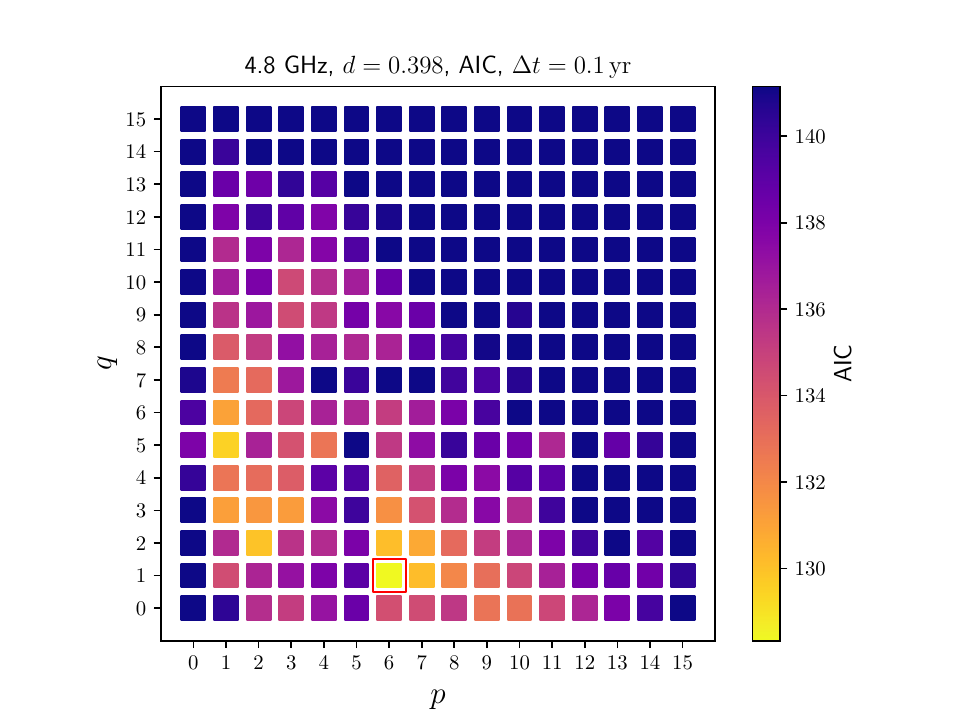} &  \includegraphics[width=0.33\textwidth]{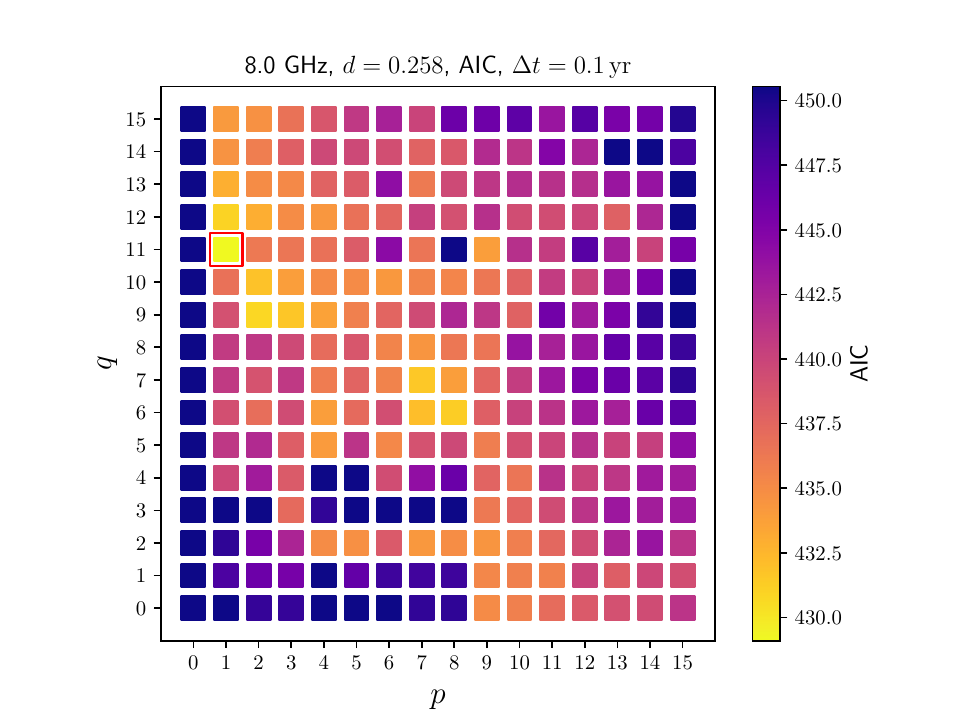} & \includegraphics[width=0.33\textwidth]{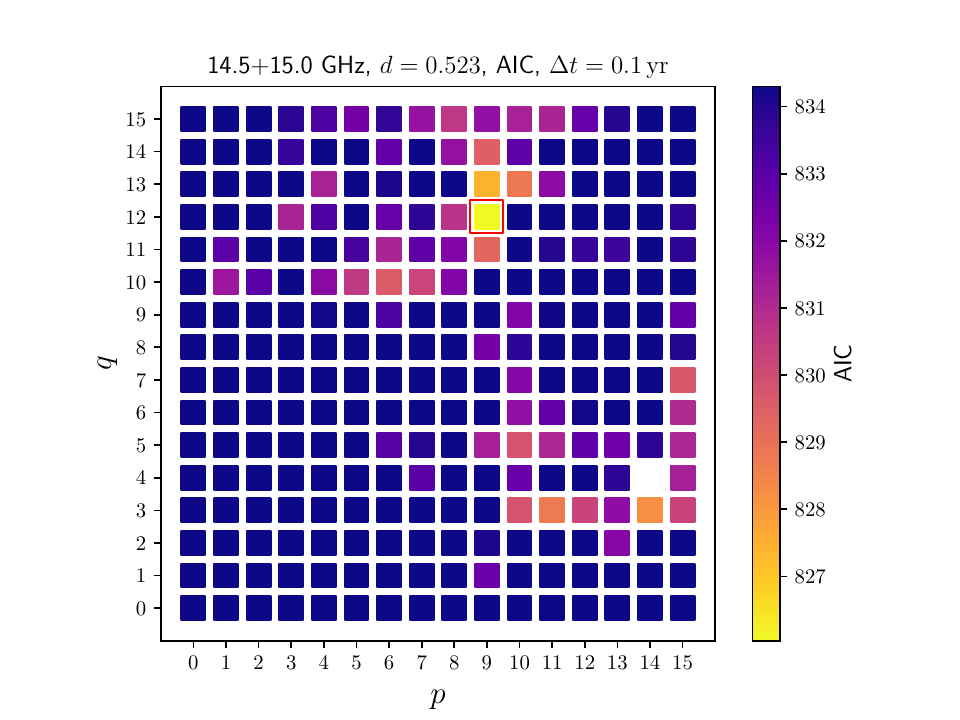} \\
     \includegraphics[width=0.33\textwidth]{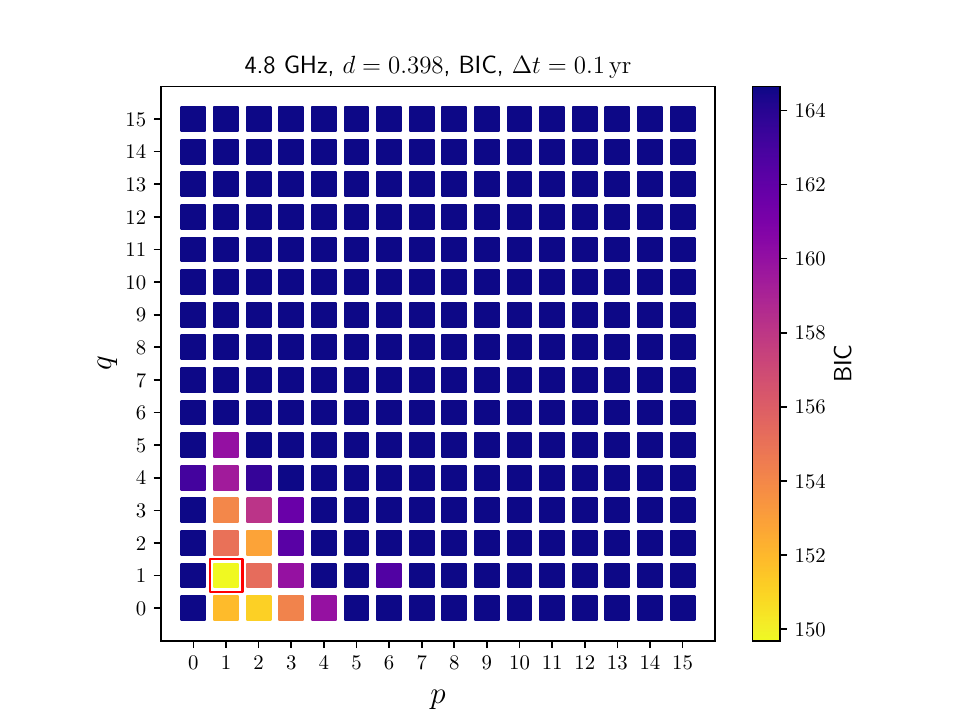} &  \includegraphics[width=0.33\textwidth]{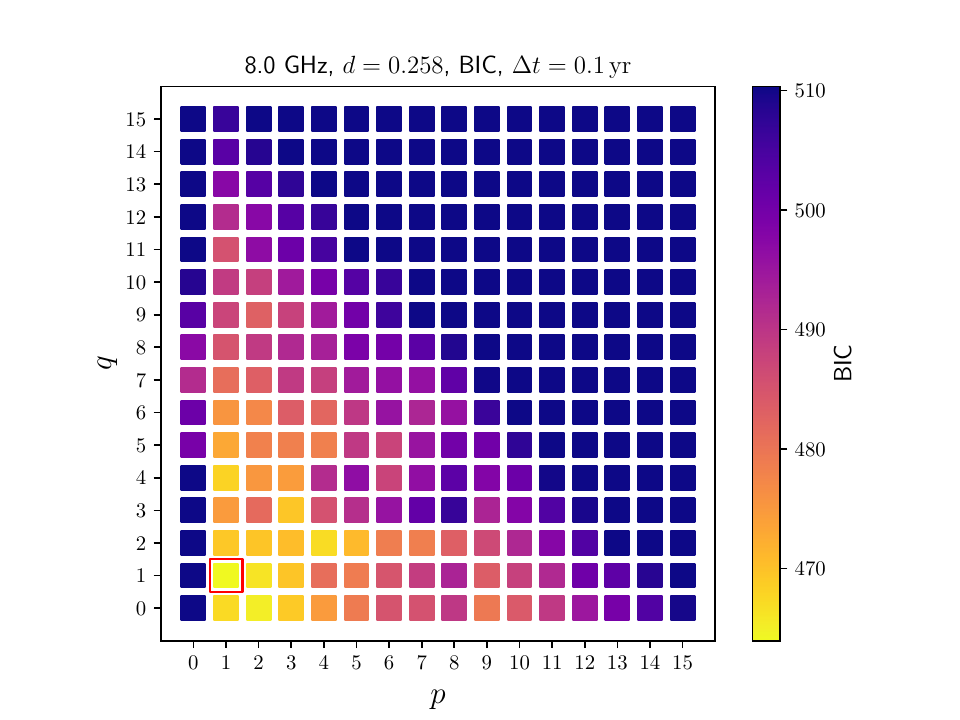} & \includegraphics[width=0.33\textwidth]{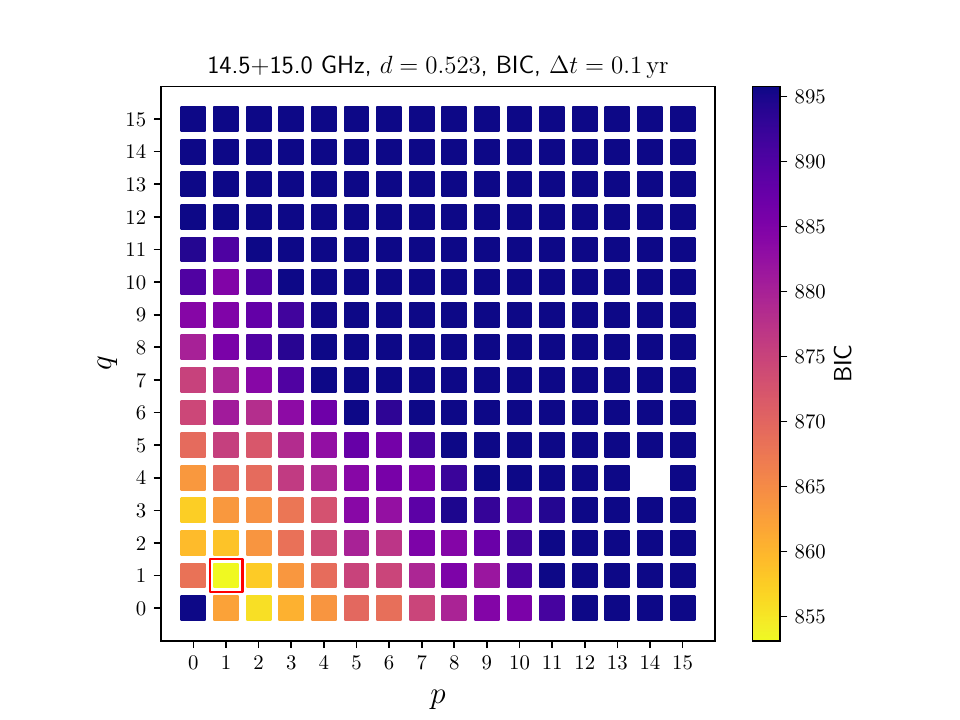} \\
     \includegraphics[width=0.33\textwidth]{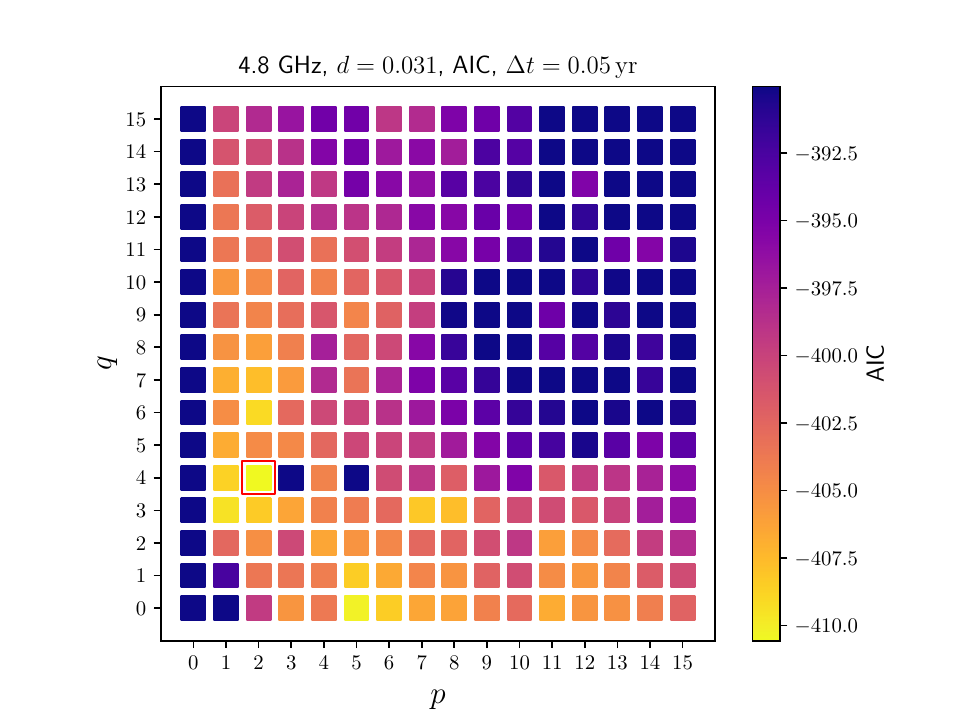}  &  \includegraphics[width=0.33\textwidth]{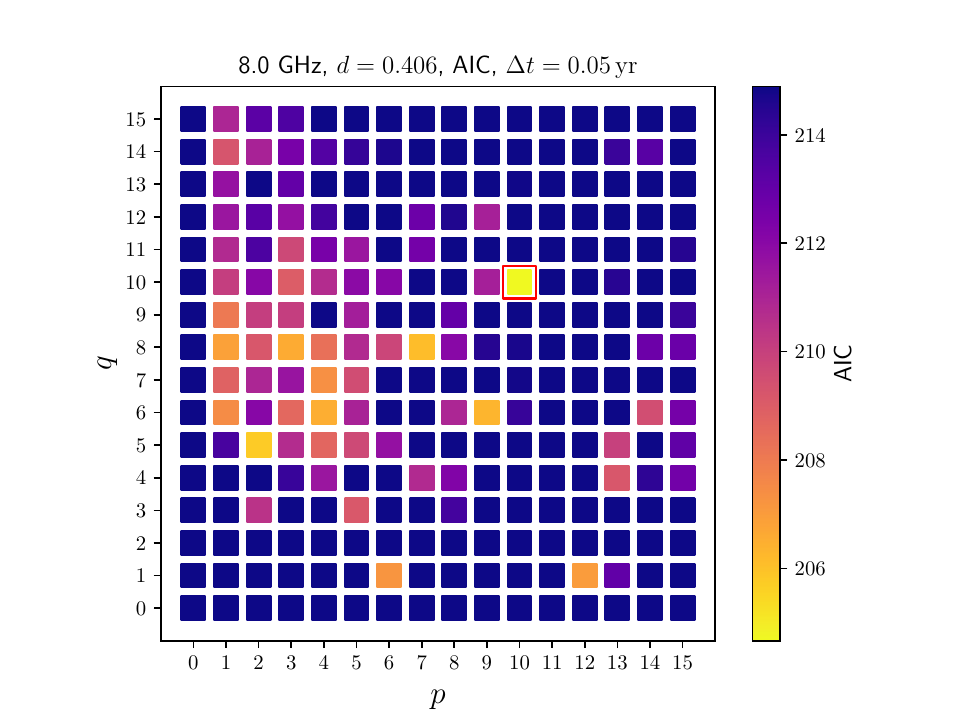}   &  \includegraphics[width=0.33\textwidth]{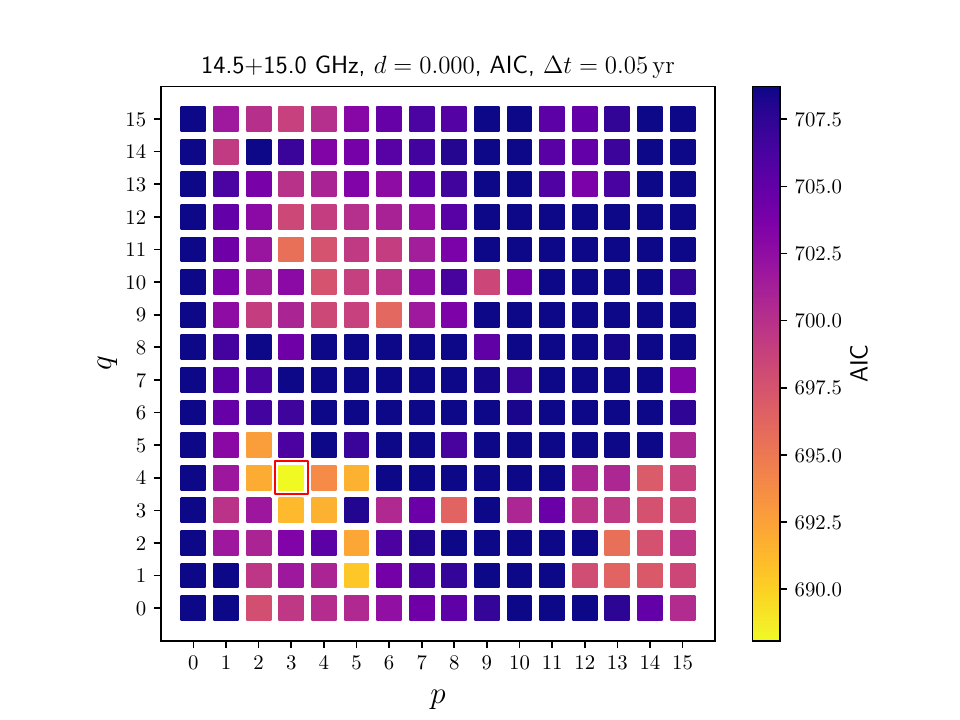}  \\
     \includegraphics[width=0.33\textwidth]{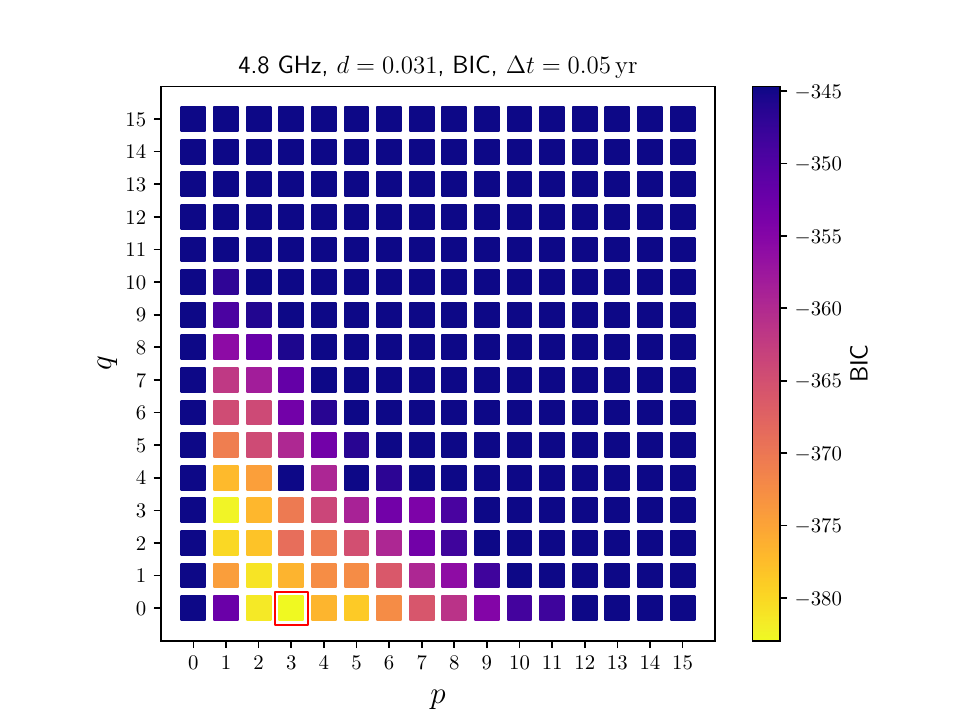}  & \includegraphics[width=0.33\textwidth]{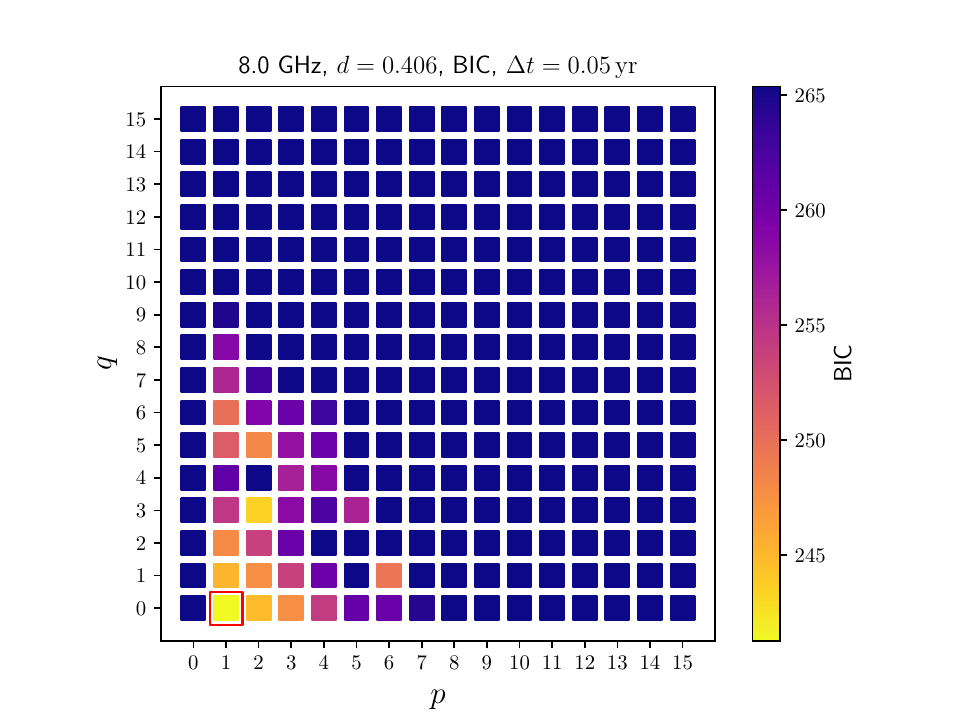}   & \includegraphics[width=0.33\textwidth]{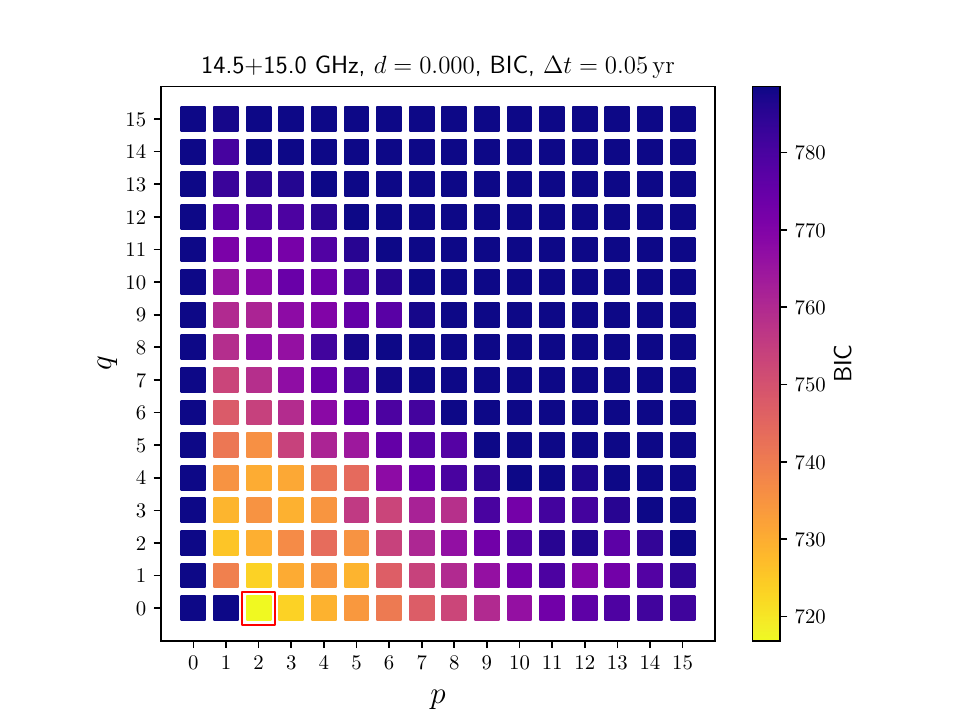}      
    \end{tabular}
    \caption{Distribution of Akaike and Bayesian Information Criteria (AICs and BICs) for ARIMA models fitted to 4.8, 8.0, and 14.5+15.0 GHz light curves (from the left to the right columns) that are fractionally differenced to reach stationarity. AIC and BIC minima are denoted by red-framed squares. The degree of fractial differencing is listed in the title of each figure. The upper two rows correspond to the radio light curves interpolated to the regular time-step of 0.1 years, while the lower two rows correspond to the interpolation time-step of 0.05 years. The white squares in the right upper two panels denote $p$ and $q$ values, for which the ARIMA model fit did not converge.}
    \label{fig_ARIMA_pdq_scan3}
\end{figure}

In Fig.~\ref{fig_ARIMA_pdq_scan3}, we fit the fractionally differenced radio light curves with ARIMA models, which again indicates that $p=1$ and $q=1$ for the minimum BIC cases for the 0.1-year time step. For the 0.5-year time step, $p$ is equal to 3, 1, and 2 for 4.8, 8.0, and 14.5$+$15.0 GHz light curves, respectively, while $q=0$ for the minimum BIC cases.

\begin{table}[h!]
    \centering
    \caption{Summary of the radio light curve analysis using ARIMA models. The light curves were interpolated to a regular time-step of 0.1 years.}
    \resizebox{1.\textwidth}{!}{%
    \begin{tabular}{c|c|c|c}
    \hline
    \hline
         & 4.8 GHz  & 8.0 GHz  & 14.5$+$15.0 GHz\\
    \hline     
    Augmented Dickey-Fuller test - no differencing  &  ADF$=-1.719$ ($p=0.421$) &  ADF$=-1.675$ ($p=0.444$)  &  ADF$=-1.280$ ($p=0.638$)  \\
    Augmented Dickey-Fuller test - 1st differencing  &  ADF$=-10.626$ ($p=5.357\times 10^{-19}$)& ADF$=-10.543$ ($p=8.539\times 10^{-19}$)         & ADF$=-7.525$ ($p=3.705\times 10^{-11}$)  \\
    Fractional difference degree & 0.398 & 0.258 & 0.523 \\
    AIC$_{\rm min}$ for $d=0$ & 104.2, ARIMA(6,0,0)   & 420.4, ARIMA(1, 0, 11)           &  814.1, ARIMA(9,0,9) \\
    BIC$_{\rm min}$ for $d=0$ & 123.7, ARIMA(1,0,1)  &  443.1, ARIMA(1,0,1)           & 850.1, ARIMA(1,0,1)  \\
    AIC$_{\rm min}$ (frac. diff.) & 128.3, ARIMA(6,0.398,1)   & 429.1, ARIMA(1, 0.258, 11)           &  826.0, ARIMA(9, 0.523, 12) \\
    BIC$_{\rm min}$ (frac. diff.) & 149.7, ARIMA(1,0.398, 1)  &  464.0, ARIMA(1, 0.258,1)           & 853.1, ARIMA(1, 0.523, 1)  \\
    AIC$_{\rm min}$ for $d=1$ & 100.7, ARIMA(5,1,1)   & 414.9, ARIMA(6,1,4)
           & 809.1, ARIMA(14,1,6) \\
    BIC$_{\rm min}$ for $d=1$ & 117.7, ARIMA(2,1,1)  & 435.5, ARIMA(2,1,1)           &  836.2, ARIMA(1,1,2) \\
    Preferred ARIMA model & (2,1,1), AIC$=102.5$, BIC$=117.7$ & (2,1,1), AIC$=419.4$, BIC$=435.5$ & (1,1,2), AIC$=819.8$, BIC$=836.2$ \\
    Best-fit ARIMA parameters & $f_{t}'=(1.034 \pm 0.067)f_{t-1}'-(0.328    \pm  0.040)f_{t-2}'+$ & $f_{t}'=(1.077\pm 0.054)f_{t-1}'-(0.310 \pm 0.038)f_{t-2}'+  $          & $f_{t}'=( 0.710 \pm 0.050)f_{t-1}'+  $  \\
  & $+\epsilon_{t}-(0.811 \pm 0.060)\epsilon_{t-1}$ &  $+\epsilon_{t}-(0.860 \pm 0.052)\epsilon_{t-1}$          &  $+\epsilon_{t}-(0.502 \pm      0.063)\epsilon_{t-1}-(0.393 \pm 0.045) \epsilon_{t-2}$       \\    
    \hline
    \end{tabular}  
    }
    \label{tab_ARIMA_results1}
\end{table}

\begin{table}[h!]
    \centering
    \caption{Summary of the radio light curve analysis using ARIMA models. The light curves were interpolated to a regular time-step of 0.05 years.}
    \resizebox{1.\textwidth}{!}{%
    \begin{tabular}{c|c|c|c}
    \hline
    \hline
         & 4.8 GHz  & 8.0 GHz  & 14.5$+$15.0 GHz\\
    \hline     
    Augmented Dickey-Fuller test - no differencing  &  ADF$=-2.752$ ($p=0.066$) & ADF$=-1.746$ ($p=0.408$) & ADF$=-4.978$ ($p=2.449 \times 10^{-5}$)   \\
    Augmented Dickey-Fuller test - 1st differencing  &  ADF$=-10.012$ ($p=1.771 \times 10^{-17}$) & ADF$=-10.378$ ($p=2.174 \times 10^{-18}$)       & ADF$=-10.999$ ($p=6.751 \times 10^{-20}$)  \\ 
     Fractional difference degree & 0.031 & 0.406 & 0.000 \\
    AIC$_{\rm min}$ for $d=0$ & -411.9, ARIMA(2, 0, 4)   &  138.8, ARIMA(8, 0, 11)  & 688.1, ARIMA(3, 0, 4)   \\
    BIC$_{\rm min}$ for $d=0$ & -384.2, ARIMA(3, 0, 0)  &  170.9, ARIMA(3, 0, 0)  & 716.9, ARIMA(2, 0, 0)  \\
   AIC$_{\rm min}$ (frac. diff.) &  -410.6, ARIMA(2, 0.031, 4)   &  204.7, ARIMA(10, 0.406, 10)  & 688.1, ARIMA(3, 0.000, 4)   \\
    BIC$_{\rm min}$ (frac. diff.) & -383.0, ARIMA(3, 0.031, 0)  &  241.3, ARIMA(1, 0.406, 0)  & 716.9, ARIMA(2, 0.000, 0)  \\  
    AIC$_{\rm min}$ for $d=1$ & -417.0, ARIMA(1, 1, 4)    & 134.1, ARIMA(6, 1, 11) 
           & 680.0, ARIMA(3, 1, 2) \\
    BIC$_{\rm min}$ for $d=1$ & -390.0, ARIMA(1, 1, 4) & 162.0, ARIMA(3, 1, 1)
        & 700.5, ARIMA(2, 1, 1)  \\
    Preferred ARIMA model & (1,1,4), AIC$=-417.0$, BIC$=-390.0$ & (3,1,1), AIC$=138.4$, BIC$=162.0$  & (2,1,1), AIC$=681.4$, BIC$=700.5$  \\
    Best-fit ARIMA parameters & $f_{t}'=(0.838  \pm 0.053)f_{t-1}'+\epsilon_{t}-$ & $f_{t}'=(1.108 \pm 0.039 )f_{t-1}'-(0.053 \pm 0.043)f_{t-2}'+  $  & $f_{t}'=( 1.341 \pm 0.023)f_{t-1}'-(0.435 \pm 0.019)f_{t-2}'  $  \\
  & $-(0.632 \pm 0.060 )\epsilon_{t-1}-(0.039 \pm      0.041)\epsilon_{t-2}-$ &  $-(0.160 \pm 0.031)f_{t-3}'+\epsilon_{t}-(0.932 \pm  0.030)\epsilon_{t-1}$           &  $+\epsilon_t-(0.958 \pm 0.013)\epsilon_{t-1}$       \\ 
   &  $-(0.060 \pm  0.036)\epsilon_{t-3}-(0.154 \pm      0.040)\epsilon_{t-4}$          &       &     \\
    \hline
    \end{tabular}  
    }
    \label{tab_ARIMA_results2}
\end{table}

\begin{figure}[h!]
    \centering
    \includegraphics[width=0.32\textwidth]{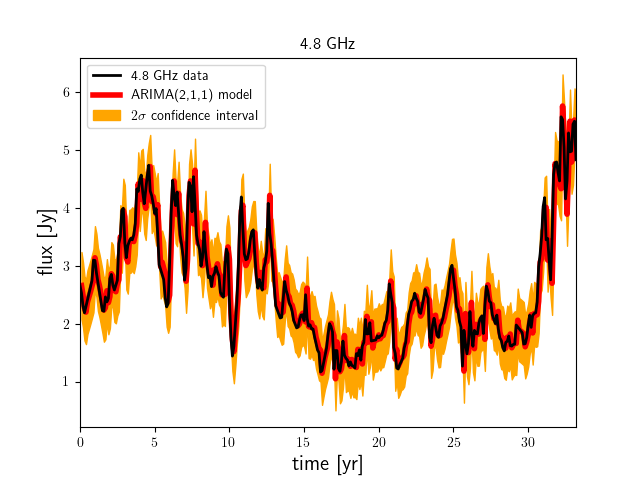}
     \includegraphics[width=0.32\textwidth]{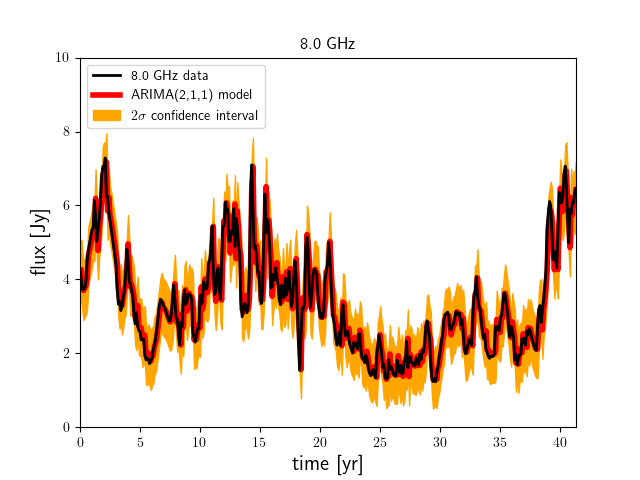}
      \includegraphics[width=0.32\textwidth]{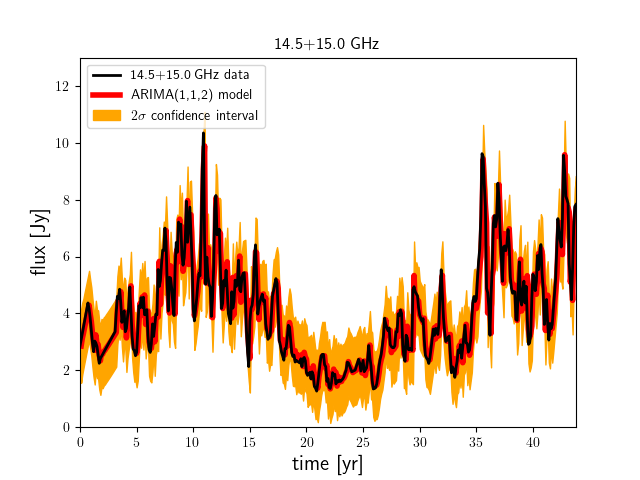} 
    \caption{ARIMA(2,1,1) model fitted to the 4.8 and 8.0 GHz light curves and ARIMA(1,1,2) model fitted to the 14.5+15.0 GHz light curve. All the light curves were initially interpolated using the regular time step of 0.1 years. The red solid line stands for the ARIMA mean model, while the orange region around it stands for the 2$\sigma$ confidence region.}
    \label{fig_ARIMA_fit}
\end{figure}

In Fig.~\ref{fig_ARIMA_fit}, we present ARIMA$(p,d,q)$ fits of preferred models to 4.8, 8.0, and 14.5+15.0 GHz light curves (from the left to the right panels; the regular time step is 0.1 years). For 4.8 and 8.0 GHz datasets, the preferred model based on the minimum BIC value is ARIMA(2,1,1), while for 14.5+15.0 GHz dataset, the preferred model is ARIMA(1,1,2).

\begin{figure}[h!]
    \centering
    \includegraphics[width=0.32\textwidth]{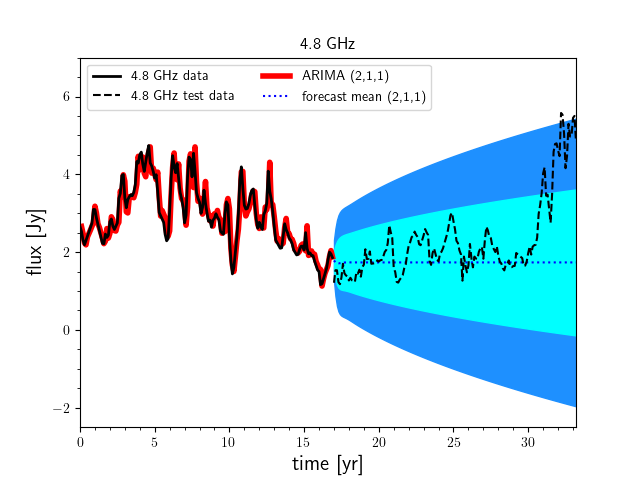}
    \includegraphics[width=0.32\textwidth]{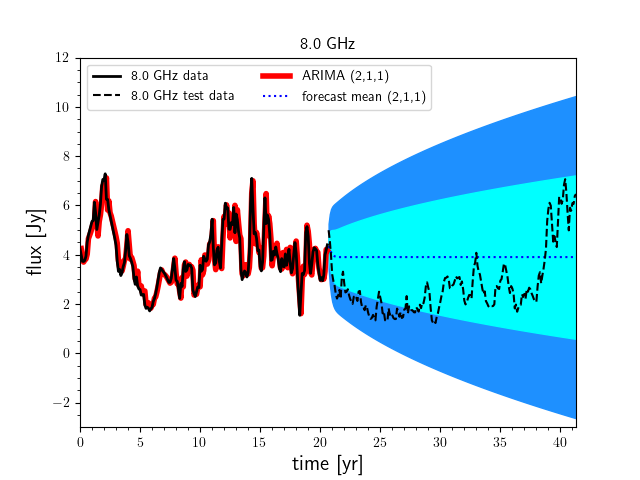}
       \includegraphics[width=0.32\textwidth]{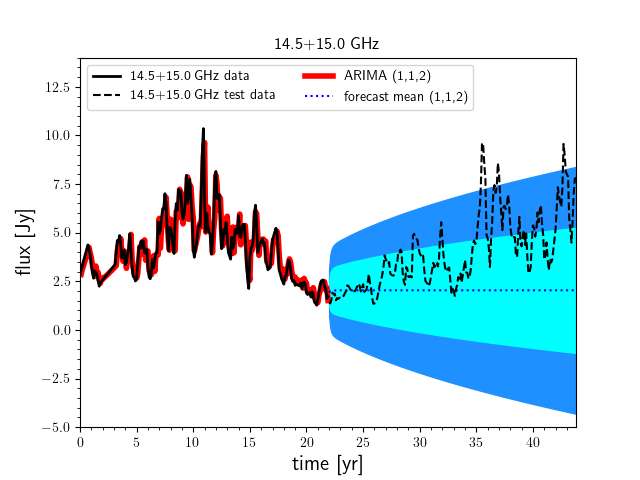}
    \caption{Forecasting ARIMA(2,1,1) models for 4.8 and 8.0 GHz light curves and ARIMA(1,1,2) for the 14.5+15.0 GHz light curve. All the light curves were initially interpolated using the regular time step of 0.1 years. The model was trained using approximately the first half of radio light curves (from left to right: 4.8, 8.0, and 14.5+15.0 GHz). The red solid line stands for the mean ARIMA model. The forecast mean value is represented by a blue dotted line, whereas the forecast 1$\sigma$ and 2$\sigma$ confidence regions are within cyan and dodgeblue regions, respectively.}
    \label{fig_ARIMA_forecast}
\end{figure}

In Fig.~\ref{fig_ARIMA_forecast}, we present forecasts of the trained ARIMA$(p,d,q)$ models. The models were first fit to an approximately first half of light curves and then the forecasts of the ARIMA means, 1$\sigma$ and 2$\sigma$ confidence intervals were performed for the second half of the light curves. The test light curve data are denoted by dashed lines. 

\begin{figure}[h!]
    \centering
    \includegraphics[width=0.32\textwidth]{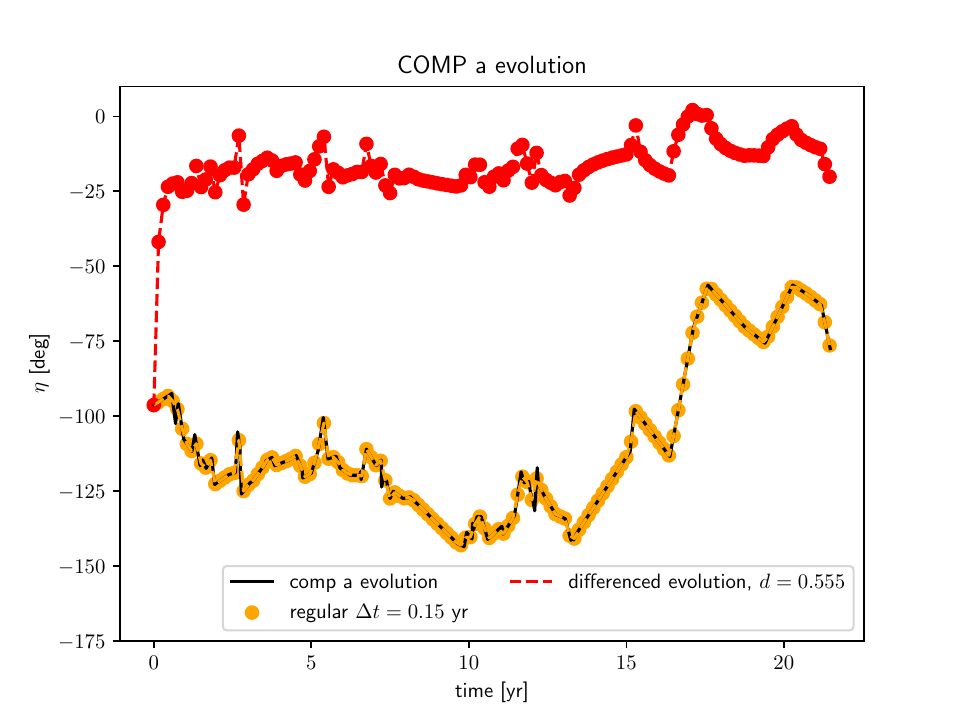}
    \includegraphics[width=0.32\textwidth]{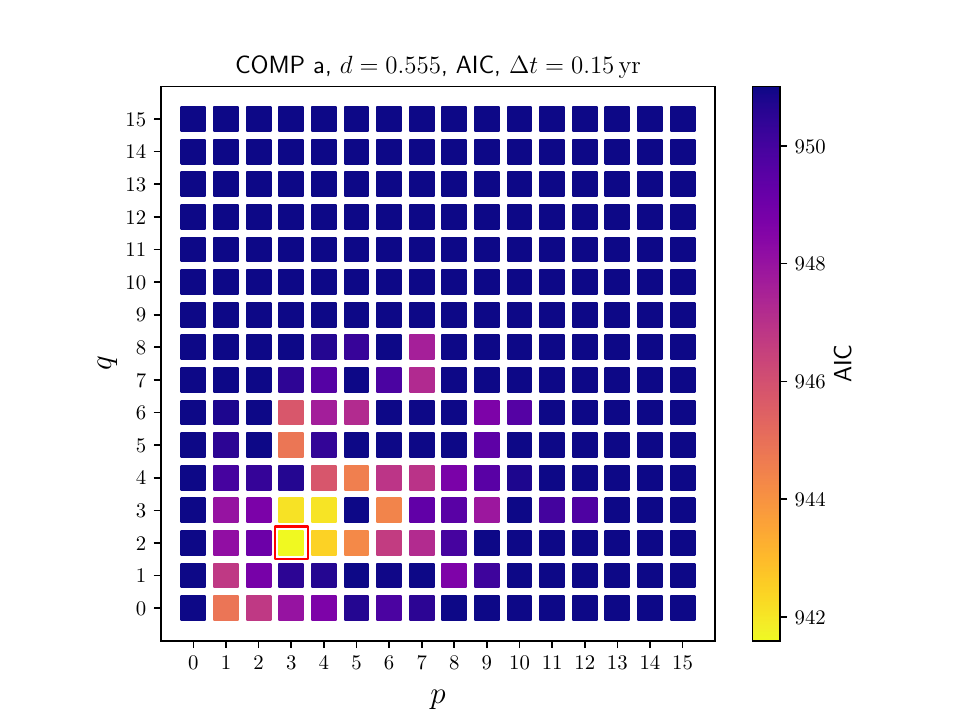}
    \includegraphics[width=0.32\textwidth]{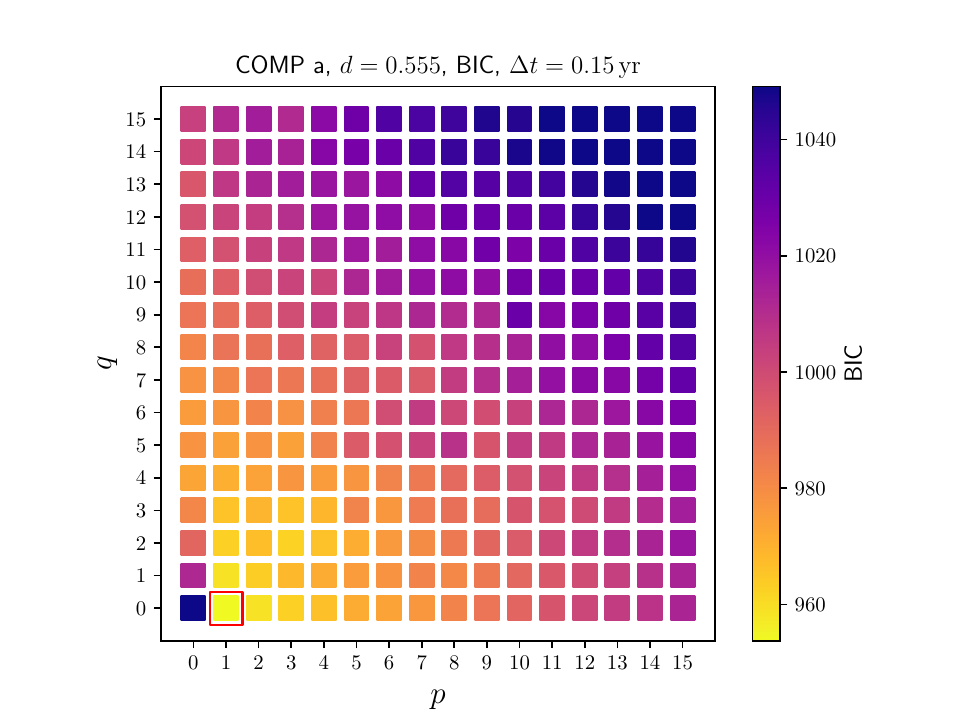}
    \caption{Temporal evolution of stationary component a. Left panel: Temporal evolution of component a (black line) is interpolated to the regular grid with the time step of $\Delta t=0.15$ years (orange points). The evolution is fractionally differenced with $d=0.555$ to reach the stationarity according to the ADF statistic. Middle panel: Distribution of AICs for ARIMA model fitting to the differenced evolution of the component a. The minimum AIC is reached for $p=3$ and $q=2$. Right panel: Distribution of BICs for ARIMA model fitting. The minimum value is reached for $p=1$ and $q=0$.}
    \label{fig_compa_AIC_BIC}
\end{figure}

In addition to the ARIMA modelling of the radio light curves, we apply the model to the position-angle evolution of the stationary component a, see Fig.~\ref{fig_compa_AIC_BIC}. The original evolution with 92 measurements is interpolated to the regular grid with the time step of $0.15$ years. This evolution is not stationary and we have to apply the fractional differencing with $d=0.555$ to remove the long-term trend, see the left panel of Fig.~\ref{fig_compa_AIC_BIC}. Subsequently, we fit a series of ARIMA(p,q) models to the differenced evolution, see the middle and the right panels of Fig.~\ref{fig_compa_AIC_BIC} for the AIC and BIC distribution. The minimum AIC value is reached for $p=3$ and $q=2$ and the minimum BIC value is reached for $p=1$ and $q=0$, i.e. just the autoregressive polynomial seems to be necessary. Overall, the $p$ and $q$ order is rather small, implying the timescale for the stationary stochastic process in the range of $\sim 0.15-0.45$ years.  

\begin{figure}[h!]
    \centering
    \includegraphics[width=0.45\textwidth]{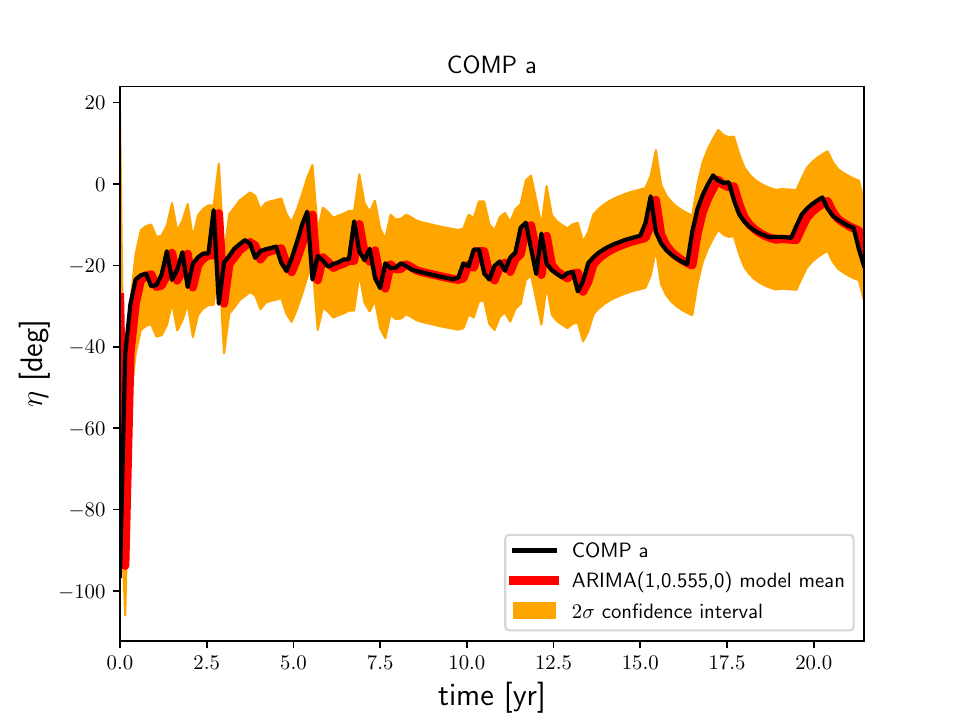}
    \includegraphics[width=0.45\textwidth]{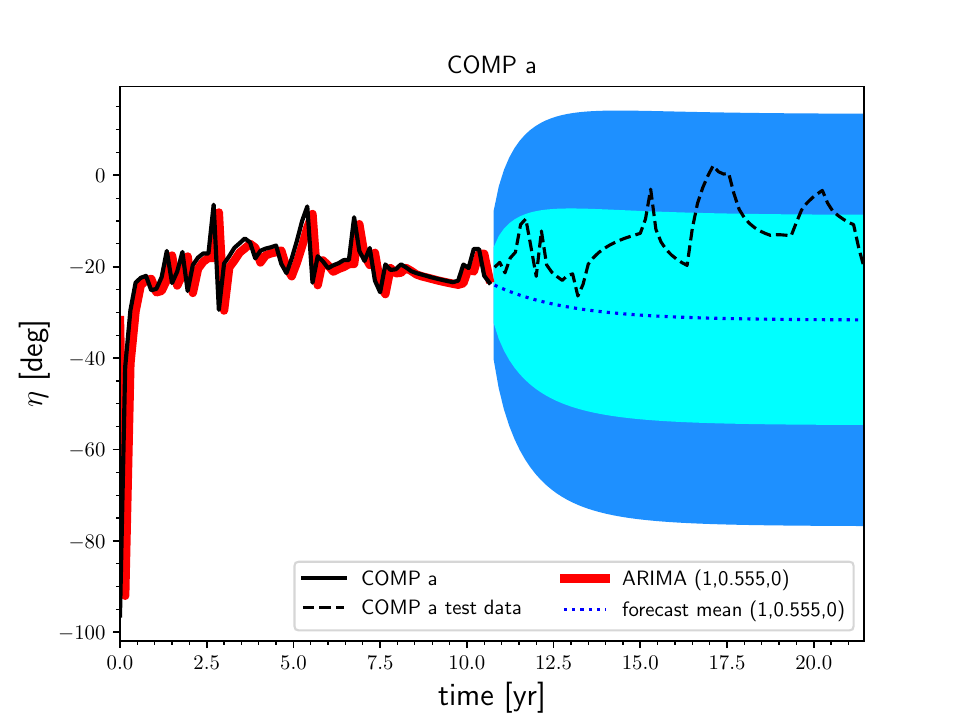}
    \caption{The optimal ARIMA model for the evolution for the stationary component a. Left panel: The optimal ARIMA model with $p=1$ and $q=0$ fitted to the differenced evolution of the component a. The red line stands for the model mean evolution while the orange shaded area represents the 2$\sigma$ confidence interval. Right panel: Forecast of the optimal ARIMA(1,0.555,0) model. The model is first fitted to the first half of the evolution and then forecast for the rest of the evolution. The forecast mean evolution deviates from the observed evolution.}
    \label{fig_compa_ARIMA_fit_forecast}
\end{figure}

We fit the ARIMA(1,0.555,0) model, which corresponds to the minimum BIC value, to the differenced temporal evolution of the position angle, see Fig.~\ref{fig_compa_ARIMA_fit_forecast}. The best-fit model is $\eta(t)= (0.963 \pm 0.020)\eta(t-1)+(-27.851 \pm 16.033)+\epsilon(t)$, i.e. having just the autoregressive term. Next, we first fit ARIMA(1,0.555,0) to the first half of the evolution and perform the forecast for the rest, see the right panel in  Fig.~\ref{fig_compa_ARIMA_fit_forecast}. The forecast mean clearly deviates from the measured position angles, while within $1\sigma$ confidence region, the measured detrended evolution is nearly consistent with the forecast model.

\end{document}